\pdfoutput=1
\documentclass{aastex62}

\newcommand{\beq}{\begin{equation}}           % begin equation
\newcommand{\eeq}{\end{equation}}             % end equation
\newcommand{\bfi}{\begin{figure}}           % begin figure
\newcommand{\efi}{\end{figure}}             % end figure

\shorttitle{Triple-Lens Microlensing}

\shortauthors{Dan\v{e}k \& Heyrovsk\'y}

\begin{document}

\title{Triple-lens Gravitational Microlensing: Critical Curves for Arbitrary Spatial Configuration}
	
\author[0000-0001-7081-3857]{Kamil Dan\v{e}k}
\affiliation{Institute of Theoretical Physics, Faculty of Mathematics and Physics, Charles University,\\ V Hole\v{s}ovi\v{c}k\'ach 2, 18000~Praha 8, Czech Republic}
\author[0000-0002-5198-5343]{David Heyrovsk\'y}
\affiliation{Institute of Theoretical Physics, Faculty of Mathematics and Physics, Charles University,\\ V Hole\v{s}ovi\v{c}k\'ach 2, 18000~Praha 8, Czech Republic}
\email{kamil.danek@utf.mff.cuni.cz, david.heyrovsky@mff.cuni.cz}

\begin{abstract}

Since the first observation of triple-lens gravitational microlensing in 2006, analyses of six more events have been published by the end of 2018. In three events the lens was a star with two planets; four involved a binary star with a planet. Other possible triple lenses, such as triple stars or stars with a planet with a moon, are yet to be detected. The analysis of triple-lens events is hindered by the lack of understanding of the diversity of their caustics and critical curves. We present a method for identifying the full range of critical curves for a triple lens with a given combination of masses in an arbitrary spatial configuration. We compute their boundaries in parameter space, identify the critical-curve topologies in the partitioned regions, and evaluate their probabilities of occurrence. We demonstrate the analysis on three triple-lens models. For three equal masses the computed boundaries divide the parameter space into 39 regions yielding nine different critical-curve topologies. The other models include a binary star with a planet, and a hierarchical star--planet--moon combination of masses. Both have the same set of 11 topologies, including new ones with doubly nested critical-curve loops. The number of lensing regimes thus depends on the combination of masses -- unlike in the double lens, which has the same three regimes for any mass ratio. The presented approach is suitable for further investigations, such as studies of the changes occurring in nonstatic lens configurations due to orbital motion of the components or other parallax-type effects.

\end{abstract}


\section{Introduction}
\label{sec:Intro}

The majority of observed Galactic gravitational microlensing events are caused by single-star lenses passing close to the line of sight to source stars in the Galactic bulge \citep{paczynski96}. Second in terms of frequency are events due to binary-star lenses \citep{schneider_weiss86}. These lead to a higher diversity of light-curve shapes, with particularly abrupt changes in flux occurring when the source star crosses the caustic of the lens.

Similar in their character are microlensing events with the lens consisting of a star with a planet. First detected in 2003 \citep{bond_etal04}, these events are the prime targets of the ongoing microlensing surveys, which have discovered 75 planets so far\footnote{As of 2019/04/30 at http://exoplanetarchive.ipac.caltech.edu}. The main appeal of microlensing as a planet-finding technique is that its sensitivity extends down to Earth-mass planets, it is highest for planets at astronomical-unit-scale separations from their host stars, and it is independent of host-star type.

The first observed event unambiguously caused by a triple lens was OGLE-2006-BLG-109 \citep{gaudi_etal08,bennett_etal10}, with the lens consisting of a host star and two planets. Two more microlensing events involving a star + two planets lens have been published so far: OGLE-2012-BLG-0026 \citep{han_etal13,beaulieu_etal16} and OGLE-2014-BLG-1722 \citep{suzuki_etal18}. In addition, four events involving a binary-star lens with a planet have been published: OGLE-2007-BLG-349 with a circumbinary planet \citep{bennett_etal16}, OGLE-2008-BLG-092 \citep{poleski_etal14}, OGLE-2013-BLG-0341 \citep{gould_etal14}, and OGLE-2016-BLG-0613 \citep{han_etal17b}.

The events with a binary-star or star+planet lens can be described by two-point-mass microlensing. The properties of this lens model have been studied in detail and are well understood \citep[e.g.,][]{schneider_weiss86,erdl_schneider93,witt_petters93,dominik99}. There are three lensing regimes with corresponding critical-curve topologies and caustic structures occurring for different separations of the point-mass lens components. The understanding of the variety and parameter dependence of possible light curves greatly facilitates the analysis of such events.

The analysis of triple-lens events is hindered by the lack of similar insight. The great diversity provided by the underlying three-point-mass lens model has not been explored fully yet: the overall range of different lensing regimes, critical-curve topologies, and caustic structures remains unknown. Note that the general triple lens includes not only the observationally detected star+two-planet or binary-star+planet systems but also lenses formed by two other types of systems: triple stars, and stars with a planet with a moon. Triple lenses thus also provide an interesting channel for potential exo-moon detection.

Much of the published theoretical work on triple lenses concentrates on the detectability of microlensing by specific types of lenses: stars with two planets \citep[e.g.,][]{gaudi_etal98,bozza99,han_etal01,ryu_etal11,song_etal14}, binary stars with a planet \citep[e.g.,][]{lee_etal08,han08a,luhn_etal16,han_etal17a}, or stars with a planet with a moon \citep[e.g.,][]{han_han02,han08b,liebig_wambsganss10}. In addition, the close and wide triple-lens limits were studied analytically by \cite{bozza00a,bozza00b}.

In our previous work we developed techniques for the analysis of general $n$-point-mass lenses based on the properties of the lens-equation Jacobian \citep{danek_heyrovsky15a}. In \cite{danek_heyrovsky15b} we applied these methods to simple two-parameter triple-lens models and mapped the critical curves and caustics in their respective parameter spaces. Based on these results, we extend the approach further in this work and map the different lensing regimes for a given combination of three masses in an arbitrary spatial configuration. In this way we may approach a similar level of understanding of triple lenses to that achieved by the earliest analyses of \cite{schneider_weiss86} for two-point-mass lenses.

We start in Section~\ref{sec:3-point} by reviewing the main properties of the triple lens, its critical curve, and the transitions between different critical-curve topologies. We introduce our parameterization of the spatial configuration of the lens in Section~\ref{sec:Parameter_choice}, choosing one parameter to define its size and two to define its shape. In Section~\ref{sec:Close-to-wide} we demonstrate how to map topology changes for a given configuration shape as a function of its size. This approach leads to the partitioning of the parameter space into regions with different critical-curve topologies. We describe the identification of the topologies within the individual regions in Section~\ref{sec:Mapping_topologies} and the computation of the relative probabilities of their occurrence in Section~\ref{sec:Probabilities}.

Our main results can be found in the next three sections, in which we analyze three triple-lens models with different combinations of masses: the Equal Masses model in Section~\ref{sec:EqM}, the Planet in Binary model in Section~\ref{sec:PiB}, and the Hierarchical Masses model in Section~\ref{sec:HiM}. For each model we present the boundary surfaces partitioning its parameter space, the sequence of critical-curve topologies from the close to the wide triple regime, and their relative probabilities. We summarize the results and discuss their implications in Section~\ref{sec:Summary_conclusions}.

\section{The Triple Lens and Its Critical Curve}
\label{sec:3-point}

We model the three components of the lens by point masses and assume the thin-lens approximation, in which all three lie at a single distance from the observer. Following \cite{witt90}, we identify the plane of the sky with the complex plane and mark positions in the plane by complex numbers. Using the angular Einstein radius of the total mass as a length unit, we can write the lens equation relating the position of a background source $\zeta$ and the position $z$ of its image as
\beq
\zeta=z-\sum^{3}_{j=1}{\frac{\mu_j }{\overline{z}-\overline{z_j}}}\,,
\label{eq:lenseq-3_complex}
\eeq
where $z_j$ and $\mu_j$ are the positions and fractional masses, respectively, of the individual components of the lens, and bars over variables denote their complex conjugation. The fractional masses are expressed in terms of the total mass; hence, they add up to unity, $\sum^{3}_{j=1}\mu _j=1 $.

Depending on the source position and lens configuration, Equation~(\ref{eq:lenseq-3_complex}) produces 4, 6, 8, or 10 images. Varying the position of a point-like source leads to pairs of images appearing and disappearing along the critical curve of the lens \citep[e.g.,][]{schneider_etal92}. In mathematical terms, the critical curve is the set of points $z_{\rm cc}$ at which the Jacobian of the lens Equation~(\ref{eq:lenseq-3_complex}) vanishes. Since the Jacobian can be expressed as
\beq
{\rm det}\,J\,(z)= 1-\left|\,\sum^{3}_{j=1}{\frac{\mu_j }{(z-z_j)^2}}\,\right|^{\,2},
\label{eq:Jacobian}
\eeq
the critical curve can be computed from
\beq
\sum^{3}_{j=1}{\frac{\mu_j }{(z_{\rm cc}-z_j)^2}}=e^{-2\,{\rm i}\,\phi}\, ,
\label{eq:critical-3point}
\eeq
where $\phi\in[0,\pi)$ is a phase parameter \citep{witt90,danek_heyrovsky15a}. The source positions corresponding to the critical curve form the caustic of the lens $\zeta_{\rm c}$, obtained by setting $z=z_{\rm cc}$ in Equation~(\ref{eq:lenseq-3_complex}).

The caustic consists of a set of closed cusped loops in the source plane, which may intersect or self-intersect\footnote{Intersections occur in $n$-point-mass lenses with $n\geq3$.}. The critical curve consists of the same number of closed smooth loops in the image plane. These loops may be separated, each enclosing a different region of the image plane, or some loops may be located (``nested'') inside other loops. The number of loops and their relative spatial positions define the topology of the critical curve. Transitions between topologies involve self-intersecting loops, as discussed further below. In all other cases the loops of the critical curve do not intersect.

The critical curve of a point-mass lens has one topology: a single loop forming the Einstein ring of the lens. The critical curve of the two-point-mass lens \citep{schneider_weiss86,erdl_schneider93} has three different topologies: an outer loop with two inner nested loops for small separations of the components (in the ``close'' regime), a single loop for intermediate separations (``intermediate'' regime), and two separate loops for large separations (``wide'' regime). Each of the topologies thus defines a particular lensing regime with its specific characteristics.

For a general $n$-point-mass lens, varying the lens parameters $z_j$ and $\mu_j$ alters the shape and location of the loops of the critical curve. Its topology remains unaffected, except when separate parts of the critical curve come into contact. A lens in such a special configuration has a critical curve with one or more self-intersecting loops. With a further change of parameters the self-intersections disappear, leading to the merger or splitting of the involved loops and, thus, a different topology of the critical curve.

The combinations of parameters at which such contact occurs define the topology boundaries in the parameter space. The special transitional topologies along the boundaries form a link between topologies of different lensing regimes and thus occupy lower-dimensional regions of the parameter space. In the following we do not discuss them further and concentrate purely on topologies occupying nonzero volume in the studied parameter space. Examples of transitional topologies for a simple triple-lens model can be seen in Figure 5 of \cite{danek_heyrovsky15b}.

The full set of topology boundaries partitions the parameter space into topology regions. All lenses with parameter combinations from such a region share the same critical-curve topology. For the two-point-mass lens there are three topology regions, each with a unique topology. However, in the case of the triple lens a given topology may correspond to one or more separate topology regions within the full parameter space, as shown in \cite{danek_heyrovsky15b} and in the present work. Hence, the number of different topologies of a given lens model is always less than or equal to the number of topology regions.

Classification of critical-curve topologies and the mapping of their topology regions in parameter space thus starts from identifying the topology boundaries. The mathematical condition defining these topology-changing lens configurations requires the critical curve to pass through a saddle point of the Jacobian \citep[e.g.,][]{schneider_weiss86}. Saddle points are stationary points of the Jacobian that have a negative Hessian, the determinant of second derivatives \citep[see Section 3.2 of][]{danek_heyrovsky15a}. The corresponding condition $\partial^2_z\bar{\zeta}=0$ can be combined with Equation~(\ref{eq:lenseq-3_complex}) to yield the saddle-point equation
\beq
\sum^{3}_{j=1}{\frac{\mu_j }{(z-z_j)^3}}=0\,,
\label{eq:saddle-3point}
\eeq
which can be converted to a polynomial of degree 6 with roots $z=z_{\rm sadd}$. As discussed by \cite{danek_heyrovsky15a}, the number of different saddles may be lower in special cases, since some of the solutions of Equation~(\ref{eq:saddle-3point}) may represent higher-order maxima rather than saddles. In addition, the polynomial obtained from Equation~(\ref{eq:saddle-3point}) may have multiple roots, which correspond to higher-order saddle points\footnote{As shown by \cite{danek_heyrovsky15b}, triple-lens Jacobians may have monkey saddles, at which three loops of the critical curve connect.}. As a result, the three-point-mass lens has at most six different saddle points. The topology boundaries then consist of sets of lens parameters that allow a common solution $z = z_{\rm cc}=z_{\rm sadd}$ of Equations~(\ref{eq:critical-3point}) and (\ref{eq:saddle-3point}).

Several approaches to solve the problem were summarized in \cite{danek_heyrovsky15a,danek_heyrovsky15b}. In this work we analyze critical-curve topologies by exploiting the Jacobian contour correspondence pointed out by \cite{danek_heyrovsky15a} for lenses consisting of $n$ point masses. According to the correspondence, a positive-value contour of the Jacobian has the shape of the critical curve of a lens with the same components placed closer together. A negative-value contour has the shape of the critical curve of a lens with the same components placed farther apart. The curves differ only in their scale: in the former case the contour is larger than the corresponding critical curve; in the latter case the contour is smaller than the corresponding critical curve. The scaling of this correspondence can be expressed by
\beq
z_\lambda(\mu_j,z_j)=z_{\rm cc}(\mu_j,z_j\sqrt[4]{1-\lambda})\,/\,\sqrt[4]{1-\lambda}\,,
\label{eq:correspondence}
\eeq
where $z_\lambda(\mu_j,z_j)$ is the ${\rm det}\,J=\lambda$ contour and $z_{\rm cc}(\mu_j,z_j)$ is the critical curve of a lens with fractional masses $\mu_j$ and positions $z_j$. The full set of contours $\lambda\in(-\infty,1)$ thus illustrates the critical curves of all similar rescaled lens configurations with component separations ranging from $\infty$ to $0$.

For our purposes this property can be utilized as follows: if we identify the Jacobian contour passing through a given saddle point by setting $z=z_{\rm sadd}$ in Equation~(\ref{eq:Jacobian}), we can find a lens configuration on a topology boundary by shifting the component positions \mbox{$z_j\to z_j\sqrt[4]{1-{\rm det}\,J(z_{\rm sadd})}$}. For a given shape of the triangle defined by the lens components with given masses we can thus easily determine all rescaled configurations that lie on topology boundaries. Since there are at most six different saddle points, there will be at most six different Jacobian contours passing through them, taking into account that several saddle points may lie on the same contour. As a result, the critical curve of a given triple-lens shape may undergo up to six topology changes as we gradually reposition the components from the close regime (resembling a single combined-mass lens) to the wide regime (leading asymptotically to three independent lenses).

\section{Parameter Choice}
\label{sec:Parameter_choice}

The static triple lens has eight degrees of freedom. In the notation of Section~\ref{sec:3-point} these are, for example, the real and imaginary parts of the positions of the three components $z_j$, plus any two of the fractional masses $\mu_j$. In this work we concentrate on the topology of the critical curve. Therefore, we may reduce the number of free parameters to five by setting the origin of the complex plane (e.g., at the centroid of the configuration) and its orientation. The conversion to such a preferred frame is described in Appendix~\ref{sec:Appendix-conversion-preferred}.

Instead of attempting to visualize topology boundaries in a 5D parameter space, we fix the two fractional masses and explore the dependence on the three spatial configuration parameters in detail. Such a separation is natural, since one may concentrate on astrophysically different types of triple lenses (triple stars, binary stars with a planet, stars with two planets, etc.). Moreover, a significant fraction of the microlensing events with planets or multiple stars involves their orbital motion and parallax effects, both of which can be described by changing the positions of fixed-mass components. From the numerical point of view, mapping critical-curve topologies in a 3D parameter space as described below is sufficiently robust to permit determination of the full range of topologies that may occur for the given combination of masses.

The configuration of the lens can be described by the properties of the projected triangle formed by the components in the plane of the sky. In order to utilize the Jacobian contour correspondence described above, it is convenient to use one scaling parameter to describe the size and two parameters to describe the shape of the triangle. We chose the perimeter $p$ of the triangle as the scaling parameter. In the limit $p\to 0$ all components approach a single point, while in the $p\to\infty$ limit at least one of the components is asymptotically separated from the others. Note that since $p$ is the perimeter in units of the total Einstein radius, increasing $p$ may involve increasing the physical separations of the components or decreasing the total Einstein radius -- and vice versa. Of our five parameters, $p$ is the only one affected by changes in the total lens mass (with fixed fractional masses) and the line-of-sight distances to the lens and source.

The most intuitive way to characterize the shape of the triangle would be to use two of its angles. Unfortunately, such a parameterization fails to distinguish between different collinear configurations with the third component positioned along the line connecting the first two \citep{danek_heyrovsky15b}. All such configurations would correspond to the same set of angles $\{0, 0, \pi\}$. Instead, we use any two lengths of sides $a$, $b$, $c$ of the triangle expressed as fractions of the perimeter: $a_p\equiv a/p$, $b_p\equiv b/p$, $c_p\equiv c/p$. Note that all three are connected by $a_p+b_p+c_p=1$. Due to the triangle inequality, each of the fractional side lengths ranges from 0 to 1/2. As a consequence, the sum of any two fractional side lengths ranges from 1/2 to 1. When one of the side lengths is equal to 1/2, the triangle is reduced to a collinear configuration with the two outer components separated by 1/2 and the position of the third component specified by the other two ``side'' lengths. When two of the side lengths are equal to 1/2, the third is equal to 0, so that the positions of two components coincide and the triple lens is effectively reduced to a two-point-mass lens. The conversion between these triangular parameters and the preferred-frame component positions is described in Appendix~\ref{sec:Appendix-conversion-triangular}. Their relation to parameters typically used in the analysis of triple-lens microlensing events can be found in Appendix~\ref{sec:Appendix-conversion-microlensing}.

In order to visualize the division of the 3D parameter space, we may use projected plots of the boundary surfaces, or --- for more detail --- a series of 2D sections. In our choice of parameters we use $p={\rm const.}$ sections, with $p\in(0,\infty)$. Such a section represents triangles of all different shapes with fixed perimeter $p$. In order to retain the full symmetry of the model, we use ternary plots to depict these sections. The ternary plot is an equilateral-triangle-shaped plot with three slanted axes representing three variables (in our case $a_p$, $b_p$, $c_p$) that add up to unity. In most applications, each variable runs from 0 to 1, with 0 corresponding to a side of the plot and 1 to the opposite vertex. However, in our case each variable runs from 0 to 1/2. The axis labeling that maintains the unit sum in this case runs from 0 at a vertex to 1/2 at the opposite side of the plot.

As shown in Figure~\ref{fig:LittleTriangles}, any point in such a ternary plot unambiguously determines the shape of the lens configuration, with all shapes being represented. Each of the small red triangles illustrates the shape corresponding to the ternary point at its centroid. Sides $a$, $b$, and $c$ denote the right, left, and bottom sides of the small triangles, respectively. We labeled the ticks on the axes in twelfths in order to include the equilateral configuration (the central point of the plot with $a_p=b_p=c_p=1/3=4/12$), as well as the axis limits ($1/2=6/12$), and to provide better orientation than if we just used sixths. The medians of the ternary plot correspond to isosceles triangles, the sides correspond to collinear configurations, and the vertices correspond to two-point-mass lenses (i.e., degenerate triple lenses with two components coinciding). By adding the perimeter $p\in(0,\infty)$ as an axis perpendicular to the ternary plot, we represent the parameter space of a triple lens with fixed masses by a semi-infinite ternary prism. This can be seen in the figures starting from the top left panel of Figure~\ref{fig:EqMassSeq} and Figure~\ref{fig:EqMassSur}.

\section{Tracking Critical-curve Changes from Close to Wide Regime}
\label{sec:Close-to-wide}

For a given combination $a_p$, $b_p$ defining the shape of the triple-lens configuration, we compute $c_p=1-a_p-b_p$ and use Equation~(\ref{eq:abp-to-zi}) to compute the component positions for a unit perimeter $p=1$. We then identify the Jacobian saddle points from Equation~(\ref{eq:saddle-3point}) and compute the corresponding perimeter values $p=\sqrt[4]{1-{\rm det}\,J(z_{\rm sadd})}$ at the topology boundaries. As described in Section~\ref{sec:3-point}, generally there are six such values for a given ternary point, although the number may be lower --- mostly in symmetric configurations. There may be as few as two values, e.g., for three equal masses in an equilateral-triangle configuration \citep[TE model with mass parameter $\mu=1/3$ in][]{danek_heyrovsky15b} or for a linear configuration with two equal masses symmetrically bracketing a third component with sufficiently large mass \citep[LS model with mass parameter $\mu=1/9$ or $\mu\geq0.2$ in][]{danek_heyrovsky15b}. Varying the perimeter of a triple lens with a fixed configuration shape may thus yield from three to seven different topologies of the critical curve. For comparison, note that the two-point-mass lens has two boundaries separating three topologies for any mass ratio \citep{erdl_schneider93}.

We illustrate the sequence of topology changes with perimeter in Figure~\ref{fig:EqMassSeq} on the example of a lens with equal-mass components and fractional sides $(a_p$, $b_p$, $c_p)=(0.25$, $0.35$, $0.4)$. The positions of the lens components are marked by crosses in the image plane in the bottom left panel, with axes marked in units of the perimeter. The vertical dot-dashed line in the ternary prism in the top left panel corresponds to configurations with perimeters $p\in[\,0,8\,]$. Its intersections with the small colored polygons indicate the six topology boundaries at $p=\{1.776$, $1.917$, $2.519$, $3.471$, $4.454$, $6.346\}$. The seven configurations marked by bullets at $p=\{1.50$ ,$1.86$, $2.22$, $3.00$, $4.00$, $5.40$, $8.00\}$ lie in the intervals defined by the intersections.

The blue curve in the bottom left panel is the critical curve for the lowest bullet, $p=1.50$. It has the typical close-triple topology of an outer loop, which corresponds to the total-mass Einstein ring for $p\to 0$, surrounding four small inner loops (the smallest central loop is barely visible). The first six black Jacobian contours inward from the outer loop each pass through one of the Jacobian saddle points (marked by plus signs). They represent the transitional critical curves corresponding to the six intersection perimeters from the top left panel. The added innermost contour shows the critical curve in the wide-triple limit with $p=14$; the three contours added outward from the blue outer loop with $p=\{1.25,\,1.06,\,0.92\}$ are the outer loops of close-triple critical curves. Their inner loops are nested inside the small blue loops.

The sequence of the seven different critical-curve topologies corresponding to the bullets in the top left panel is shown in the right panels together with the corresponding caustics. The critical curves (blue) are plotted over the six transitional contours (black) from the bottom left panel. The axes are marked similarly in units of the perimeter, so that the lens components stay at the same positions in the seven plots. In the case of the caustics (red), however, the axes are marked in units of Einstein radii in order to include all loops of the caustic. Their tick spacing is $0.5$, and the major ticks indicate the position of the lens configuration centroid.

Going from $p=1.50$ to $p=1.86$, the bottom inner loop of the critical curve connects with the outer loop, so that the topology has one outer loop surrounding three inner loops. Increasing the perimeter further to $p=2.22$, the left inner loop connects with the outer loop, leading to an outer-plus-two-inner-loops topology. Next, the right inner loop connects with the outer, as seen in the $p=3.00$ outer-plus-one-inner-loop topology. By $p=4$ the last inner loop has connected with the outer loop, forming a single-loop topology. The final two changes correspond to the detachment of loops corresponding to the individual lens components. In this case first the left component splits off, with a two-separate-loops topology for $p=5.40$. Finally, the two right components split, leading to the wide-triple topology of three separate loops as seen for $p=8.00$. Asymptotically, for $p\to\infty$, these loops correspond to the Einstein rings of three independent lenses.

The changes in the caustics can be seen in the right column, with loops of the caustic gradually connecting and disconnecting in step with the changes in the critical curve. Unlike in the case of the critical curve, loops of the caustic may self-intersect (here in all seven cases), separate loops may overlap (here in the $p=3.00$ case), and pairs of cusps may appear and disappear by additional caustic metamorphoses that do not affect the critical-curve topology.

The number of cusps on a particular loop of the caustic can be found by counting the number of intersections of the corresponding loop of the critical curve with the orange cusp curve \citep{danek_heyrovsky15a} in the bottom left panel. For example, for the blue $p=1.50$ critical curve we see that all four small inner loops have three intersections each. Hence, the corresponding loops of the caustic have three cusps each. The outer loop of the critical curve has six intersections; thus, the corresponding central loop of the caustic has six cusps. The generic close-triple form of the central loop with four cusps is reached for a lower value of the perimeter. As indicated by the cusp curve, a critical curve for $p$ slightly lower than the plotted $1.06$ contour would not intersect the top-right-corner branch of the cusp curve, leaving only four intersections on the outer loop. Two of the cusps on the central loop of the caustic would then disappear by a reverse swallow-tail metamorphosis. For more on the cusp curve and triple-lens caustic metamorphoses see \cite{danek_heyrovsky15a,danek_heyrovsky15b}.

\section{Mapping Critical-curve Topologies in the Parameter Space}
\label{sec:Mapping_topologies}

The example shown in Figure~\ref{fig:EqMassSeq} illustrates the sequence of topologies for a single point on the ternary-plot base, i.e., a single triangular shape of the equal-mass lens configuration. However, other points and other mass combinations may have a different sequence of topologies from the close to the wide regime, as shown by \cite{danek_heyrovsky15b}. By performing the same analysis for all points in our ternary grid discussed above, we obtain the six boundary perimeter values varying as a function of ternary plot position. As a result, they form six boundary surfaces $p\,(a_p, b_p)$ vertically dividing the ternary prism. Ordered by perimeter values from smallest to largest, these continuous surfaces define the sequence of transitions starting from the boundary above the close-triple regime and ending with the boundary below the wide-triple regime.

However, the surfaces are not smooth at their intersections, since they retain their sequential ordering. In order to identify individual smooth surfaces through their intersections self-consistently, we find their permutation minimizing the absolute values of second derivatives of $p$ with respect to $a_p$ or $b_p$ for each surface. This step is important for identifying the 3D regions of parameter space corresponding to different topologies. Hence, in addition to the boundary values, we keep track of the permutations maintaining smooth surfaces. Note that the identification is complicated by the fact that instead of involving six independent surfaces, the boundaries are formed by a lower number of self-intersecting surfaces. Therefore, the ordering has to be checked carefully.

We map the topology boundaries in the parameter space using a triangular part of a $1000\times1000$ grid in ternary coordinates $(a_p,\,b_p)$ with limits defined by conditions for the sides of a triangle, $a_p+b_p \geq 1/2$, $a_p \leq 1/2$, $b_p \leq 1/2$. For each grid point $(a_p,\,b_p)$ we identify the vertical sequence of six perimeter values corresponding to the boundaries, as described in Section~\ref{sec:Close-to-wide}. We then scan the triangular part of the $1000\times1000\times6$ matrix of perimeter values and construct another triangular part of a $1000\times1000\times6$ matrix storing the permutation sequences identifying continuous surfaces as described above.

Having found the boundaries, the next step is to identify all 3D regions of the ternary-prism parameter space partitioned by them. Based on the combined geometry of the boundary surfaces, we introduce a finely spaced vertical sequence of perimeter values covering our region of interest. Each value defines a constant-perimeter ternary-plot section, on which we perform a horizontal identification of all boundary-separated areas. We start with one point and identify its 3D region by checking its preceding vertical neighbor and the matrix of boundary values. If any of the boundary values are found to lie vertically between the two points (or if we are analyzing the lowest section), the point is assumed to lie in a new 3D region. The horizontal mapping then gradually proceeds to neighboring points using a six-way flood-fill algorithm. For a given neighbor we check boundary positions to make sure it does not cross into another region, and we test for changes in boundary permutation in order to avoid flooding adjacent regions through boundary-surface intersections. Once the continuous area corresponding to the 3D region with the starting point is determined on the horizontal section, another starting point outside the area is selected. The procedure is repeated until regions are mapped for all points at the given perimeter value.

The same analysis is then carried out for the next perimeter value. After finishing the last value in the sequence, a final vertical check of adjacent 3D regions is performed. At this stage the explored parameter space is fully partitioned into a set of disjoint 3D regions. Since several nonneighboring regions may correspond to the same critical-curve topology, a sample parameter combination is used from each of the 3D regions to identify its topology. This final step completes the full 3D mapping of all critical-curve topologies occurring within the explored perimeter interval for an arbitrary spatial configuration of the studied combination of components.

As a final note, the described algorithm may also be directly used for volume integration in parameter space, for example, for comparing the rates of occurrence of different critical-curve topologies.

\section{Probability of Topology Occurrence}
\label{sec:Probabilities}

Using the ternary plot in terms of dimensionless side lengths has the disadvantage that it lacks an intuitive means for assessing the relative frequency of occurrence of different topologies. Due to its construction, there is no clear comparison between individual $p={\rm const.}$ sections. Moreover, even within a single section the probability of occurrence of a given topology does not scale directly with the area it occupies within the plot.

As a first step, we have to define what we mean exactly by probability of occurrence. To obtain a measure of relevance of the different critical-curve topologies, we compute the corresponding volumes in lens-component position space $\{z_1,\,z_2,\,z_3\}$. Such volumes directly yield the probability of occurrence of a given topology if all three lens components are randomly positioned with a homogeneous uncorrelated probability distribution in some region of the image plane.

In order to assess the relative probabilities within a given $p={\rm const.}$ section, we integrate over all configurations with perimeter $p$. The probability of occurrence of topology ${\rm T}_i$ for configurations with perimeter $p$ is
\beq
\mathcal{P}_{{\rm T}_i}(p)=\frac{\int_{[{\rm T}_i](p)}{\rm d Re}(z_1)\,{\rm d Im}(z_1)\,{\rm d Re}(z_2)\,{\rm d Im}(z_2)\,{\rm d Re}(z_3)\,{\rm d Im}(z_3)} {\int_{[all\,{\rm T}'s](p)}{\rm d Re}(z_1)\,{\rm d Im}(z_1)\,{\rm d Re}(z_2)\,{\rm d Im}(z_2)\,{\rm d Re}(z_3)\,{\rm d Im}(z_3)}\;,
\label{eq:probability_p_positions}
\eeq
where the perimeter $p=|z_3-z_2|+|z_1-z_3|+|z_2-z_1|$. We can simplify the integration if we transform to our ternary-prism parameters by introducing the following set of six integration variables: the real and imaginary parts of the centroid position $z_{\rm ctr}=(z_1+z_2+z_3)/3$, the angle $\chi=\arg(z_2-z_1)$ subtended by $z_2$ from $z_1$ in the complex plane, perimeter $p$, and fractional side lengths $a_p$ and $b_p$. In terms of the positions of the components, the side lengths are $a=|z_3-z_2|$, $b=|z_1-z_3|$, and $c=|z_2-z_1|$.

The remarkably simple Jacobian of this transformation\footnote{Not to be confused with the lens-equation Jacobian discussed elsewhere in this article.} is
\beq
\frac{a\,b\,c\,p^2}{2\,S_{abc}}=\frac{2\,a_p\,b_p\,c_p\,p^3}{\sqrt{(1-2a_p)(1-2b_p)(1-2c_p)}}\,,
\label{eq:JacABC}
\eeq
where we used Heron's formula for the area $S_{abc}$ of the triangle in terms of the lengths of its sides. The dependence of this Jacobian on the shape of the triangle illustrates the relative weights of different regions of the ternary plot in the computed probability. The Jacobian has a minimum at the equilateral configuration at the center of the ternary plot ($a_p=b_p=c_p=1/3$) and diverges at the sides of the ternary plot ($a_p\to1/2$, $b_p\to1/2$, $c_p\to1/2$), except for vertices. In a vertex, the limit of the Jacobian depends on the direction of approach: along the sides it is infinite, but from within the plot it has a finite value that depends on the angle of approach.

In the new variables the probability from Equation~(\ref{eq:probability_p_positions}) simplifies to
\beq
\mathcal{P}_{{\rm T}_i}(p)=\frac{2\,p^3\int_{[{\rm T}_i](p)}a_p\,b_p\,c_p\,\left[(1-2a_p)(1-2b_p)(1-2c_p)\right]^{-1/2}\,{\rm d} b_p\,{\rm d} a_p} {2\,p^3\int_{0}^{1/2}\int_{1/2-a_p}^{1/2}a_p\,b_p\,c_p\,\left[(1-2a_p)(1-2b_p)(1-2c_p)\right]^{-1/2} \,{\rm d} b_p\,{\rm d} a_p}\;,
\label{eq:probability_p_ternary}
\eeq
where the integration over centroid position and orientation angle $\chi$ canceled out, since the topology does not depend on them. Moreover, note that the probability depends on the perimeter only through the ternary area occupied by the given topology. The integral in the denominator can be performed analytically, yielding the final expression
\beq
\mathcal{P}_{{\rm T}_i}(p)=\frac{84}{\pi}\int_{[{\rm T}_i](p)}\frac{a_p\,b_p\,(1-a_p-b_p)}{\sqrt{(1-2a_p)(1-2b_p)(2a_p+2b_p-1)}}\,{\rm d} b_p\,{\rm d} a_p\,,
\label{eq:probability_p_final}
\eeq
which we can simply integrate over the area in the ternary plot occupied by topology ${\rm T}_i$, using the flood-fill algorithm discussed in Section~\ref{sec:Mapping_topologies}.

In order to estimate the overall occurrence of a given topology, we define a similar probability integrated over perimeters from zero to some maximum value $p_{\rm max}$. Note that the weight given by Equation~(\ref{eq:JacABC}) expressed in terms of the dimensionless $a_p$, $b_p$, $c_p$ is proportional to the third power of the perimeter $p$. This reflects the fact that by increasing the perimeter of the configuration the volume in $z_j$-space increases accordingly. While the choice of $p_{\rm max}$ is arbitrary, if we set it too large, the asymptotic combination of single- and two-point-mass-lens topologies would dominate. For this reason, in the models discussed further below we cut off the integration at the (model-specific) perimeter value at which the last nonasymptotic critical-curve topology disappears.

The probability of occurrence of topology ${\rm T}_i$ in lens configurations with perimeter $p\in(0,\,p_{\rm max})$ can be expressed from Equation~(\ref{eq:probability_p_ternary}) as follows:
\beq
\mathcal{P}_{{\rm T}_i,p_{\rm max}}=\frac{\int_{0}^{p_{\rm max}}\,p^3\int_{[{\rm T}_i](p)}a_p\,b_p\,c_p\,\left[(1-2a_p)(1-2b_p)(1-2c_p)\right]^{-1/2}\,{\rm d} b_p\,{\rm d} a_p\,{\rm d} p} {\int_{0}^{p_{\rm max}}\,p^3\int_{0}^{1/2}\int_{1/2-a_p}^{1/2} a_p\,b_p\,c_p\,\left[(1-2a_p)(1-2b_p)(1-2c_p)\right]^{-1/2}\,{\rm d} b_p\,{\rm d} a_p\,{\rm d} p}\;.
\label{eq:probability_pmax_ternary}
\eeq
Evaluating the integral in the denominator and expressing the result in terms of the perimeter-dependent probability $\mathcal{P}_{{\rm T}_i}(p)$ from Equation~(\ref{eq:probability_p_final}), we get
\beq
\mathcal{P}_{{\rm T}_i,p_{\rm max}}=\frac{4}{p_{\rm max}^4}\int_{0}^{p_{\rm max}}p^3\,\mathcal{P}_{{\rm T}_i}(p)\,{\rm d} p\,.
\label{eq:probability_pmax_final}
\eeq
To compute the integral, we use again the flood-fill algorithm described in Section~\ref{sec:Mapping_topologies}.

We note here that while the computed probabilities provide a measure of occurrence of a given critical-curve topology, they are not equal to probabilities of observing a microlensing event by a triple-lens system with the given topology. It is beyond the scope of this work to evaluate actual probabilities of detection or the actual frequency of three-point-mass lens configurations based on further physical considerations.

\section{Model 1: Equal Masses}
\label{sec:EqM}

We first study a triple lens with equal-mass components ($\mu_1=\mu_2=\mu_3=1/3$). The model is symmetric to all permutations of the components, and thus we may expect this symmetry to be reflected in the division of the ternary-prism parameter space.

\subsection{Boundary Surfaces}
\label{sec:EqM_surfaces}

The boundary surfaces $p\,(a_p, b_p)$ computed following Sections~\ref{sec:Close-to-wide} and \ref{sec:Mapping_topologies} are shown in Figure~\ref{fig:EqMassSur}. Parts of the intersecting surfaces are colored in order of their perimeter values from the closest (purple) to the widest (red) transition. For better orientation, an animated version showing the prism rotated about the vertical axis is available in the online version of this article.

In order to explore the overall structure, we start from important lines and planes of the prism. Along the vertical edges two of the components coincide so that the triple lens is reduced to a two-point-mass lens with fractional masses $1/3$ and $2/3$. The purple and blue surfaces both intersect these edges at the perimeter corresponding to the close/intermediate transition, $p=2\,\left[\sqrt[3]{1/3}+\sqrt[3]{2/3}\right]^{-3/4}\approx 1.428$, where we used Equation~(21) from \cite{danek_heyrovsky15b}. Similarly, the cyan surface intersects these edges at the perimeter corresponding to the wide/intermediate transition, $p=2\,\left[\sqrt[3]{1/3}+\sqrt[3]{2/3}\right]^{3/2}\approx 3.923$, where we used Equation~(20) from \cite{danek_heyrovsky15b}.

The front vertical face corresponds to collinear configurations with component 3 positioned along the line from 1 to 2, so that $c_p=1/2$. The other vertical faces correspond to collinear configurations with $a_p=1/2$ (left rear) and $b_p=1/2$ (right rear). Such configurations correspond to the linear asymmetric (LA) model studied in detail by \cite{danek_heyrovsky15b}. With an added subscript LA the parameters of the model are the separation parameter $s_{\rm LA}=p/4$ and the fractional position parameter $p_{\rm LA}=2\,a_p=1-2\,b_p$ for the front face. Vice versa, our perimeter $p=4\,s_{\rm LA}$, and fractional side lengths $a_p=p_{\rm LA}/2$ and $b_p=(1-p_{\rm LA})/2$ for the front face.

The boundary intersections with the front face are shown in the left panel of Figure~\ref{fig:EqMassSur-planes}, which is adapted from the parameter-space division in the left panel of Figure~8 in \cite{danek_heyrovsky15b}. The lines corresponding to intersections of different surfaces from Figure~\ref{fig:EqMassSur} are plotted in their respective colors. Degenerate intersections shared by two different surfaces are indicated by two-color-dashed lines. The topology boundaries consist of two pairs of curves rising symmetrically from one edge to the other. One pair starts from the wide/intermediate binary transition at the edges and is formed gradually by the cyan, orange, and red surface intersection with the face. The second pair starts from the close/intermediate binary transition at the edges and is formed gradually by the purple/blue, cyan/green, and green/orange pairs of surfaces meeting at the face. All these degeneracies break down with the slightest perturbation inward from the face, in this case to $c_p<1/2$, when the configuration forms a nearly collinear triangle.

As discussed by \cite{danek_heyrovsky15b}, due to the symmetry of these configurations, there are only four distinct boundaries for most positions of the central component (here values of $a_p$), with three exceptions. For the most symmetric case with the third component placed at the midpoint between the other two ($a_p=0.25$ along the vertical midline of the face), only two boundaries are encountered between the close and wide topology limits. This is one of the configurations with the minimum number of topologies mentioned above in Section~\ref{sec:Close-to-wide}. The first transition occurs at the quadruple intersection of the purple, blue, cyan, and green surfaces at $p\approx 2.184$, and the second occurs at the orange and red intersection at $p\approx 6.711$. The other two exceptions correspond to the symmetric configurations with $a_p\approx 0.1197$ and $a_p\approx 0.3803$. Increasing the perimeter, either of these undergoes three transitions each, passing through the triple intersection of the cyan, green, and orange surfaces at $p\approx 4.672$. In terms of the Jacobian, the number of colors meeting at a given point indicates the number of Jacobian saddle points lying on the critical curve of the corresponding configuration. The configurations of the Equal Masses model with the highest number of saddles on the critical curve lie on these vertical faces, as seen here on the front face at the $(a_p,p)\approx (0.25, 2.184)$ point with four saddles.

In addition to the vertical faces, \cite{danek_heyrovsky15b} also studied configurations of the Equal Masses model on vertical midplanes along the medians of the ternary plot, which extend from each edge to the vertical midline of the opposite face of the prism. These configurations correspond to triangular isosceles configurations -- the TI model of \cite{danek_heyrovsky15b}. In this case both model parameters (denoted by subscript TI) are combinations of our two parameters. For the midplane extending from the left edge of the prism in Figure~\ref{fig:EqMassSur}, the legs of the triangle are $a=c$ and the vertex angle varies from $0$ at the edge to $\pi$ at the opposite face. These parameters are used in the TI model, so that the leg length $s_{\rm TI}=p\,a_p$ and the vertex angle $\theta_{\rm TI}=2\,\arcsin[a_p^{-1}/2-1]$. For the inverse transformation, our perimeter $p=2\,s_{\rm TI}\,[1+\sin(\theta_{\rm TI}/2)]$, and fractional side lengths $a_p=c_p=[1+\sin(\theta_{\rm TI}/2)]^{-1}/2$, $b_p=\sin(\theta_{\rm TI}/2)/[1+\sin(\theta_{\rm TI}/2)]$.

The boundary intersections with this midplane are shown in the right panel of Figure~\ref{fig:EqMassSur-planes}, which is based on the parameter-space division in the left panel of Figure~15 in \cite{danek_heyrovsky15b}, recalculated in terms of our parameters. In comparison with the left panel, the horizontal axis is $\sqrt{3}/2$ times shorter since it corresponds to the median of the ternary plot (see Figure~\ref{fig:LittleTriangles}). The lines are colored according to the intersecting surfaces from Figure~\ref{fig:EqMassSur} as in the left panel. The left edge of the plot coincides with the left edge of the collinear plot in the left panel. The right edge coincides with the $c_p=3/12$ vertical midline of the $b_p=1/2$ collinear plot, which is the same as the left panel with $c_p$ along the horizontal axis. The boundaries thus intersect the right edge at $p\approx 2.184$ and $p\approx 6.711$, as mentioned above.

Overall, the topology boundaries along this plane are more complex and have a richer structure. In particular, we point out two merging points, at which two distinct boundaries meet and continue as a single curve. In one case, the curve starting out as cyan from the left edge and continuing as green meets the orange curve approaching from above at $(a_p,p)\approx(0.4192,4.196)$. The other case occurs at $(a_p,p)\approx(0.2537,1.884)$ near the right edge, where the purple curve rising to the right from the horizontal axis meets the blue curve approaching from the left. In terms of the Jacobian, for these two configurations the critical curve passes through a second-order monkey saddle. The Jacobian for the corresponding $a_p$ values thus has only four simple saddles and one monkey saddle.

Another interesting feature is the existence of a different close-triple limit topology indicated by the boundaries at $a_p\approx 0.2637$, described and discussed in \cite{danek_heyrovsky15b}. For this value of $a_p$ the Jacobian has a second-order maximum and only five simple saddles, as indicated by the five colors of the three boundaries crossed at nonzero perimeters. In the close limit the small critical-curve loop around the second-order maximum corresponds to a four-cusped caustic loop, while those around the simple maxima correspond to the usual three-cusped caustic loops.

Due to the symmetry of the isosceles configurations, at least one of the boundaries is degenerate for any value of $a_p$. The highest number of distinct boundaries -- five -- occurs for $a_p\in (0.4192,0.5)$ with the exception of $a_p\approx 0.4300$, where the cyan / green intersection reduces the number to four. Four boundaries occur also for nearly all values $a_p\in (0.2537,0.4192]$ with the exception of $a_p\in\{0.2637,0.2725,0.2797\}$, which have three, and $a_p=1/3$, which has two. Three boundaries also occur for $a_p\in (0.25,0.2537]$. The $a_p=1/3$ case represents the most symmetric triple lens with components in an equilateral configuration. This is another of the configurations with the minimum number of topologies mentioned in Section~\ref{sec:Close-to-wide}. The first transition occurs at the triple intersection of the purple, blue, and cyan surfaces at $p\approx 2.009$; the second occurs at the triple green, orange, and red intersection at $p\approx 4.480$. Note that the two other two-boundary configurations discussed earlier can be seen here at the left edge (degenerate binary) and the right edge (symmetric collinear triple).

The number of colors at a given point indicates the number of Jacobian saddles on the critical curve of the corresponding configuration, but in this case with two exceptions. These occur at the two merging points at $(a_p,p)\approx(0.4192,4.196)$ and $(a_p,p)\approx(0.2537,1.884)$, where the two merging colors (green / orange, and purple / blue, respectively) correspond to a single monkey saddle instead of two simple saddles. Perturbing $a_p$ breaks down either monkey saddle into a pair of simple saddles lying on the same Jacobian contour (here for lower $a_p$), or on two different contours (here for higher $a_p$).

Aided by the structure on the vertical planes, we may now briefly summarize the overall division of the parameter space in Figure~\ref{fig:EqMassSur}. In the close limit the purple boundary extends down to the $p=0$ base in narrow spikes, leading to three symmetric isosceles configurations, at which the Jacobian has a double maximum and only five simple saddle points. The next surface appears at the $p\approx 1.428$ close/intermediate binary transitions at the three vertical edges. Most of the transitions occur between this perimeter value and $p\approx 6.711$, at which the red / orange intersection disappears and the remaining green, orange, and red boundaries asymptotically approach the vertical edges. In the high-perimeter regime the region close to any of the edges corresponds to an equal-mass binary lens with a distant third companion. The red surface then corresponds to the wide/intermediate boundary of the binary, and the orange and green surfaces represent the close/intermediate boundary perturbed by the third companion.

\subsection{Critical-curve Topologies}
\label{sec:EqM_topologies}

For a more detailed interpretation of the different regimes we directly study the different critical-curve topologies. In order to identify the topologies in all the 3D regions of the prism separated by the boundary surfaces, we follow the procedure described in Section~\ref{sec:Mapping_topologies} and explore the topologies on a sequence of $p={\rm const.}$ horizontal sections.

For illustration, we first present the $p=1.679$ section in Figure~\ref{fig:EqMPsection}. The black curves in the central ternary plot correspond to boundaries between different critical-curve topology regions. Here the outer parts of the figure-eight loops facing the corners of the plot are formed by the blue surface, and the rest of the curves by the purple surface from Figure~\ref{fig:EqMassSur}. These curves divide the section into 13 distinct continuous regions. However, due to the symmetry of the model, we expect at most four different critical-curve topologies.

Examples of critical curves and caustics are shown for a point inside a figure-eight loop (A), inside an elliptical loop (B), close to the ternary-plot corner (C), and in the central region of the plot (D). The critical curve of A is formed by an outer loop with three inner loops, two of which appear as mere dots with equally dot-like caustic counterparts. The critical curve of B has the same topology as A. The critical curve of C has an outer loop with two tiny inner loops (with equally tiny caustic counterparts). The critical curve of D has an outer loop with four inner loops. The tiny central loop has its caustic counterpart in a tiny loop not discernible inside the larger central loop. Overall, there are three different critical-curve topologies in 13 topology regions on this particular section.

The results for a sequence of 12 such horizontal sections of the ternary prism from Figure~\ref{fig:EqMassSur} are shown in Figure~\ref{fig:EqMTernaries}. Here the individual regions are colored according to the topology of their critical curves, following the schematic key in the bottom row. The topologies are numbered in order of the lowest perimeter value at which they appear in this Equal Masses model. The sections start with the closest $p=1.498$ in the top left corner, continuing to the right, in the middle and bottom rows, and ending at $p=6.711$ in the bottom right corner. The positions of these sections are marked in Figure~\ref{fig:EqMassSur} by the black horizontal contours on the surfaces. The first one at $p=1.498$ lies just above the close/intermediate binary boundary at the edges; the last one at $p=6.711$ lies at the top of the last disappearing (i.e., nonasymptotic) region at the intersections of the red and orange surfaces on the vertical faces of the prism.

For interpreting the sequence of sections in Figure~\ref{fig:EqMTernaries} we will keep referring to the ternary prism with the boundary surfaces in Figure~\ref{fig:EqMassSur}. The sequence starts at $p=1.498$ with two close-triple topologies: the generic T1 (outer + four inner loops) and the special T2 (outer + three inner) in the islands within the purple spikes seen in the prism. In addition, topology T3 (outer + two inner) appears in the corners starting from the close/intermediate binary transition at $p=1.428$, as seen in the prism. It is separated from T1 by thin figure-eight regions of T2 (in the prism these correspond to nonisosceles configurations in the narrow gap between the purple and blue surfaces). The same structure can be seen in the next section with $p=1.679$, which corresponds to the example from Figure~\ref{fig:EqMPsection}. With increasing perimeter T1 shrinks and T2 expands until the facing lobes of the figure-eight regions connect at the medians of the plot, as seen in the $p=1.713$ section. T3 expanded as well in the corners, and new narrow crescent-like T3 regions appear adjacent to the points where the T2 lobes connected. The T3 and the inner T2 regions expand further, as seen in the $p=1.794$ section. The next topology T4 (outer + inner) can be seen in the $p=2.103$ section. It appears in short succession at the six points where the cyan surface emerges from the blue surface in Figure~\ref{fig:EqMassSur} and at the center of the plot, where it replaced T1 at $p=2.009$. T1 finally disappears at the edges of the plot at $p=2.184$, where it is replaced by the single-loop topology T5. T2 disappears next, so that neither blue nor yellow appears in the following $p=2.398$ section.

The remaining T3, T4, and T5 topologies change in proportion, with T5 gradually dominating, as seen in the $p=2.700$ and $p=3.903$ sections. The first two of the four asymptotic topologies appear in rapid succession: T6 (outer + two inner + separate loop) at the wide/intermediate binary transition at $p=3.923$ at the edges of the prism, followed by T7 (two separate) replacing T4 along the medians of the ternary plot. Both can be seen in the $p=4.198$ section. In addition, the asymptotic T8 (outer + inner + separate) appeared at $p=4.196$ at the outer tips of the T7 region adjacent to T4. At $p=4.198$ the T8 region is still too small to be visible in the plot. It is best seen here in the $p=4.353$ section, becoming extremely narrow at higher perimeter values. Last to appear is the asymptotic T9 (three separate), which replaces T4 at the center of the plot at $p=4.480$. Topology T3 disappears at $p=4.672$ at the intersections of the cyan, green, and orange surfaces on the faces of the prism, followed instantly by T4. The $p=4.803$ section thus has only five remaining topologies. The final topology to disappear is T5 at $p=6.711$ (shown here as the final section), at the intersections of the orange and red surfaces on the faces of the prism.

Topology T9, which eventually corresponds to Einstein rings of three independent lenses, gradually dominates over the other asymptotic topologies. These describe a binary lens in the intermediate (T7) and close (T6) regime with a distant companion. Topology T8 appears in the narrow transition region between the two, where the perturbation by the distant companion causes one of the two inner loops of T6 to merge sooner than the other with the outer loop.

In total, the boundary surfaces cut the full ternary-prism parameter space into 39 disjoint regions. Due to the high degree of symmetry, some of these regions correspond to identical spatial configurations with different permutations of components. Other regions correspond to mirror-symmetric configurations, which have the same critical-curve topology. Overall, there are only 13 distinct types of regions, so that it is sufficient to check the critical-curve topology for 13 parameter-space points. The final lineup of the nine topologies of the Equal Masses model is shown in the bottom row of Figure~\ref{fig:EqMTernaries}. Interestingly, exactly the same set was found by \cite{danek_heyrovsky15b} for the isosceles configurations. All the topologies thus appear on the vertical planes cutting the prism along any median of the ternary plot, as seen in the right panel of Figure~\ref{fig:EqMassSur-planes}. Clearly, breaking the spatial symmetry of the isosceles triple does not lead to any additional critical-curve topology. Finally, we note that no other topologies were found in the remaining two nonequal-mass LS and TE models studied by \cite{danek_heyrovsky15b} either.

\subsection{Topology Probabilities}
\label{sec:EqM_probabilities}

We quantify the frequency of occurrence of the different topologies by computing the probabilities $\mathcal{P}_{{\rm T}_i}(p)$ of generating a critical curve with topology ${\rm T}_i$ by three randomly placed lens components forming a triangle with perimeter $p$. Following the description in Section~\ref{sec:Probabilities}, we evaluate the integral in Equation~(\ref{eq:probability_p_final}) for perimeters ranging from $0$ to $p_{\rm max}=6.711$, after which only asymptotic topologies remain. At any value of $p$ these probabilities add up to unity, $\sum_{i=1}^{9}\mathcal{P}_{{\rm T}_i}(p)=1$.

In the individual panels of Figure~\ref{fig:EqMPerPer} we plot the perimeter dependence of $\log_{10} \mathcal{P}_{{\rm T}_i}(p)$ for topologies T1 (top) through T9 (bottom). The sequence of appearing and disappearing topologies as a function of $p$ corresponds to the discussion of the sequence of sections in Section~\ref{sec:EqM_topologies}. Note that low probabilities $\mathcal{P}_{{\rm T}_i}(p)\lesssim 0.001$ are beyond the lower limit of the plots, so that, for example, $\mathcal{P}_{{\rm T}2}(p<0.6)$ cannot be seen even though the topology extends all the way to the left to $p=0$.

The panels show that the sequence of dominant topologies from close to wide regime is T1--T3--T5--T7--T9, with the even-numbered topologies not dominating at any perimeter value. The last topology T9 dominates beyond the plotted range in the $p\rightarrow\infty$ limit, where the probabilities of T6, T7, and T8 drop to zero. Within the horizontal limits of the plot, only T5 covers more than half of the perimeter range. T2 occurs in the narrowest interval and is concentrated in a rather sharp peak.

For an overall comparison of topologies occurring for configurations with perimeter $p\in(0,\,p_{\rm max}=6.711)$, we evaluate the probabilities $\mathcal{P}_{{\rm T}_i,p_{\rm max}}$ from Equation~(\ref{eq:probability_pmax_final}). These are normalized so that $\sum_{i=1}^{9}\mathcal{P}_{{\rm T}_i,p_{\rm max}}=1$. The values obtained for the nine topologies are shown in the ``Equal Masses" column of Table~\ref{tab:Volumes}. Within the given perimeter range, we see that topology T7 alone occurs in nearly 50\% of cases. On the lowest-probability end, topologies T1, T8, and T2 together occur in less than 1\% of cases. Sorted in descending order of probability, the topology sequence is T7 $>$ T5 $>$ T6 $>$ T9 $>$ T3 $>$ T4 $>$ T1 $>$ T8 $>$ T2.

Overall, close-type topologies are highly disfavored, which can be seen from the $\propto\,p^3$ proportionality of the Jacobian weight function in Equation~(\ref{eq:JacABC}). Beyond our $p_{\rm max}$ limit only the asymptotic topologies T6, T7, T8, and T9 extend. Hence, if we increased the cutoff to a higher perimeter value, the probabilities of all other topologies would decrease. At sufficiently high $p$, topology T9 approaches 100\% and even the other asymptotic topologies drop to 0\%. This is clearly the case for any other triple-lens model as well: at high perimeters the three-independent-lenses regime dominates.

Uncertainties of the numerically computed probabilities are mostly due to discretization in the $(a_p, b_p)$-plane and the related precision limit on the exact positions of boundaries of the topology regions. The error is thus proportional to the probability corresponding to points on the boundaries inside the prism of all the regions with the given topology. In this case, $0.05\%$ of the total number of pixels were not assigned to any topology region. By integrating over these pixels in the numerator of Equation~(\ref{eq:probability_pmax_ternary}), we find that they correspond to a probability of $0.01\%$ (out of the total $100\%$) left unattributed to any topology.

\section{Model 2: Planet in Binary}
\label{sec:PiB}

In our second model we keep two components with equal masses and add a third component with a different mass. Mirroring any configuration of such a system across an axis symmetrically separating the equal-mass components always produces the same critical-curve topology. The division of the ternary-prism parameter space should therefore be reflection-symmetric across a vertical plane along one of the ternary-plot medians.

We chose a third component much lighter than the equal-mass pair, with fractional masses $\mu_1=\mu_2=0.49995$ and $\mu_3=0.0001$. Such a model would describe an equal-mass binary star with a planetary companion (for solar-mass components of the binary $\mu_3$ would correspond to a 0.7 Saturn-mass planet). Side $c$ of the triangular configuration describes the separation of the two stars, and sides $a,\,b$ the separation of the planet from stars 2 and 1, respectively. In the ternary plot in Figure~\ref{fig:LittleTriangles} switching sides $a$ and $b$ corresponds to reflection across the vertical median leading from $(a_p, b_p, c_p)=(1/2, 1/2, 0)$ to $(1/4, 1/4, 1/2)$. This median defines the plane of symmetry discussed above.

Even though our model describes an arbitrary projected spatial configuration of the three components, it is useful to view the regions of the following plots in terms of the mutual vicinities of the components in a face-on orientation of the binary orbit. Thus, the region near the top vertex ($c_p\to 0$) of the ternary plot describes a close binary star with a distant circumbinary planet. The regions near the bottom left ($b_p\to 0$) or bottom right ($a_p\to 0$) vertices describe a planet around star 1 or 2, respectively, which has a distant binary companion.

In the following subsections we describe the boundaries, topologies, and their probabilities in the Planet in Binary model, concentrating primarily on differences from the Equal Masses model.

\subsection{Boundary Surfaces}
\label{sec:PiB_surfaces}

The boundary surfaces $p\,(a_p, b_p)$ computed according to Sections~\ref{sec:Close-to-wide} and \ref{sec:Mapping_topologies} are shown in Figure~\ref{fig:PiBSur}. Parts of the intersecting surfaces are colored in order of their perimeter values from closest (purple) to widest (red) transition, as in Figure~\ref{fig:EqMassSur} for the Equal Masses model. Due to the lower symmetry, we present three views of the ternary prism, from left to right facing ternary axes $a_p$, $b_p$, and $c_p$. For better orientation, an animated version showing the prism rotated about the vertical axis is available in the online version of this article.

The $(a_p, b_p, c_p)=(1/2, 0, 1/2)$ vertical edge (left edge in the left panel) corresponds to the planet coinciding with star 1. Similarly, the $(0, 1/2, 1/2)$ vertical edge (right edge in the left panel) corresponds to the planet coinciding with star 2. In both cases the triple lens is reduced to a two-point-mass lens with fractional masses $0.49995$ and $0.50005$. The purple and blue surfaces both intersect these edges at the perimeter corresponding to the close/intermediate transition, $p=2\,\left[\sqrt[3]{0.49995}+\sqrt[3]{0.50005}\right]^{-3/4}\approx 1.414$. The cyan surface intersects these edges at the perimeter corresponding to the wide/intermediate transition, $p=2\,\left[\sqrt[3]{0.49995}+\sqrt[3]{0.50005}\right]^{3/2}\approx 4.000$. The $(1/2, 1/2, 0)$ vertical edge (left edge in the middle panel) corresponds to both stars coinciding. In this case the triple lens is reduced to a two-point-mass lens with fractional masses $0.0001$ and $0.9999$. The purple and blue surfaces intersect this edge at $p=2\,\left[\sqrt[3]{0.0001}+\sqrt[3]{0.9999}\right]^{-3/4}\approx 1.933$, and the cyan surface intersects it at $p=2\,\left[\sqrt[3]{0.0001}+\sqrt[3]{0.9999}\right]^{3/2}\approx 2.141$.

In the left panel the front vertical face corresponds to collinear configurations with the planet positioned along the line from star 1 to star 2, so that $c_p=1/2$. The left rear vertical face ($a_p=1/2$) corresponds to collinear configurations with star 1 along the line from star 2 to the planet, and the right rear ($b_p=1/2$) has star 2 along the line from star 1 to the planet. The $(1/4, 1/4, 1/2)$ configuration in the middle of the front face corresponds to a special case of the linear symmetric (LS) model studied in detail by \cite{danek_heyrovsky15b}. With an added subscript LS, the separation parameter of the model $s_{\rm LS}=p/4$ and the central fractional mass $\mu_{\rm LS}=\mu_3=0.0001$. The boundary intersections with the midline of the front face can be seen in the left panel of Figure~2 in \cite{danek_heyrovsky15b} as the intersections of the plotted curves with a $\mu_{\rm LS}=0.0001$ vertical line. Three topology changes occur along the line: the first just above $p=\sqrt{2}\approx 1.414$, corresponding to the intersection with the purple and blue surfaces in the prism, the second just below $p=4$ (intersection with cyan and green surfaces), and the third just above $p=4$ (intersection with orange and red surfaces).

Another special case that can be found in \cite{danek_heyrovsky15b} is the equilateral configuration at the center of the prism, $(a_p, b_p, c_p)=(1/3, 1/3, 1/3)$. The triangular equilateral (TE) model has the following parameters: triangle side $s_{\rm TE}=p/3$ and one vertex mass $\mu_{\rm TE}=\mu_3=0.0001$. The boundary intersections with the central axis of the prism can be seen in the left panel of Figure~11 in \cite{danek_heyrovsky15b} along a $\mu_{\rm TE}=0.0001$ vertical line. In this case four topology changes occur, as indicated by the intersections of the line with the plotted curves. The first three follow in rapid succession: two just below $p=3/\sqrt{2}\approx 2.121$ (the first corresponding to the intersection with the purple and blue surfaces; the second with the cyan surface), and one just above $p=3/\sqrt{2}\approx 2.121$ (green and orange surfaces). The last transition just below $p=6$ corresponds to the intersection with the red surface.

Despite the lower symmetry of the Planet in Binary model, the overall structure of its parameter space in Figure~\ref{fig:PiBSur} has similar features to that in the case of the Equal Masses model. In the close limit, the purple boundary extends down to the $p=0$ base in three narrow spikes, at each of which the Jacobian has a double maximum and only five simple saddle points. In this case one of these configurations is isosceles, with $(a_p,\,b_p,\,c_p)\approx(0.2863, 0.2863, 0.4274)$, and the other two are reflection-symmetric, $(0.3099, 0.2827, 0.4074)$ and $(0.2827, 0.3099, 0.4074)$. Unlike in the Equal Masses model, the spikes lie close together so that their surfaces intersect and form new topology regions before the purple and blue surfaces from the front face of the left panel cross them. The purple and blue surfaces appear at the close/intermediate binary transitions at $p\approx 1.414$ at the front two vertical edges, and nearly instantaneously across the entire front face.

The close/intermediate (purple and blue) and wide/intermediate (cyan) planet+star transitions at the rear vertical edge appear at $p\approx 1.933$ and $p\approx 2.141$, respectively. The cyan surface corresponding to the wide/intermediate binary transitions at the front two vertical edges appears nearly simultaneously at $p\approx 4.000$ on the entire front face. The final nonasymptotic region disappears at $p\approx 5.593$, at which the red / orange intersection reaches the rear faces of the prism. The remaining green, orange, and red boundaries asymptotically approach the vertical edges. In the high-perimeter regime the regions close to the front two edges in the left panel correspond to a star with a planet and a distant stellar companion. The region close to the rear vertical edge corresponds to a close binary with a distant planetary companion. In all cases, the red surface then corresponds to the wide/intermediate boundary of the close companions, and the orange and green surfaces represent the close/intermediate boundary perturbed by the distant third component.

\subsection{Critical-curve Topologies}
\label{sec:PiB_topologies}

The critical-curve topologies of the Planet in Binary model can be mostly understood as binary-lens critical curves with additional loops due to the planet. From the properties of the Jacobian it can be seen that a third lens component adds one pole and typically two maxima and three saddle points to the two-point-mass lens Jacobian surface. If the mass of the third component is much smaller than the first two masses, all the additional features of the Jacobian occur in a small region around the third component. If it is located sufficiently far from any loop of the two-point-mass lens critical curve, it simply adds extra loops to the critical curve.

The number of additional loops depends on the sign of the two-point-mass Jacobian at the position of the third component: (a) if it is positive, the critical curve has one extra loop around the planet; (b) if it is negative, the critical curve has two extra loops around two triple-lens Jacobian maxima in the vicinity of the planet. The situation becomes more complicated if the planet is located close to the two-point-mass lens critical curve, especially in case the critical curve is near the close/intermediate topology boundary.

Following the procedure described in Section~\ref{sec:Mapping_topologies}, we present in Figure~\ref{fig:PiBTernaries} the topology maps for a sequence of eight $p={\rm const.}$ sections of the ternary prism from Figure~\ref{fig:PiBSur}. The ternary plot regions are colored according to the topology of their critical curves, following the schematic key in the bottom row. In the key the topologies are ordered by the lowest perimeter value at which they appear in the Planet in Binary model. We preserve the numbering of the topologies found in the Equal Masses model.

The sections start with the closest $p=1.016$ in the top left corner, which is entirely in the close-triple regime before the purple spikes in Figure~\ref{fig:PiBSur} intersect. The sequence ends at $p=5.640$ in the bottom right corner, just above the last disappearing (i.e., nonasymptotic) region at the intersections of the red and orange surfaces on the rear vertical faces of the prism in the left panel of Figure~\ref{fig:PiBSur} (better seen in the other two panels). The positions of the sections are marked in Figure~\ref{fig:PiBSur} by the black horizontal contours on the surfaces.

The sequence starts at $p=1.016$ with the same two close-triple topologies as in the Equal Masses model: T1 for nearly all configurations and T2 within the purple spikes. The sequence of transitions differs already in the next step: the purple spikes seen in the prism intersect, and topology T3 appears as the corresponding yellow regions come into contact. Next, the T3 regions connect at the midpoint between the spikes, forming an entirely new topology T10 not found in any of the previously studied models. This peculiar topology differs from T3 in having an additional small loop inside one of the inner critical-curve loops; hence, it has an outer + two inner + doubly nested loop structure.

Next, topology T3 appears at the lower edge of the plot from the close/intermediate binary transition at $p=1.414$, and the corresponding nearly horizontal boundary gradually sweeps through the ternary plot to its top vertex. Note, however, that the boundary consists of two extremely close surfaces: initially purple and blue at the front face, finally green and orange at the rear vertical edge of the prism. This pair of surfaces is separated owing to the planetary-mass perturbation. Therefore, the boundary seen in the ternary plots is accompanied by an extremely narrow horizontal band (not discernible in the plots) with different topologies. Initially there is a T2 band separating T3 from T1. As it crosses the region of the spikes, this double boundary changes T2 via T3 (narrow) to T4, T3 via T4 to T5, and T10 via another entirely new topology T11 to T7. Topology T11 differs from T10 in lacking the simple inner loop, having an outer + inner + doubly nested loop structure. The regions occupied by T11 are too small to be visible on any of the sections; hence, we mark it in the key with a dashed-line sketch. The described situation corresponds to the second $p=1.694$ section, where the sequence of topologies in the narrow boundary band along $c_p\approx 5/12$ from the left edge to the center of the plot is as follows: T2 (between T3 and T1), T3 (between T4 and T2), T4 (between T5 and T3), and T11 (between T7 and T10).

The close/intermediate (purple and blue surfaces) and wide/intermediate (cyan surface) transitions due to the planet and combined-mass star in the circumbinary regime close to the rear vertical edge of the prism occur at $p\approx 1.933$ and $p\approx 2.141$, respectively. The corresponding sequence of topologies can be seen at the top of the ternary plot for the $p=2.270$ section: from the vertex inward T6--T3--T2--T1. Note that in this model topology T6 appears after T7 (i.e., at higher perimeter), unlike in the Equal Masses model. In the same section we see that the horizontal double boundary has crossed the egg-shaped central region and topologies T10 and T11 have disappeared. In the $p=2.315$ section the upper T3 and T2 regions are already connected with their counterparts at the center. By $p=2.484$ the egg-shaped central region has expanded over the horizontal double boundary, with topology T8 appearing in the narrow band between T7 and T6. The two T6 regions connect (as seen in $p=2.512$), followed by the gradual disappearance of T1, T2, and the upper T3 regions at the sides of the plot (as seen in $p=3.162$).

Topology T6 appears at the lower vertices of the plot from the wide/intermediate binary transition at $p\approx 4.000$ (cyan surface at front face of prism), and the corresponding second nearly horizontal boundary gradually sweeps through the ternary plot to its top vertex. Unlike the previous horizontal boundary, this one is simple rather than double. The final T9 topology appears close to the lower edge of the plot, and the progression of the boundary gradually eliminates topologies T3, T4, and T5 at the sides of the plot. T5 disappears at $p\approx5.593$ at the intersections of the orange and red surfaces on the rear faces of the prism. The last $p=5.640$ section includes only the same four asymptotic topologies T6, T7, T8, and T9 that we found already in the Equal Masses model. The three-Einstein-rings topology T9 gradually dominates over the other topologies, just as in any triple lens. The regions close to the vertices of the plot correspond to the close (T6), intermediate (T7), and wide (T9) regimes of different pairs of the components, with a distant third companion. Thus, the region near the top vertex corresponds to a binary star (components 1 and 2), and those near the bottom vertices to a star with a planet (components 1 and 3 at the left vertex, components 2 and 3 at the right vertex). Just as in the Equal Masses model, topology T8 appears in the narrow transition region between the close and intermediate regimes, where the perturbation by the distant companion causes one of the two inner loops of T6 to merge sooner than the other with the outer loop.

The final lineup of the 11 topologies of the Planet in Binary triple lens is shown in the bottom row of Figure~\ref{fig:PiBTernaries}. Clearly, this is the richest triple-lens model that has been studied so far. It includes all nine topologies found for the Equal Masses model in Section~\ref{sec:EqM}, which in turn include even all the topologies found in the varying-mass LS and TE models studied by \cite{danek_heyrovsky15b}. In addition, the present model includes two new peculiar topologies with doubly nested loops, T10 and T11.

Topology T10 can be constructed by placing a low-mass companion inside an inner loop of the critical curve of a close binary. This forces a small negative-Jacobian region inside a positive-Jacobian region close to a binary-lens Jacobian maximum, hence the doubly nested loop. Topology T11 then occurs when the loop around the other binary-lens Jacobian maximum connects with the outer loop sooner than the loop with the doubly nested loop, due to the perturbing influence of the planetary companion. While T10 occurs in a substantial region of parameter space, T11 occurs in a narrow boundary separating it from the adjacent T7 region. We illustrate a sample critical-curve transition involving topologies T10 and T11 together with the corresponding caustics in Appendix~\ref{sec:Appendix-T10T11}.

\subsection{Topology Probabilities}
\label{sec:PiB_probabilities}

We compute the probabilities $\mathcal{P}_{{\rm T}_i}(p)$ of generating a critical curve with topology ${\rm T}_i$ by three randomly placed lens components forming a triangle with perimeter $p$. We evaluate the respective integral in Equation~(\ref{eq:probability_p_final}) for perimeters ranging from $0$ to $p_{\rm max}=5.640$. This value lies just above the upper end of the last nonasymptotic topology T5 in the prism. At any value of $p$ the probabilities add up to unity, $\sum_{i=1}^{11}\mathcal{P}_{{\rm T}_i}(p)=1$.

In the individual panels of Figure~\ref{fig:PiBPerPer} we plot the perimeter dependence of $\log_{10} \mathcal{P}_{{\rm T}_i}(p)$ for topologies T1 through T11 in order of their first appearance in the Planet in Binary model. The sequence of appearing and disappearing topologies corresponds to the discussion in Section~\ref{sec:PiB_topologies}. Regions of low probability $\mathcal{P}_{{\rm T}_i}(p)\lesssim 0.001$ not visible in the plot occur here, for example, for T2, which extends down to $p=0$; for T11, which is entirely outside the plot so that only its range of occurrence is indicated by dashed lines; or for T8, which occurs continuously from $p\approx 2.4$ to the higher perimeter values at which it is visible in the plot.

The sequence of dominant topologies from close to wide regime is T1--T3--T7--T9. Unlike in the Equal Masses model, topology T5 is suppressed here so that there is no perimeter interval in which it would dominate. The last topology T9 dominates beyond the plotted range in the $p\rightarrow\infty$ limit. In comparison with the Equal Masses model, the close-limit topologies T1 and T2 extend to higher perimeters (ending around $p\approx 2.95$); the intermediate T3 and T4 have an expanded range, appearing earlier and disappearing later (even though T4 reaches lower probabilities); T10 and T11 are new, occurring in narrow perimeter intervals; T5 has a reduced range, appearing and disappearing earlier; and all asymptotic topologies appear earlier, especially T7, which appears before T6 already around $p\approx 1.65$.

Within the horizontal limits of the plot, T3 occurs in the widest perimeter range. Only T10, T11, and T9 cover less than half of the perimeter range, with T11 occurring in the narrowest interval. Generally, the main difference between the models is the expanded perimeter range of most topologies in the Planet in Binary model. In fact, all the topologies except T9, T10, and T11 occur at any perimeter $p\in(2.4,2.95)$. Thus, in this perimeter range the close-limit T1 and T2 coexist with the wide-limit asymptotic T7, T6, and T8. We recall that in the Equal Masses model there is no such overlap: the close-limit topologies are broadly separated from the wide-limit topologies.

For an overall comparison of topologies occurring for configurations with perimeter $p\in(0,\,p_{\rm max}=5.64)$, we evaluate the probabilities $\mathcal{P}_{{\rm T}_i,p_{\rm max}}$ from Equation~(\ref{eq:probability_pmax_final}) normalized so that $\sum_{i=1}^{11}\mathcal{P}_{{\rm T}_i,p_{\rm max}}=1$. The values obtained for the 11 topologies are shown in the ``Planet in Binary" column of Table~\ref{tab:Volumes}. Within the given perimeter range, we see that even in this model topology T7 alone occurs in nearly 50\% of cases. On the lowest-probability end, topologies T4, T8, T2, T10, and T11 together occur in less than 1\% of cases. Sorted in descending order of probability, the topology sequence is T7 $>$ T6 $>$ T9 $>$ T3 $>$ T5 $>$ T1 $>$ T4 $>$ T8 $>$ T2 $>$ T10 $>$ T11.

Comparing the topologies occurring in both models, T1 shows the highest relative increase in probability from the Equal Masses model, due to its expanded perimeter range. On the other hand, the highest relative decrease can be seen in T5, which maintains a wide perimeter range but achieves lower probabilities.

As in the previous model, the numerical error is proportional to the probability corresponding to points on the boundaries inside the prism of all the regions with the given topology. Some of the regions in this model have complex changes of shape with perimeter; some are limited to narrow gaps between close surfaces. In total, $0.03\%$ of pixels were not assigned to any topology region, leaving a probability of $0.04\%$ not attributed to any topology. Due to the more intricate partitioning of the parameter space of the Planet in Binary model, this probability may also serve as an upper bound on the probability of potential topologies undetected for numerical reasons.

\section{Model 3: Hierarchical Masses}
\label{sec:HiM}

The components in our third model each have a different mass, so that we do not expect any symmetry in the parameter-space division. We chose hierarchically distributed masses in the ratio $1:0.01:0.0001$. In terms of fractional masses, $\mu_1=1/1.0101\approx 0.99,$ $\mu_2=0.01/1.0101\approx 0.0099,$ and $\mu_3=0.0001/1.0101\approx 0.0001$. Such a system may represent a star with a planet with a moon or a star with two planets.

Even though the model describes a projected spatial configuration of the components, it is useful to view the regions of the following ternary plots in terms of the relative positions of the components as if they were in face-on orbits. Thus, the region near the top vertex ($c_p\to 0$) describes a star with a close-in massive planet and a distant lower-mass planet. The region near the bottom left ($b_p\to 0$) vertex describes a star with a close-in lower-mass planet and a distant massive planet. The region near the bottom right ($a_p\to 0$) vertex describes a star with a planet with a moon.

Based on the last case, we will refer to components 1, 2, and 3 in the following for simplicity as the star, planet, and moon, respectively. In these terms, side $c$ of the triangular configuration describes the separation of the planet from the star, side $a$ the separation of the moon from the planet, and side $b$ the separation of the moon from the star.

In the following subsections we describe the boundaries, topologies, and their probabilities in the Hierarchical Masses model, concentrating primarily on differences from the previous two models.

\subsection{Boundary Surfaces}
\label{sec:HiM_surfaces}

The boundary surfaces $p\,(a_p, b_p)$ computed according to Sections~\ref{sec:Close-to-wide} and \ref{sec:Mapping_topologies} are shown in Figure~\ref{fig:HiMSur}. Parts of the intersecting surfaces are colored in order of their perimeter values from closest (purple) to widest (red) transition, as in Figure~\ref{fig:EqMassSur} and Figure~\ref{fig:PiBSur} for the previous two models. Due to the lack of symmetry, we present three views of the ternary prism, from left to right facing ternary axes $a_p$, $b_p$, and $c_p$. For better orientation, an animated version showing the prism rotated about the vertical axis is available in the online version of this article.

The $(a_p, b_p, c_p)=(1/2, 0, 1/2)$ vertical edge (left edge in the left panel) corresponds to the moon coinciding with the star (in terms of projected positions, as always), so that the triple lens is reduced to a two-point-mass lens with fractional masses $0.01/1.0101$ and $1.0001/1.0101$. The purple and blue surfaces both intersect this edge at the perimeter corresponding to the close/intermediate transition, $p= 2\,\left[\sqrt[3]{0.0099}+\sqrt[3]{0.9901}\right]^{-3/4}\approx 1.732$. The cyan surface intersects this edge at the perimeter corresponding to the wide/intermediate transition, $p= 2\,\left[\sqrt[3]{0.0099}+\sqrt[3]{0.9901}\right]^{3/2}\approx 2.667$.

The $(0, 1/2, 1/2)$ vertical edge (right edge in the left panel) corresponds to the moon coinciding with the planet, reducing the triple lens to a two-point-mass lens with fractional masses $0.0101/1.0101$ and $1/1.0101$. The purple and blue surfaces both intersect the edge at the close/intermediate transition, $p= 2\,\left[\sqrt[3]{0.009999}+\sqrt[3]{0.990001}\right]^{-3/4}\approx 1.731$. The cyan surface intersects the edge at the wide/intermediate transition, $p= 2\,\left[\sqrt[3]{0.009999}+\sqrt[3]{0.990001}\right]^{3/2}\approx 2.669$.

The $(1/2, 1/2, 0)$ vertical edge (left edge in the middle panel) corresponds to the planet coinciding with the star, reducing the triple lens to a two-point-mass lens with fractional masses $0.0001/1.0101$ and $1.01/1.0101$ . The purple and blue surfaces intersect this edge at $p= 2\,\left[\sqrt[3]{0.000099}+\sqrt[3]{0.999901}\right]^{-3/4}\approx 1.933$, and the cyan surface intersects it at $p= 2\,\left[\sqrt[3]{0.000099}+\sqrt[3]{0.999901}\right]^{3/2}\approx 2.140$. In this case the combination of masses and the corresponding boundaries are nearly identical to the same vertical edge in the Planet in Binary model, as seen in the middle panel of Figure~\ref{fig:PiBSur}.

In the left panel the front vertical face corresponds to collinear configurations with the moon positioned along the line from the planet to the star, so that $c_p=1/2$. The left rear vertical face ($a_p=1/2$) corresponds to collinear configurations with the star along the line from the planet to the moon, and the right rear ($b_p=1/2$) has the planet along the line from the star to the moon.

Despite the total lack of symmetry of the Hierarchical Masses model, the overall structure of its parameter space in Figure~\ref{fig:HiMSur} has similar features to those in the previous two models. The close limit has the three purple spikes extending down to the $p=0$ base, where the Jacobian has a double maximum and only five simple saddle points. In this model their locations are $(a_p,\,b_p,\,c_p)\approx(0.0415,0.4845,0.4740)$, $(0.0417,0.4717,0.4866)$, and $(0.0626,0.4673,0.4701)$. Note that all three of these special configurations lie in the $a_p\to 0$ star+planet+moon region of the parameter space, where the two lighter components lie close together, far from the heaviest component. Just as in the Planet in Binary case, the boundary surfaces of the spikes intersect and form new topology regions before the purple and blue surfaces from the front face of the left panel cross them. The purple and blue surfaces appear at the close/intermediate transitions at $p\approx 1.731$ at the front right vertical edge, and nearly instantaneously across the entire front face.

The close/intermediate (purple and blue) and wide/intermediate (cyan) transitions at the rear vertical edge appear at $p\approx 1.933$ and $p\approx 2.140$, respectively. The cyan surface corresponding to the wide/intermediate transition at the front left vertical edge appears at $p\approx 2.667$ and nearly simultaneously on the entire front face. The final nonasymptotic region disappears at $p\approx 4.799$, at which the red / orange intersection reaches the rear left face of the prism. The remaining green, orange, and red boundaries asymptotically approach the vertical edges.

In the high-perimeter regime the region close to the front left vertical edge in the left panel corresponds to a star with a close-in lower-mass planet and a distant massive planet. The region close to the front right edge corresponds to a planet with a moon and a distant host star. The region close to the rear vertical edge corresponds to a star with a close-in massive planet and a distant lower-mass planet. In all cases, the red surface then corresponds to the wide/intermediate boundary of the close companions, and the orange and green surfaces represent the close/intermediate boundary perturbed by the distant third component.

\subsection{Critical-curve Topologies}
\label{sec:HiM_topologies}

The critical-curve topologies of the Hierarchical Masses model can be mostly understood as an Einstein ring of the star as a single lens with additional loops due to the two lower-mass components. If the components are located sufficiently far from the Einstein ring and sufficiently far apart, each of them just adds extra loops to the critical curve. If we first add the planet (component 2), the number of additional loops depends on the sign of the single-lens Jacobian at the position of the planet: (a) if it is positive (outside the Einstein ring), the critical curve has one extra loop around the planet; (b) if it is negative (inside the Einstein ring), the critical curve has two extra loops around the Jacobian maxima in the vicinity of the planet. Adding next the moon (component 3), the number of additional loops depends on the sign of the two-point-mass-lens Jacobian of the star+planet at the position of the moon: (a) if it is positive, the critical curve has one extra loop around the moon; (b) if it is negative, the critical curve has two extra loops around the triple-lens Jacobian maxima in the vicinity of the moon. Placing either of the lighter components close to the Einstein ring of the star, and in particular placing the planet and moon close together, leads to more complicated situations and topologies.

Following the procedure described in Section~\ref{sec:Mapping_topologies}, we present in Figure~\ref{fig:HiMTernaries} the topology maps for a sequence of eight $p={\rm const.}$ sections of the ternary prism from Figure~\ref{fig:HiMSur}. The ternary plot regions are colored according to the topology of their critical curves, following the schematic key in the bottom row. In the key the topologies are ordered by the lowest perimeter value at which they appear in the Hierarchical Masses model. We preserve the numbering of the topologies found in the previous two models.

The section with the closest configurations $p=1.417$ in the top left corner is entirely in the close-triple regime before the purple spikes in Figure~\ref{fig:HiMSur} intersect. The sequence ends at $p=4.799$ in the bottom right corner when the last nonasymptotic region disappears at the intersection of the red and orange surfaces on the left rear vertical face of the prism in the left panel of Figure~\ref{fig:HiMSur} (better seen in the middle panel). The positions of the sections are marked in Figure~\ref{fig:HiMSur} by the black horizontal contours on the surfaces.

The sequence starts at $p=1.417$ with the same two close-triple topologies as in the previous models: T1 for nearly all configurations and T2 within the purple spikes. The next transitions follow the sequence seen in the Planet in Binary model: the purple spikes intersect and T3 appears as the corresponding T2 regions come into contact, followed by T10 as the T3 regions come into contact. Another T3 region appears next at the lower edge of the plot from the close/intermediate boundary at $p=1.731$, and the corresponding nearly horizontal boundary gradually sweeps through the ternary plot to its top vertex. Just as in the Planet in Binary model, this boundary is formed by two extremely close surfaces in Figure~\ref{fig:HiMSur} separated owing to perturbation by the moon: initially purple and blue at the front face in the left panel, finally green and orange at the rear vertical edge of the prism (as seen in the middle panel). Therefore, the boundary seen in the ternary plots is accompanied by an extremely narrow horizontal band (not discernible in the plots) with different topologies. As it sweeps through the plot from the edge through the region of the spikes, the double boundary changes T1 via T2 (narrow) to T3, followed by T2 via T3 to T4 (this situation corresponds to the second $p=1.782$ section), T3 via T4 to T5, and T10 via T11 to T7. Just as in the Planet in Binary model, the regions occupied by T11 are too small to be visible on any of the sections. The T10 and T11 regions disappear very quickly -- both are absent in the next $p=2.022$ section.

The close/intermediate (purple and blue surfaces) and wide/intermediate (cyan surface) transitions at the rear vertical edge of the prism occur at $p\approx 1.933$ and $p\approx 2.140$, respectively. In the ternary plot a double boundary separating T1 via a narrow band of T2 from T3 appears first at the top vertex, spreading downward along the right ($b_p=1/2$) side. This double boundary arises from the narrow separation of the purple and blue surfaces owing to the stronger perturbation by the planet. However, unlike in the case of the horizontal double boundary, this double boundary does not extend at first along the full length of the side. Close to the bottom right vertex it initially swerves to the right side ($p=2.022$ section), and then the broadened T2 layer connects with the T2 region extending from the structure close to the vertex ($p=2.065$ section). The double boundary passes through the structure by gradually connecting to its topology regions ($p=2.243$ section). After crossing the structure, it becomes gradually more parallel to the right side, except close to the bottom of the plot.

The $p=2.243$ section already shows the wide/intermediate boundary extending from the top vertex along the right side. The corresponding sequence of topologies going from the vertex inward is the same as in the Planet in Binary model: T6--T3--T2--T1. The behavior of the boundary is similar to the preceding double boundary. It gradually connects regions in the parallel bands with the regions of the bottom right structure and turns near-parallel to the right side after crossing the structure. In the same section we see that the double boundary from the right side has crossed the structure, at the top of which topology T8 appeared in the narrow band between T7 and T6. In addition, the small T1 and T2 regions along the right side of the plot have disappeared.

Topology T6 appears nearly instantaneously along the bottom edge from the wide/intermediate transition at $p\approx 2.667$ (cyan surface at front face of prism), and the corresponding second nearly horizontal boundary gradually sweeps through the ternary plot to its top vertex. The last T9 topology appears close to the lower edge of the plot, and the progression of the boundary changes T3 regions to T6, T4 to T8, T5 to T7, and T7 to T9, as seen in section $p=2.743$. The progressing double boundaries eliminate topologies T1 and T2 at the left side of the plot around $p\approx 3.65$. The $p=3.655$ section still has T3 and T5 separated by T4. All three disappear at the left side: first T3, nearly instantly followed by T4, and then T5. After T5's disappearance at $p\approx4.799$ at the last intersection of the orange and red surfaces seen in the middle panel of Figure~\ref{fig:HiMSur}, only the four asymptotic topologies T6, T7, T8, and T9 remain, as shown in the last $p=4.799$ section.

Gradually, the three-Einstein-rings T9 topology dominates over the other topologies, just as in any triple lens. As in the previous models, the regions close to the vertices of the plot correspond to the close (T6), intermediate (T7), and wide (T9) regimes of different pairs of the components, with a distant third companion. Thus, the region near the top vertex corresponds to a star with a massive planet (components 1 and 2), near the bottom left vertex a star with a lower-mass planet (components 1 and 3), and near the bottom right vertex a planet with a moon (components 2 and 3). Just as in the previous models, topology T8 appears in the narrow transition region between the close and intermediate regimes, where the perturbation by the distant companion causes one of the two inner loops of T6 to merge sooner than the other with the outer loop.

The general structure of the topology regions in the parameter space of the Hierarchical Masses model (and by extension for any triple-lens model with two low-mass components) is fairly straightforward to understand. The overall division is given by the close/intermediate and wide/intermediate boundaries owing to the star+planet and star+moon two-point-mass lenses. These appear in the ternary plot along the bottom and right sides, respectively, and sweep through the plot to the respective opposite vertices. The sequence in which these boundaries appear is generic: close/intermediate owing to the heavier of the two companions, close/intermediate owing to the lighter, wide/intermediate owing to the lighter, and wide/intermediate owing to the heavier. In this scenario the main topologies are T1, T3, T5, T6, T7, and T9, with T2, T4, and T8 occurring only in the narrow bands along the close/intermediate boundaries.

This overall picture breaks down only near the bottom right star+planet+moon vertex, where the two low-mass components are located close together. All more complex changes, transitions, and the unusual topologies T10 and T11 occur in this region, limited here roughly by $a_p<1/6$.

The final lineup of the 11 topologies of the Hierarchical Masses triple lens shown in the bottom row of Figure~\ref{fig:HiMTernaries} exactly matches the lineup found in the Planet in Binary triple lens. Both models are thus equally rich in terms of the variety of possible critical-curve topologies.

\subsection{Topology Probabilities}
\label{sec:HiM_probabilities}

As in the previous models, we compute the probabilities $\mathcal{P}_{{\rm T}_i}(p)$ of generating a critical curve with topology ${\rm T}_i$ by three randomly placed lens components forming a triangle with perimeter $p$. We evaluate the respective integral in Equation~(\ref{eq:probability_p_final}) for perimeters ranging from $0$ to $p_{\rm max}=4.799$, at the upper end of the last nonasymptotic topology region (T5) in the prism. At any value of $p$ the probabilities add up to unity, $\sum_{i=1}^{11}\mathcal{P}_{{\rm T}_i}(p)=1$.

In the individual panels of Figure~\ref{fig:HiMPerPer} we plot the perimeter dependence of $\log_{10} \mathcal{P}_{{\rm T}_i}(p)$ for topologies T1 through T11 in order of their first appearance in the Hierarchical Masses model. The sequence of appearing and disappearing topologies, which is the same as in the Planet in Binary model, corresponds to the discussion in Section~\ref{sec:HiM_topologies}. Regions of low probability $\mathcal{P}_{{\rm T}_i}(p)\lesssim 0.001$ not visible in the plot occur here, for example, for T2, which extends down to $p=0$ and up to $p\approx 3.65$; for T4, which extends up to $p\approx 4.59$; for T11, which is entirely outside the plot so that only its range of occurrence is indicated by dashed lines; or for T8, which appears already around $p\approx 2.2$.

The sequence of dominant topologies from close to wide regime is T1--T3--T6--T9. Unlike in the Planet in Binary model, topology T7 is suppressed here relative to T6 so that there is no perimeter interval in which it would dominate. In comparison with the Planet in Binary model, the close-limit topologies T1 and T2 extend to even higher perimeters (ending around $p\approx 3.65$); the intermediate T3 and T4 have a reduced range, appearing later and disappearing earlier; T10 and T11 occur in narrower perimeter intervals with lower probabilities; T5 has a reduced range, appearing later and disappearing earlier; the asymptotic T7 and T6 appear at similar perimeters to those in the Planet in Binary model, but their relative probabilities are switched for $p>3$; and the asymptotic T8 and T9 topologies both appear earlier, especially T9, which appears already around $p\approx 2.68$.

Within the limits of the plot, T2 occurs in the widest perimeter range (followed immediately by T1). Just as in the Planet in Binary model, only T10, T11, and T9 cover less than half of the perimeter range, with T11 occurring in the narrowest interval. Generally, the main difference between the two models is the later disappearance of the close-limit topologies and earlier appearance of the final wide-limit topology T9 in the Hierarchical Masses model. As a result, there is even more overlap: all the topologies except T10 and T11 occur at any perimeter $p\in(2.68,3.65)$. Thus, in this perimeter range both close-limit topologies coexist with all wide-limit asymptotic topologies.

For an overall comparison of topologies occurring for configurations with perimeter $p\in(0,\,p_{\rm max}=4.799)$, we evaluate the probabilities $\mathcal{P}_{{\rm T}_i,p_{\rm max}}$ from Equation~(\ref{eq:probability_pmax_final}) normalized so that $\sum_{i=1}^{11}\mathcal{P}_{{\rm T}_i,p_{\rm max}}=1$. The values obtained for the 11 topologies are shown in the ``Hierarchical Masses" column of Table~\ref{tab:Volumes}. Within the given perimeter range, we see that in this model topology T6 replaces T7 as the most frequent topology, occurring in nearly 38\% of cases. On the lowest-probability end, topologies T4, T8, T2, T10, and T11 together occur in $\sim 0.2$\% of cases. Sorted in descending order of probability, the topology sequence is T6 $>$ T9 $>$ T7 $>$ T3 $>$ T1 $>$ T5 $>$ T4 $>$ T8 $>$ T2 $>$ T10 $>$ T11.

Comparing the topologies with the previous model, T1 shows the highest relative increase in probability, due to its expanded perimeter range. On the other hand, the highest relative decrease can be seen in T10, which dropped in terms of perimeter range as well as probabilities. Note that the T11 probabilities were too low to be reliably estimated in either model.

As in the previous models, the numerical error is proportional to the probability corresponding to points on the boundaries inside the prism of all the regions with the given topology. Some of the regions in this model have complex changes of shape with perimeter; some are limited to narrow gaps between close surfaces. In total, $0.03\%$ of pixels were not assigned to any topology region, leaving a probability of $0.03\%$ not attributed to any topology. As in the Planet in Binary model, due to the intricate partitioning of the parameter space, this probability may also serve as an upper bound on the probability of potential topologies undetected for numerical reasons.

\section{Summary and Conclusions}
\label{sec:Summary_conclusions}

We set out to explore the lensing regimes of a triple lens consisting of a given combination of point masses in an arbitrary spatial configuration. We parameterized the configuration by the perimeter of the projected triangle formed by the components, and by two fractional side lengths, as shown in Section~\ref{sec:Parameter_choice}. This permitted us to describe the shape of an arbitrary triangle by a point in a ternary plot. By adding the perimeter as a vertical coordinate perpendicular to the horizontal plot, our full lens configuration parameter space forms a semi-infinite ternary prism.

Based on our previous work \citep{danek_heyrovsky15a,danek_heyrovsky15b}, we describe in Sections~\ref{sec:Close-to-wide} and \ref{sec:Mapping_topologies} a method for computing for a given shape of the triangle the set of perimeter values, at which the lensing regime (described here by the topology of the critical curve) changes. By computing these values for all triangles in the ternary plot, we obtain a set of boundary surfaces dividing the parameter space into regions with different topologies. We then use a series of horizontal $p={\rm const.}$ sections to map the topologies in these regions. In this way we obtain a full 3D map of the critical-curve topologies as a function of spatial configuration of the lens components. As shown in Section~\ref{sec:Probabilities}, the parameter-space division also permits us to compute the relative probabilities of occurrence of the different topologies.

The analysis of the lensing regimes of the two-point-mass lens performed by \cite{schneider_weiss86} and \cite{erdl_schneider93} showed that for any mass ratio there were the same three regimes corresponding to three critical-curve topologies. In this work we performed a similar analysis for three sample triple-lens models defined by their fractional-mass combinations: for three equal masses (the ``Equal Masses'' model); for two equal masses with a low-mass third component (the ``Planet in Binary'' model); and for a hierarchical combination of dominant-, lower-, and lowest-mass components (the ``Hierarchical Masses'' model).

The boundary surfaces computed for the Equal Masses model in Section~\ref{sec:EqM} are shown in Figure~\ref{fig:EqMassSur}. These surfaces partition the parameter space into 39 disjoint regions, which we mapped to find a full set of nine different topologies of the critical curve, sketched at the bottom of Figure~\ref{fig:EqMTernaries}. These topologies also correspond to the full set of topologies found previously in the simpler two-parameter triple-lens models of \cite{danek_heyrovsky15b}.

A similar exploration of the Planet in Binary model in Section~\ref{sec:PiB} (boundary surfaces in Figure~\ref{fig:PiBSur}) and the Hierarchical Masses model in Section~\ref{sec:HiM} (boundary surfaces in Figure~\ref{fig:HiMSur}) showed that both of these models share the same set of 11 different topologies, sketched at the bottom of Figures~\ref{fig:PiBTernaries} and \ref{fig:HiMTernaries}. This demonstrates that unlike in the two-point-mass lens, the number of different lensing regimes of the triple lens depends on the combination of masses of the components. In addition to the nine topologies of the Equal Masses model, these models have two new T10 and T11 topologies, both of which involve doubly nested critical-curve loops (i.e., a loop in a loop in a loop). An example of transitions leading to these topologies is described in Appendix~\ref{sec:Appendix-T10T11}.

The parameter-space division of the Hierarchical Masses model in Figures~\ref{fig:HiMSur} and \ref{fig:HiMTernaries} indicates that for a triple lens with two low-mass components the critical curve may be typically described by a superposition of independent effects due to either component. However, when the two components lie closer together, this simple picture breaks down and more complicated topologies and topology changes occur. This is the case close to the bottom right vertex of the ternary plots in Figure~\ref{fig:HiMTernaries}, which corresponds to lensing by a star with a planet with a moon.

A similar parameter-space mapping may be performed for any other combination of component masses. However, carrying out a full analysis for a general triple lens would require constructing such partitioned ternary prisms as a function of two fractional masses, e.g., $\mu_1\in(0,1)$, $\mu_2\in(0,1)$, $\mu_1+\mu_2<1$. Alternatively, one would need a deeper insight into the variations of critical-curve topologies and their regions in parameter space as a function of the combination of masses. We cannot draw general conclusions based on the results for three component-mass combinations. For example, the close-limit structure of all three models seems to be generic, with topology T1 for nearly any shape and T2 within the purple boundary-surface spikes. However, other mass combinations permit a T3 close-limit topology, as seen in the LS model of \cite{danek_heyrovsky15b} with $\mu_{\rm LS}=1/9$. We conclude that it remains to be seen if the 11 topologies T1--T11 occurring in the studied models constitute the full set of critical-curve topologies for an arbitrary triple lens -- or if there are any other yet undiscovered topologies.

We computed the probabilities of occurrence of different topologies for randomly positioned components with an upper limit on the perimeter. The results in Table~\ref{tab:Volumes} show that the dominant topologies generally include the T7, T6, and T9 wide-limit topologies, plus a combination of T5, T3, or T1. The five remaining topologies have lower probabilities, lowest in the case of the new T10 and T11. The performed analysis permits the computation of more specific probabilities taking into account the orbital configurations of specific triple-lens systems. In addition, the approach chosen in this work is suitable for studying topology changes in nonstatic lens configurations involving orbital motion or other parallax-type microlensing effects.

The structure of observed microlensing light curves is primarily given by the structure of the amplification map, and the lens caustic in particular. The structure of the caustic is primarily linked to the structure of the critical curve. For example, the number of its loops can thus be directly read off from the partitioned parameter-space regions and their topologies. All the computed topology boundaries correspond to beak-to-beak metamorphoses of the caustic. In this caustic metamorphosis two approaching folds touch and separate in the perpendicular direction as two facing cusps (``beaks'').

However, the number of cusps of the triple-lens caustic may also change without affecting the critical-curve topology: in the swallow-tail or butterfly metamorphoses \citep{danek_heyrovsky15a,danek_heyrovsky15b}. In the swallow-tail metamorphosis, a bend appears on a fold caustic and develops into a swallow-tail-like feature formed by a pair of cusps and a self-intersection. In the butterfly metamorphosis, a cusp on the caustic broadens to a nonzero tangent angle and develops into a butterfly-like feature formed by an additional pair of cusps and three self-intersections. While the described metamorphoses increase the number of cusps by two, any of them may proceed in reverse as well, in which case the number of cusps decreases by two.

The partitioning of the parameter space by cusp number would thus involve additional boundary surfaces corresponding to image-plane swallow-tail or butterfly points on the critical curve. For the Equal Masses model \cite{danek_heyrovsky15b} computed the intersections of these boundaries with the vertical faces of the ternary prism (LA model in the right panel of their Figure 8), and with the vertical planes along the ternary-plot medians (TI model in the right panel of their Figure 15). While the former appear simple, the latter indicate that the overall structure of these additional surfaces might be quite intricate. The full analysis of the cusp structure of triple-lens caustics thus remains beyond the scope of this work.

\acknowledgements

We thank the anonymous referee for helpful comments and suggestions. Work on this project was supported by Czech Science Foundation grant GACR 16-17282S.

\vspace{2cm}

\appendix

\section{Conversion of Lens Configuration Parameters}
\label{sec:Appendix-conversion}

\subsection{General and Preferred-frame Positions}
\label{sec:Appendix-conversion-preferred}

The topology of the critical curve does not depend on the choice of origin and orientation of axes of the complex plane used for describing the plane of the sky. Therefore, in this work we illustrate the positions of the lens components $z_1$, $z_2$, $z_3$ for simplicity in a preferred frame defined by setting the origin at the centroid of the components ($z_1+z_2+z_3=0$) and rotating the axes so that the line connecting $z_1$ and $z_2$ is parallel to the real axis (i.e., ${\rm Im}[z_1]={\rm Im}[z_2]$) and $z_2$ lies to the right of $z_1$ (i.e., ${\rm Re}[z_1]<{\rm Re}[z_2]$).

The transformation from positions in a general frame $z_{10}$, $z_{20}$, $z_{30}$ to the preferred frame is then given by
\begin{eqnarray}
\nonumber z_1 &=& (2 z_{10}-z_{20}-z_{30})\, e^{-{\rm i}\,\chi} /\,3\\
          z_2 &=& (2 z_{20}-z_{10}-z_{30})\, e^{-{\rm i}\,\chi} /\,3\\
\nonumber z_3 &=& (2 z_{30}-z_{10}-z_{20})\, e^{-{\rm i}\,\chi} /\,3\;,
\end{eqnarray}
where $\chi=\arg(z_{20}-z_{10})$. The geometry is illustrated in Figure~\ref{fig:parameters}, where the configurations in the general and preferred frames are shown in the left and middle panels, respectively. Note that the inverse transformation is undefined, since the general positions have six degrees of freedom while in the preferred frame there are only three degrees of freedom, sufficient to describe the general triangular configuration. We also point out that the positions used in defining the probability in Equation~(\ref{eq:probability_p_positions}) should be understood as general positions rather than preferred-frame positions. Nevertheless, the independence of the probability on the choice of origin and axes leads to three degrees of freedom canceling out in the next step to Equation~(\ref{eq:probability_p_ternary}).

\subsection{Triangular Parameters}
\label{sec:Appendix-conversion-triangular}

The parameters used in this work to describe the triangular configuration are the perimeter $p$ (used as the parameter defining the size of the configuration) and two of the three fractional side lengths $a_p$, $b_p$, $c_p$ (used as the two parameters defining the shape of the configuration). In terms of the component positions,
\begin{eqnarray}
\nonumber   p &=& |z_3-z_2|+|z_1-z_3|+|z_2-z_1|\\
\nonumber a_p &=& |z_3-z_2| /\,p\\
          b_p &=& |z_1-z_3| /\,p\\
\nonumber c_p &=& |z_2-z_1| /\,p\;,
\end{eqnarray}
where the fractional lengths are connected by $a_p+b_p+c_p=1$, and the formulae are valid even if the preferred-frame positions are replaced by the general ones.

For deriving the inverse transformation we select parameters $b_p$ and $c_p$. The results can be easily converted to any other pair of fractional side lengths. First, we point out that the inversion has an inherent degeneracy. The parameter combination $p$, $b_p$, $c_p$ does not identify whether the third component lies above or below the first two in the complex plane. The two possible triangles are mirror images, so that their critical-curve topology and other properties are the same. Nevertheless, when computing the component positions, one or the other option should be specified.

We start with an intermediate triangle with vertices $0$, $p\,c_p$, $p\,b_p\,e^{{\rm i}\psi}$, which we shift so that its centroid lies at the origin. The angle $\psi$ from $c$ to $b$ at $z_1$ can be computed in terms of $b_p$ and $c_p$ by combining two identities for $a_p$, the third fractional side:
\beq
1-b_p-c_p=a_p=\sqrt{b^2_p+c^2_p-2 b_p c_p\cos{\psi}}\,.
\label{eq:a_p-elimination}
\eeq
Since we obtain $\cos{\psi}$, the two permitted solutions are $\psi\in[0,\pi]$ and $2\pi-\psi$. The resulting expressions for the transformation are
\begin{eqnarray}
\label{eq:abp-to-zi}
\nonumber z_1 &=& -p\,\left[2(b_p+c_p-b_p c_p)+2 c^2_p-1 \pm{\rm i}\sqrt{(1-2 b_p)(1-2 c_p)(2 b_p+2 c_p-1)}\right] /\,(6 c_p)\\
          z_2 &=& -p\,\left[2(b_p+c_p-b_p c_p)-4 c^2_p-1 \pm{\rm i}\sqrt{(1-2 b_p)(1-2 c_p)(2 b_p+2 c_p-1)}\right] /\,(6 c_p)\\
\nonumber z_3 &=& p\,\left[2(b_p+c_p-b_p c_p)-c^2_p-1 \pm{\rm i}\sqrt{(1-2 b_p)(1-2 c_p)(2 b_p+2 c_p-1)}\right] /\,(3 c_p)\;,
\end{eqnarray}
where the plus or minus sign before the square root places $z_3$ above or below $z_1$ and $z_2$, respectively.

\subsection{Microlensing Parameters}
\label{sec:Appendix-conversion-microlensing}

In the analysis of triple microlensing events the configuration is typically parameterized by two side lengths $s_2$, $s_3$ from the first to the two other components and the angle $\psi$ subtended by them \citep[e.g.,][]{han_etal17b}. The right panel of Figure~\ref{fig:parameters} illustrates the geometry in terms of these parameters. With component 1 representing usually the most massive component, the parameter definitions in terms of the preferred-frame positions and the triangular parameters used in this work are
\begin{eqnarray}
\nonumber s_2 &=& |z_2-z_1| = p\,c_p\\
          s_3 &=& |z_3-z_1| = p\,b_p\\
\nonumber \psi &=& \arg(z_3-z_1) = \pi\pm\{-\pi+\arccos[\,b^{-1}_p+c^{-1}_p-(2b_pc_p)^{-1}-1]\}\;,
\end{eqnarray}
where the last equality comes from Equation~(\ref{eq:a_p-elimination}), and the plus or minus sign is used when component 3 is above or below components 1 and 2, respectively.

The inverse transformation to the preferred-frame positions can be derived by shifting the centroid of a triangle with vertices $0$, $s_2$, $s_3\,e^{{\rm i}\,\psi}$ to the origin. We get
\begin{eqnarray}
\nonumber z_1 &=& -(s_2+s_3\,e^{{\rm i}\,\psi}) /\,3\\
          z_2 &=& (2s_2-s_3\,e^{{\rm i}\,\psi}) /\,3\\
\nonumber z_3 &=& (2s_3\,e^{{\rm i}\,\psi}-s_2) /\,3\;.
\end{eqnarray}
The transformation to the triangular parameters is given by
\begin{eqnarray}
\nonumber   p &=& s_2+s_3+(s^2_2+s^2_3-2s_2s_3\cos{\psi})^{1/2}\\
\nonumber a_p &=& (s^2_2+s^2_3-2s_2s_3\cos{\psi})^{1/2}\,/\,p\\
          b_p &=& s_3 /\,p\\
\nonumber c_p &=& s_2 /\,p\;.
\end{eqnarray}
Finally, in microlensing notation fractional masses are usually replaced by relative masses with respect to the most massive component,
\begin{eqnarray}
          q_2 &=& M_2/M_1=\mu_2/\mu_1\\
\nonumber q_3 &=& M_3/M_1=\mu_3/\mu_1\;.
\end{eqnarray}
The fractional masses are then obtained from
\begin{eqnarray}
\nonumber \mu_1 &=& 1/(1+q_2+q_3)\\
          \mu_2 &=& q_2/(1+q_2+q_3)\\
\nonumber \mu_3 &=& q_3/(1+q_2+q_3)\;.
\end{eqnarray}

\section{Critical-curve Topologies T10 and T11}
\label{sec:Appendix-T10T11}

We illustrate here the critical-curve transitions leading to topologies T10 and T11 with doubly nested loops, newly discovered in the Planet in Binary and Hierarchical Masses models (see Sections~\ref{sec:PiB_topologies} and \ref{sec:HiM_topologies}, respectively). We also demonstrate the corresponding changes in the caustic. As an example, we select the Planet in Binary model with fractional masses $(\mu_1,\,\mu_2,\,\mu_3)=(0.49995,0.49995,0.0001)$ in an isosceles configuration $(a_p,\,b_p,\,c_p)=(0.31276,0.31276,0.37448)$, which lies along the median leading from the top vertex to the bottom side of the ternary plots in Figure~\ref{fig:PiBTernaries}.

The Jacobian contours, cusp curve, and morph curve of the configuration are shown in Figure~\ref{fig:T11curves}. The blue contour marks the critical curve for perimeter $p=1.89$. In the top row we present a view of the full critical curve; in the bottom row we present a detailed view of the vicinity of the planet. Overall, the critical curve resembles a binary-lens critical curve in the close regime, with an outer + two inner loops, two poles (at the positions of the binary components), two Jacobian maxima (inside the inner loops), and three saddle points (along the imaginary axis). However, placing the planet inside the top inner loop leads to additional triple-lens features seen in the detail: an additional (doubly nested) loop surrounding the planet, an extra pole, two extra maxima (on both sides of the planet), and three extra saddle points (surrounding the planet). Note also that the addition of the planet to a binary lens near its close/intermediate transition causes the top inner loop to connect with the outer loop later than the bottom inner loop. Since the bottom loop is already connected at this perimeter, the critical curve has the T11 topology (outer + inner + doubly nested loops).

The orange cusp curve, which is added in the middle and right columns, generally has the simple structure of the binary-lens cusp curve \citep[see the left panel of Figure 2 in][]{danek_heyrovsky15a}. The detail shows the additional figure-eight structure close to the planet, passing through the two additional off-axis saddle points and branching at the additional maxima. Counting its intersections with the critical curve, we find that the caustic loop corresponding to the outer critical-curve loop has five cusps, the caustic loop corresponding to the inner loop has three cusps, and the caustic loop corresponding to the doubly nested loop has four cusps.

The green morph curve, which is added in the right column, also has the general structure of a binary-lens morph curve \citep[see the right panel of Figure 2 in][]{danek_heyrovsky15a}. The detail shows the additional disconnected loopy structure around the planet. The morph curve has only two intersections with the cusp curve away from the saddle points and lens positions, both on the imaginary axis: one above the planet, the other below the planet. Both occur at branching points of the cusp curve, and thus they are butterfly points. Their presence indicates that the caustic of this lens configuration undergoes two butterfly metamorphoses with changing perimeter. For details see Section~6 in \cite{danek_heyrovsky15a}.

Figure~\ref{fig:T10T11sequence} illustrates the sequence of critical curves (global in column 1, detail in column 2) and caustics (global in column 3, detail of lower part in column 4) of the configuration as the perimeter increases from the bottom to the top row. At $p=1.80$ in the bottom row the critical curve has the T3 topology (outer + two inner loops). The central caustic loop with four cusps corresponds to the outer loop of the critical curve, the top loop with three cusps corresponds to the bottom inner loop, and the bottom loop with seven cusps corresponds to the top inner loop. The number of cusps of the bottom loop of the caustic can be checked by counting the intersections of the cusp curve in the bottom middle panel of Figure~\ref{fig:T11curves}
with the Jacobian contour corresponding to the $p=1.80$ critical curve (the second contour to the right and left of the off-axis saddle points).

The top inner loop of the critical curve wraps around the planet and connects at the saddle point above the planet, generating the doubly nested loop around the planet. This can be seen in the next $p=1.88$ row, where the caustic has the T10 topology (outer + two inner + doubly nested loops). At the same time the bottom loop of the caustic disconnected a small loop with four cusps (corresponding to the doubly nested loop) via a beak-to-beak metamorphosis, leaving the bottom loop at first with five cusps. A reverse butterfly metamorphosis at the top of the loop then reduces the cusp number on the bottom loop to three, as seen in the right column. This occurs when the critical curve passes through the butterfly point above the planet in the bottom right panel of Figure~\ref{fig:T11curves}.

The $p=1.89$ row shows the situation from Figure~\ref{fig:T11curves}, in which the bottom inner loop connected with the outer loop, changing the topology to T11 (outer + inner + doubly nested loops). The top and central loops of the caustic connected via a reverse beak-to-beak metamorphosis, forming a loop with five cusps, as demonstrated above. By the top $p=1.95$ row the top inner loop of the critical curve connected with the outer loop as well (at the topmost saddle point), changing the topology to T7 (two separate loops) by releasing the doubly nested loop as a separate critical-curve loop around the planet. The bottom loop of the caustic connects with the large loop via another reverse beak-to-beak metamorphosis, forming a loop with six cusps and leaving a separate planetary loop with four cusps.

The entire sequence shown in Figure~\ref{fig:T10T11sequence} occurs in a narrow perimeter interval located between the $p=1.690$ and $p=2.270$ sections in the top row of Figure~\ref{fig:PiBTernaries}. Increasing the perimeter to higher values would lead to the final transition, corresponding to the crossing of the red surface in Figure~\ref{fig:PiBSur}. In it the large loop of the critical curve disconnects at the central saddle point into two separate loops of either of the heavier components, thus achieving the asymptotic T9 topology (three separate loops). The large loop of the caustic would split vertically in a beak-to-beak metamorphosis into two four-cusped loops.

Similarly, reducing the perimeter from $p=1.80$ to lower values leads to one final transition, in which two extra critical-curve loops around the two off-axis maxima close to the planet disconnect simultaneously from the top inner loop, leading to the close-limit T1 topology (outer + four inner loops). In Figure~\ref{fig:PiBSur} this corresponds to crossing an intersection of the blue and purple surfaces. The corresponding bottom loop of the caustic undergoes two simultaneous beak-to-beak metamorphoses splitting off two small three-cusped loops and leaving a loop initially with five cusps. For a lower perimeter value the critical curve then passes through the butterfly point below the planet in the bottom right panel of Figure~\ref{fig:T11curves}, and the five-cusped caustic loop undergoes a reverse butterfly metamorphosis to a three-cusped loop.

The entire topology sequence from the close to the wide limit for the presented configuration thus is T1--T3--T10--T11--T7--T9. Due to the isosceles symmetry (including the masses), there are only six topologies and five transitions, since the purple and blue boundary surfaces in Figure~\ref{fig:PiBSur} are crossed simultaneously. The entire sequence including the changes in caustic structure can be determined from the contours and curves presented in Figure~\ref{fig:T11curves}.

\clearpage
\bfi
{\centering
\vspace{2cm}
\includegraphics[width=12 cm]{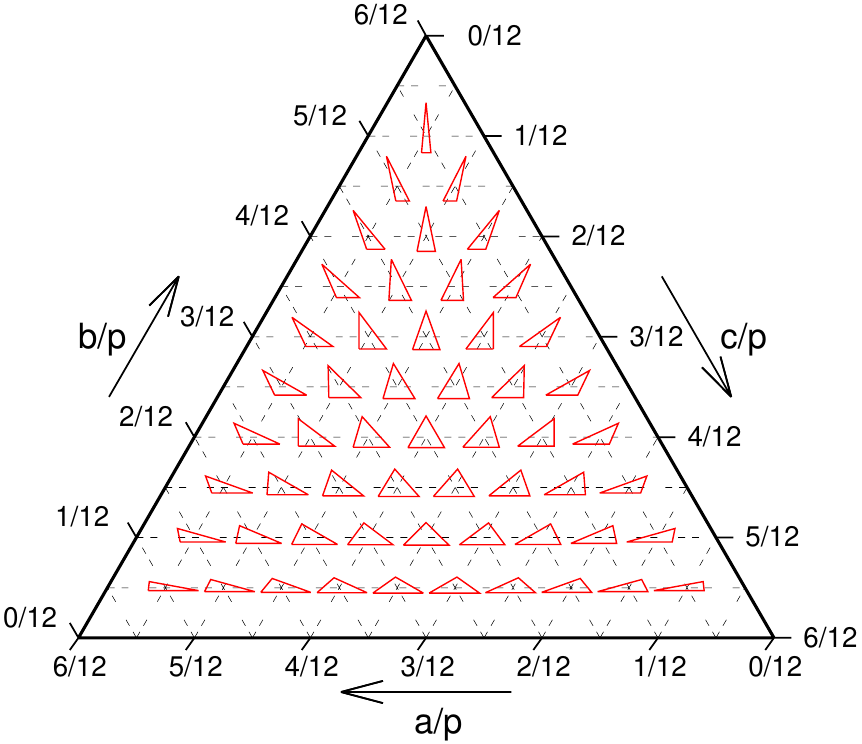}
\caption{Ternary plot of triple-lens configuration shapes parameterized by side lengths as fractions of the perimeter $p$ of the projected triangle formed by the lens components. Red triangles indicate the shapes corresponding to the intersection points of the dashed parameter grid. Their right, left, and bottom sides are marked $a,b$, and $c$, respectively. For more details see Section~\ref{sec:Parameter_choice}.}}
\label{fig:LittleTriangles}
\efi

\clearpage
\bfi
{\centering
\begin{tabular}{cc}
  \begin{tabular}{c}
    \\
    \hspace{0.3 cm} \includegraphics[width=8 cm]{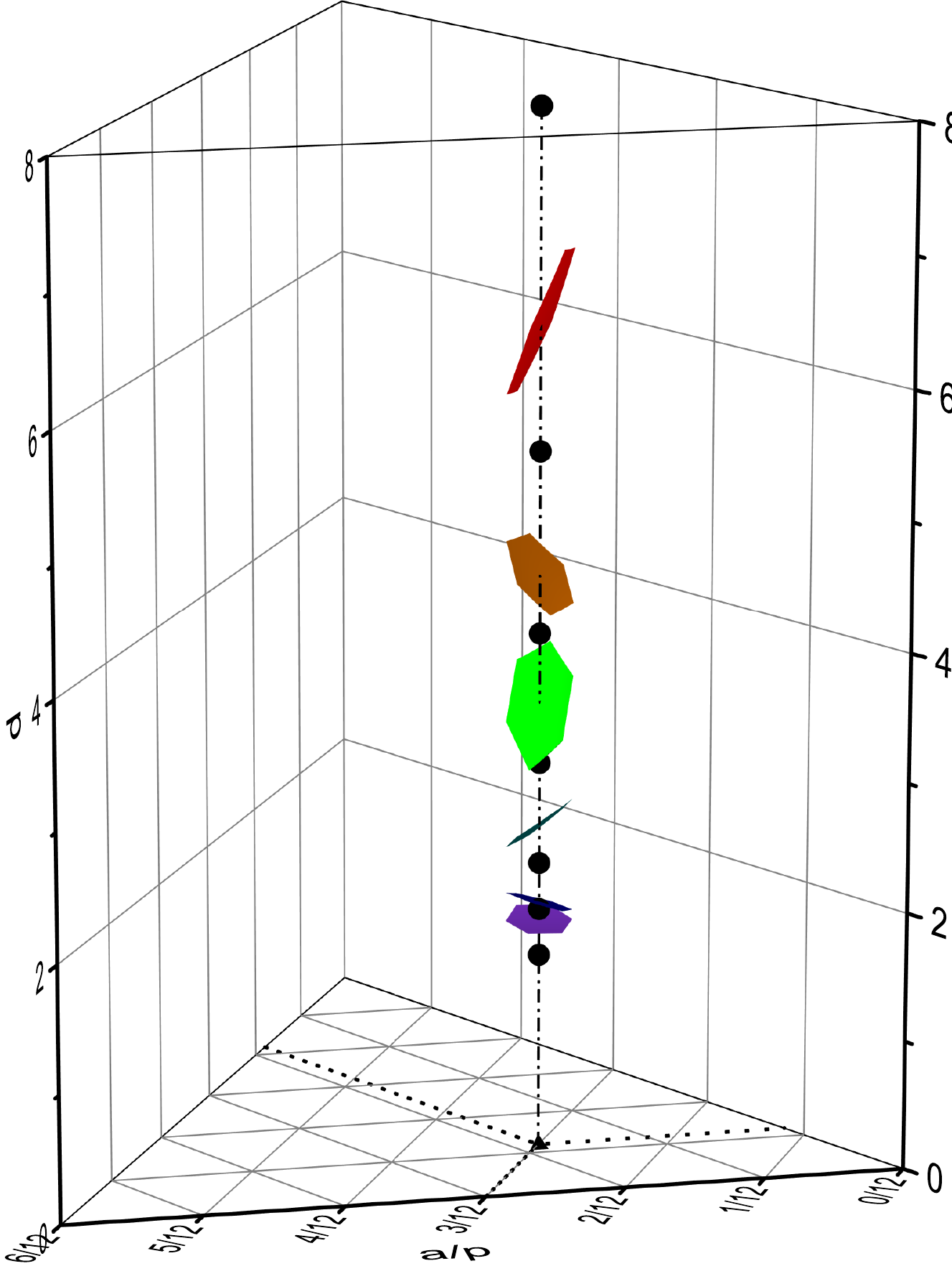} \\
    \\
    \hspace{-0.5 cm} \includegraphics[width=8 cm]{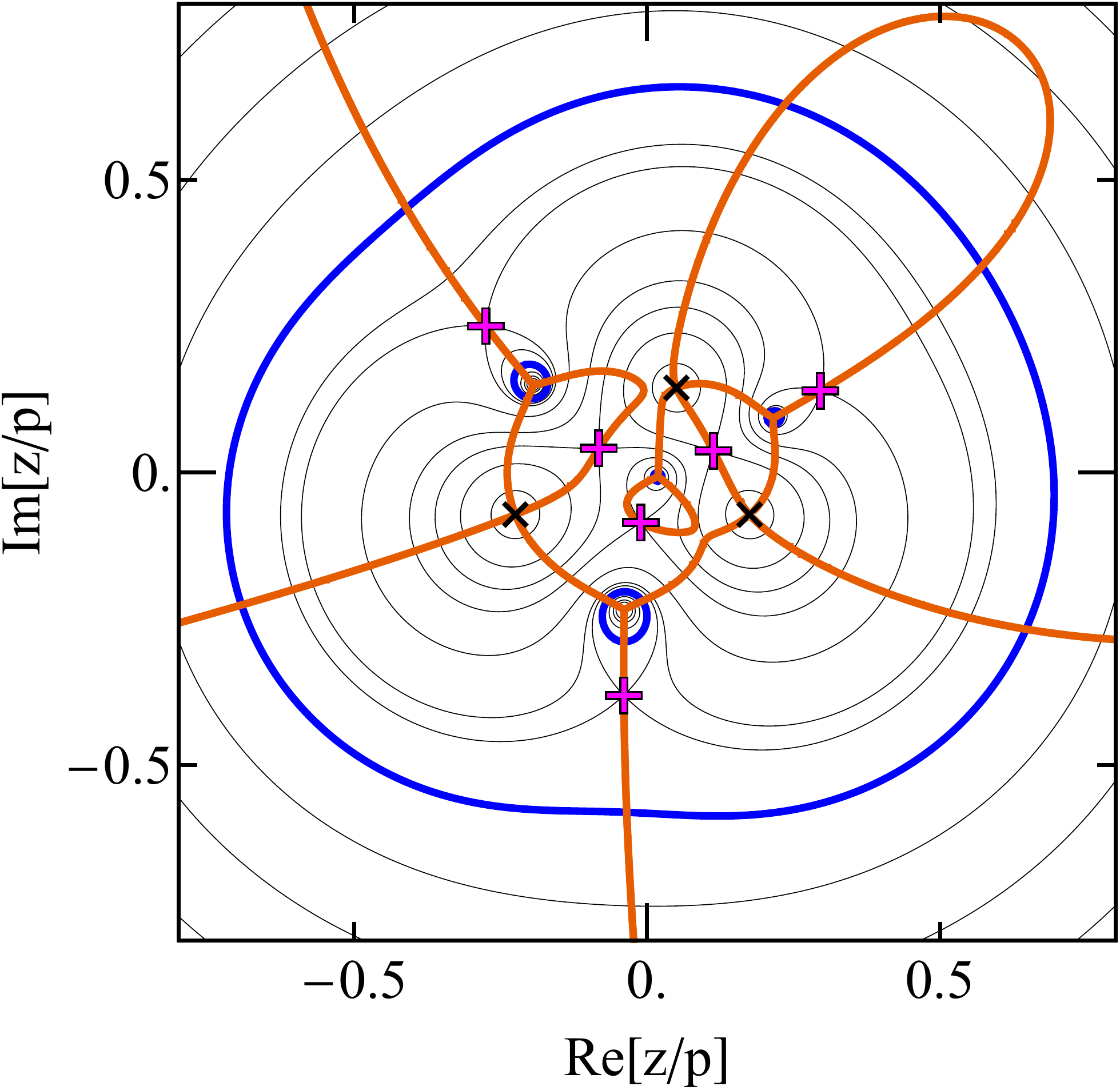}
  \end{tabular}
  &
  \begin{tabular}{c}
     \\
    \includegraphics[width=6.0 cm]{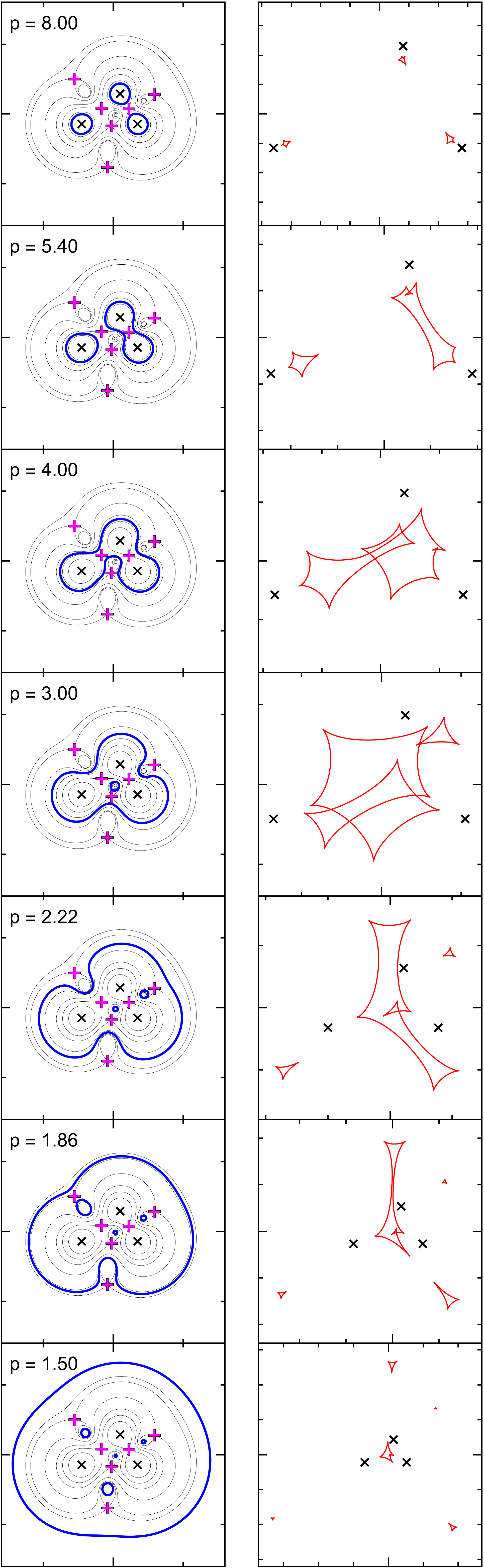} \\
  \end{tabular}
\end{tabular}
\caption{Critical-curve topology changes with perimeter $p$ for an equal-mass triple lens with fractional side lengths $(a_p,\,b_p,\,c_p)=(0.25, 0.35, 0.4)$. Top left: six boundaries (polygons) separating seven topologies along the corresponding vertical line in ternary-prism parameter space. Bottom left: image-plane plot with axes in units of $p$, lens positions (black crosses), and Jacobian saddles (magenta plus signs). Jacobian contours correspond to critical curves for $p=$ 0.92, 1.06, 1.25, 1.50 (blue), 1.776, 1.917, 2.519, 3.471, 4.454, 6.346, and 14 (from corners to lenses). The cusp curve (orange) marks positions of cusp images on contours. Right: critical curves (blue) and caustics (red) for $p$ values marked by bullets in the top left panel. See Section~\ref{sec:Close-to-wide} for details.}}
\label{fig:EqMassSeq}
\efi

\clearpage
\bfi
{\centering
\vspace{1.5 cm}
\includegraphics[width=12.0 cm]{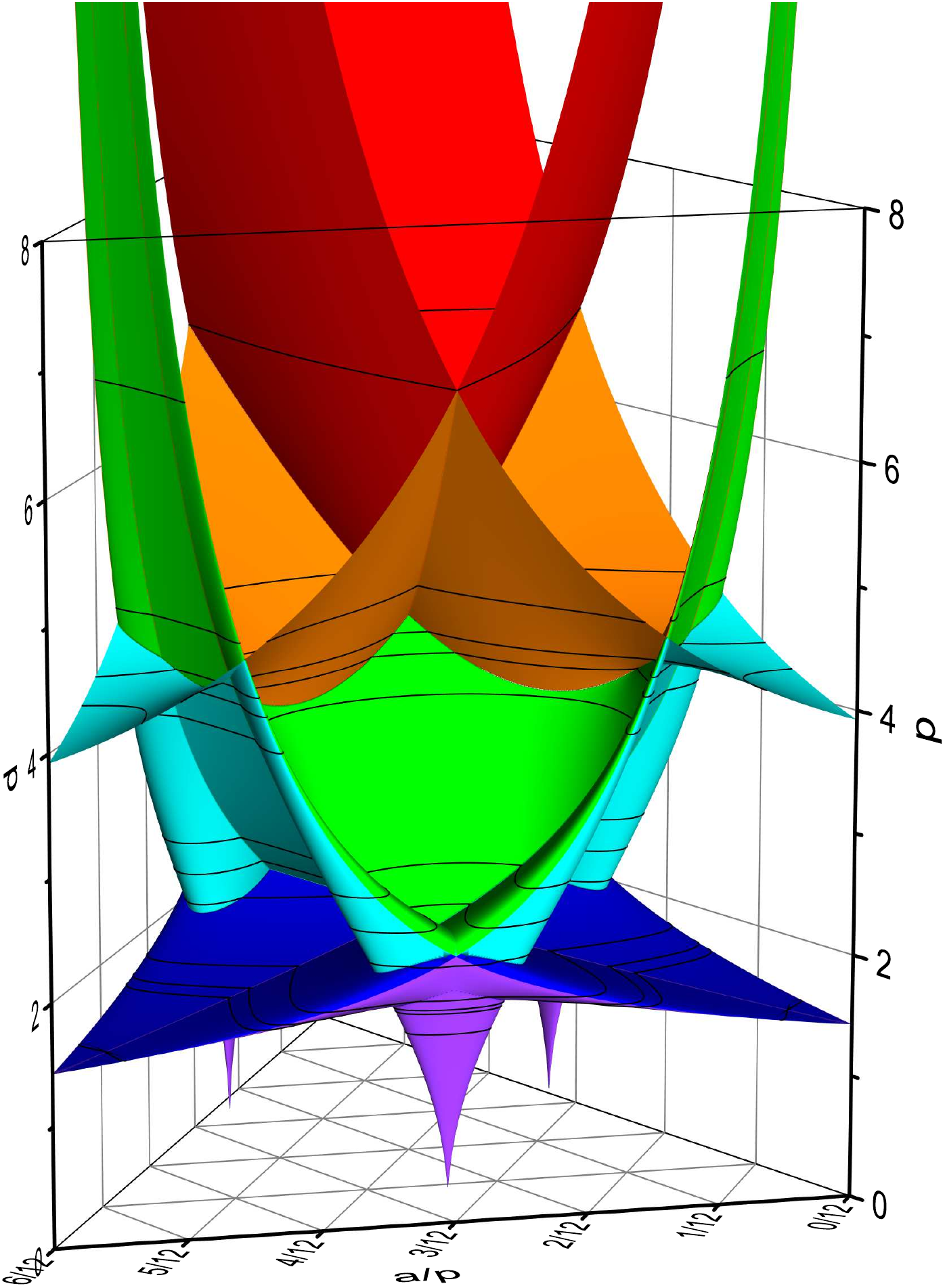}
\caption{Equal Masses: boundaries in the lens configuration parameter space separating different critical-curve topologies. For an arbitrary spatial configuration the shape is defined by the horizontal position in the ternary plot (see Figure~\ref{fig:LittleTriangles}); the size is defined by the vertically plotted perimeter $p$. The boundaries are colored in order from closest to widest transition for any shape: purple, blue, cyan, green, orange, red. Black contours identify the horizontal sections shown in Figure~\ref{fig:EqMTernaries}. An animated version of this figure is available in the online version of this article, showing the prism rotated full-circle around its vertical axis in $\pi/18$ steps.}}
\label{fig:EqMassSur}
\efi

\clearpage
\bfi
{\centering
\vspace{1.5 cm}
\includegraphics[clip,scale=0.35]{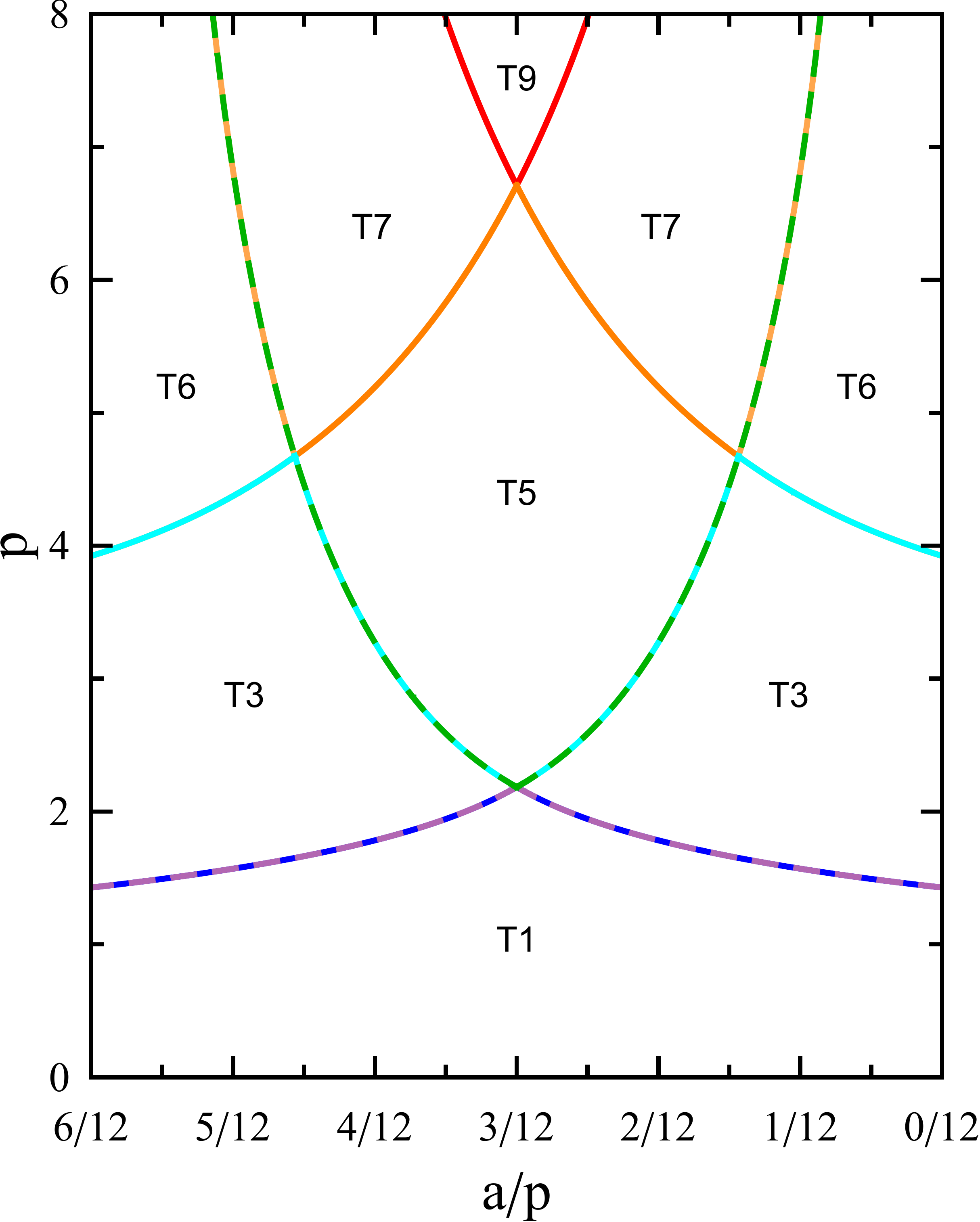}
\hspace{0.5cm}
\includegraphics[clip,scale=0.35]{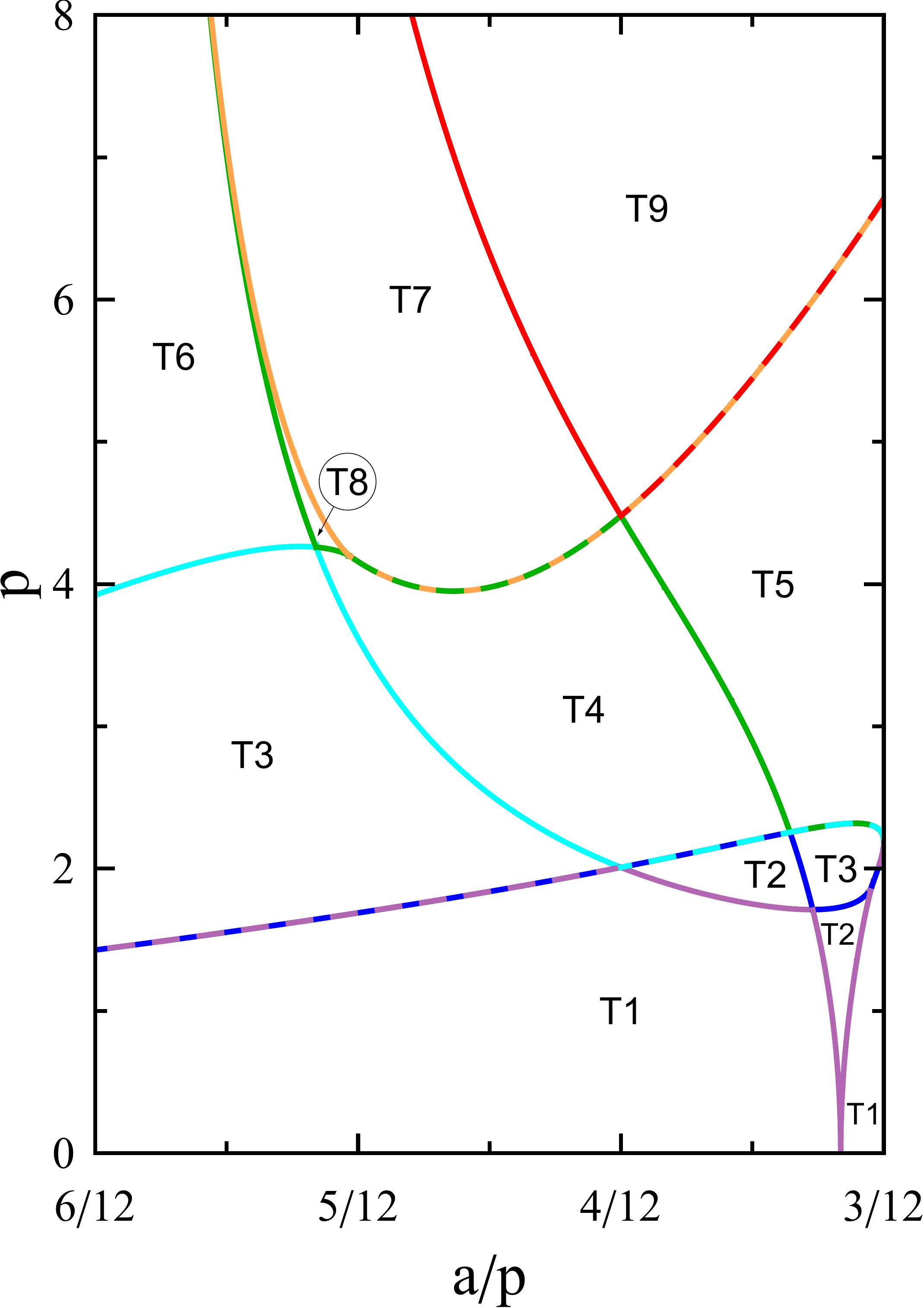}
\caption{Equal Masses: topology boundaries on vertical planes of the ternary prism from Figure~\ref{fig:EqMassSur} studied in detail by \cite{danek_heyrovsky15b}. Left panel: front face of prism (corresponding to collinear configurations); right panel: symmetry plane of prism from left edge to vertical midline of opposite face (corresponding to isosceles configurations). Labels T1--T9 mark the critical-curve topology in each region, as shown at the bottom of Figure~\ref{fig:EqMTernaries}.}}
\label{fig:EqMassSur-planes}
\efi

\clearpage
\bfi
{\centering
\vspace{2cm}
\includegraphics[width=\textwidth]{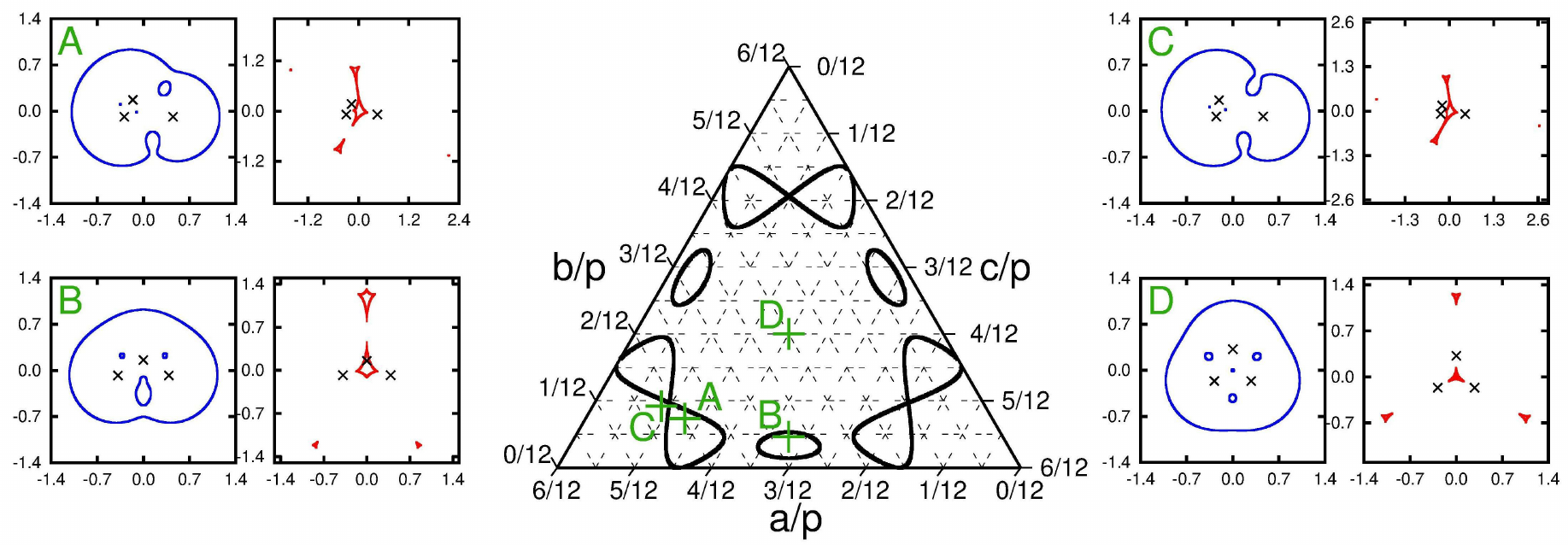}
\caption{Equal Masses: topologies on a horizontal section of the ternary prism from Figure~\ref{fig:EqMassSur} at perimeter $p=1.679$ with sample critical curves and caustics. Middle panel: ternary plot of the section, with black curves indicating topology boundaries. Side panels: critical curves (blue), caustics (red), and lens positions (black crosses) corresponding to configurations marked by points with coordinates $(a_p,\,b_p,\,c_p)$ in the ternary plot: A $(0.393,\,0.168,\,0.439)$, B $(0.269,\,0.269,\,0.462)$, C $(0.426,\,0.151,\,0.423)$, D $(1/3,\,1/3,\,1/3)$. See Section~\ref{sec:EqM_topologies} for more details.}}
\label{fig:EqMPsection}
\efi

\clearpage
\bfi
{\centering
\vspace{1cm}
\includegraphics[width=16.8 cm]{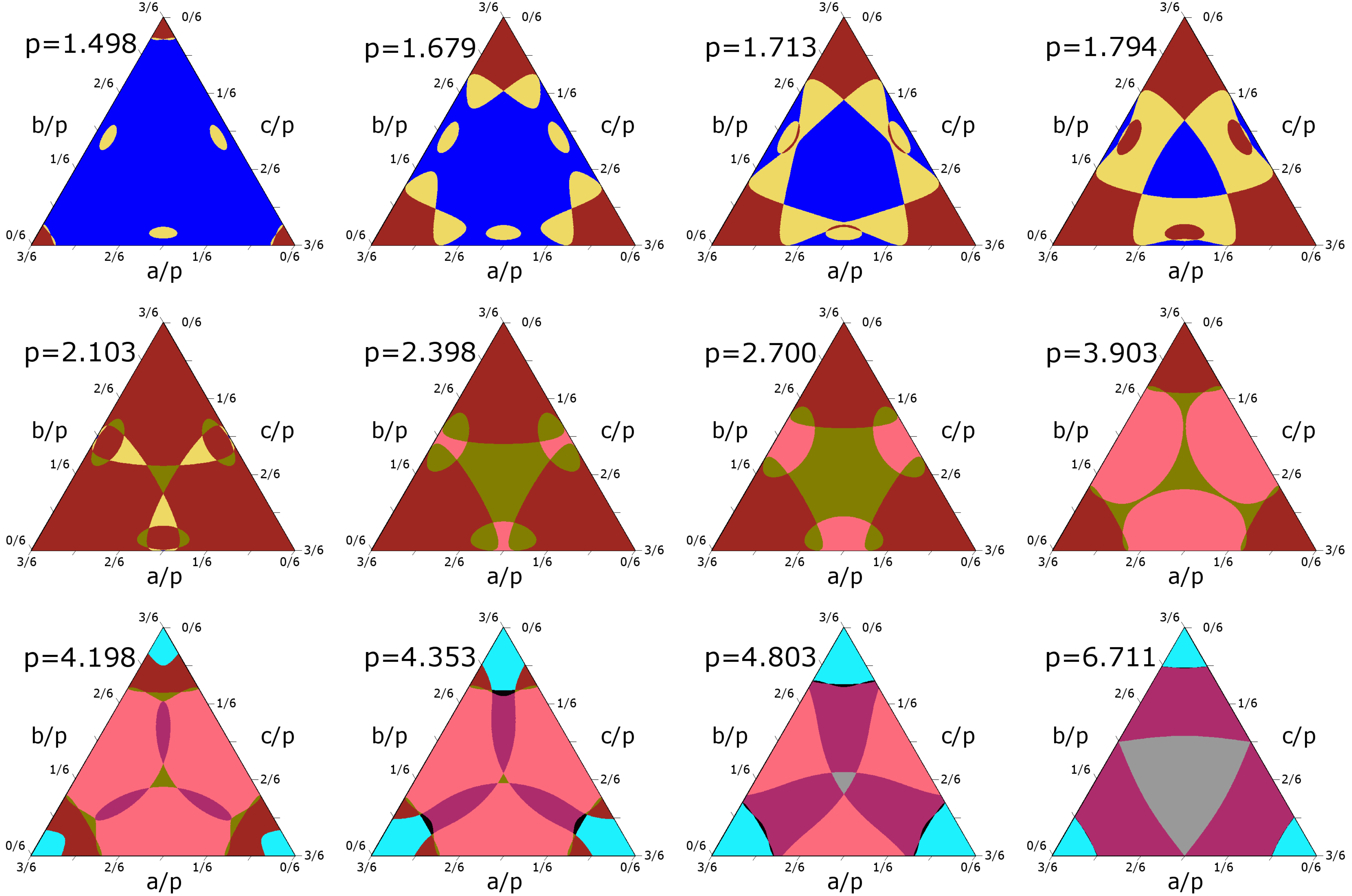}
\vspace{2mm}
\\
\begin{tabular}{ccccccccc}
\includegraphics[height=2.10 cm]{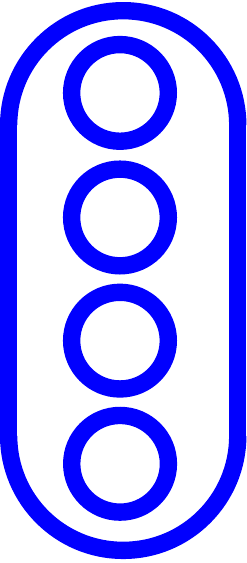}
&
\includegraphics[height=2.10 cm]{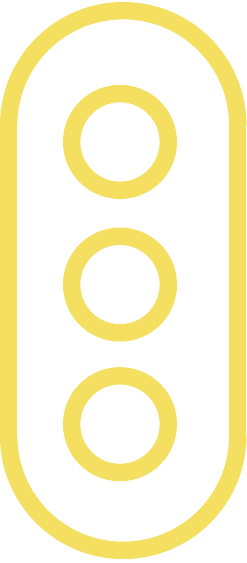}
&
\includegraphics[height=2.10 cm]{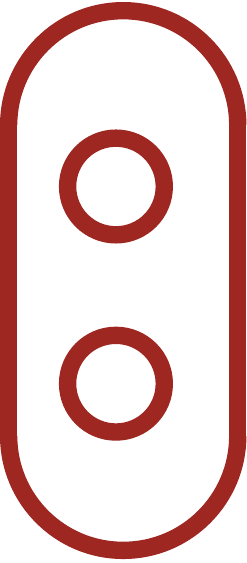}
&
\includegraphics[height=2.10 cm]{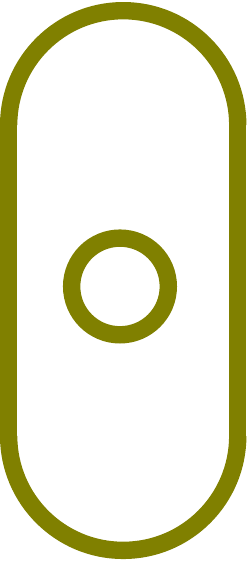}
&
\includegraphics[height=2.10 cm]{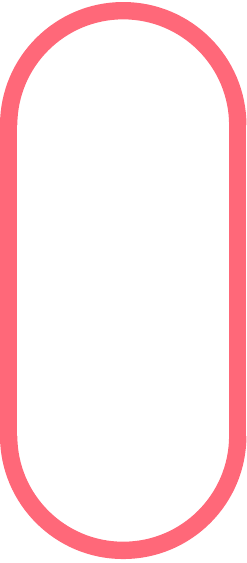}
&
\includegraphics[height=2.10 cm]{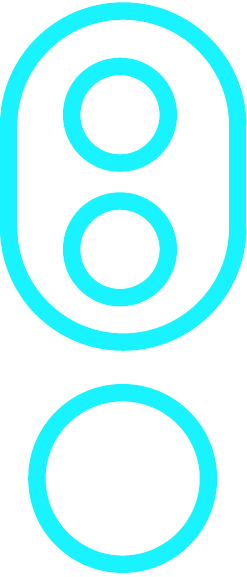}
&
\includegraphics[height=2.10 cm]{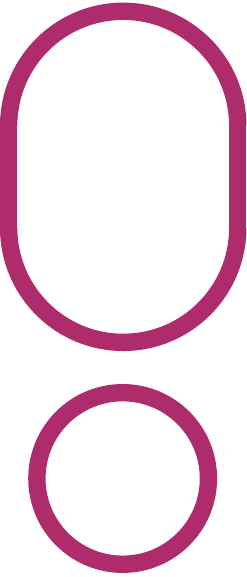}
&
\includegraphics[height=2.10 cm]{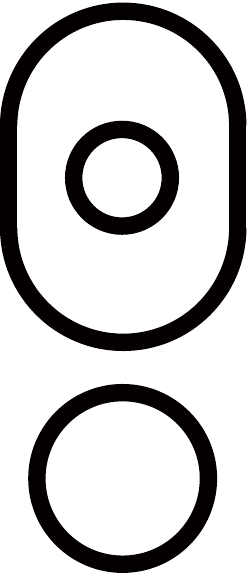}
&
\includegraphics[height=2.10 cm]{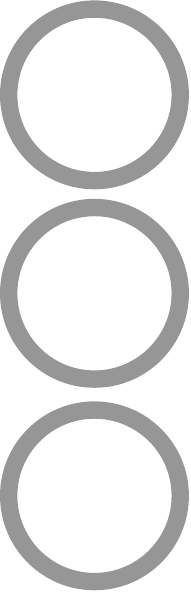}
\\
T1&T2&T3&T4&T5&T6&T7&T8&T9
\end{tabular}
\caption{Equal Masses: critical-curve topologies on a sequence of horizontal ternary-plot sections of the prism from Figure~\ref{fig:EqMassSur} at perimeters marked next to the individual plots. Colors correspond to the nine different topologies T1--T9 sketched in the bottom row. The plot sequence is ordered by perimeter from the closest $p=1.498$ at top left to the widest $p=6.711$ at bottom right.}}
\label{fig:EqMTernaries}
%\vspace{-0.1cm}
\efi

\clearpage
\bfi
{\centering
\vspace{1cm}
\includegraphics[width=8.6 cm]{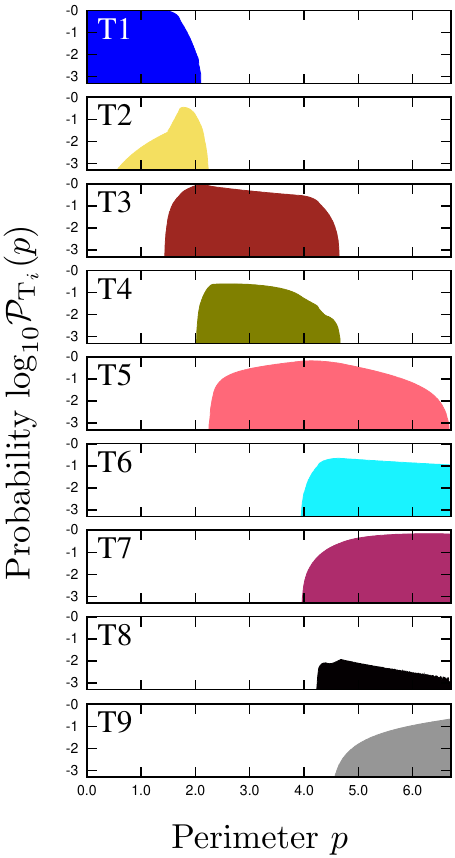}
\caption{Equal Masses: probability $\mathcal{P}_{{\rm T}_i}(p)$ of occurrence of topology ${\rm T}_i$ at perimeter $p$. Topologies are marked and color-coded following the key in the bottom row of Figure \ref{fig:EqMTernaries}.}}
\label{fig:EqMPerPer}
\efi

\clearpage
\bfi
{\centering
\vspace{1.5 cm}
\includegraphics[clip,scale=0.35]{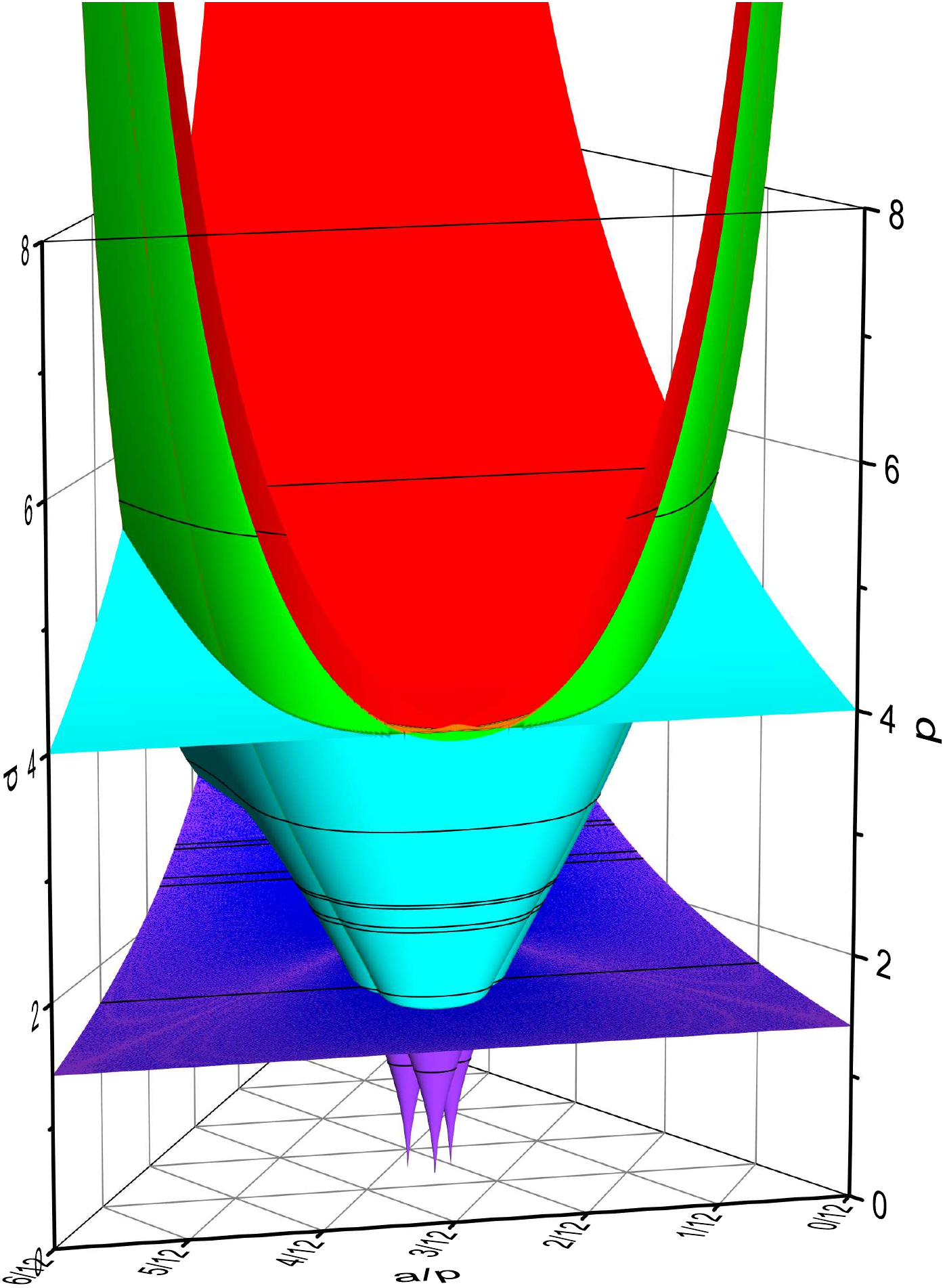}
\includegraphics[clip,scale=0.35]{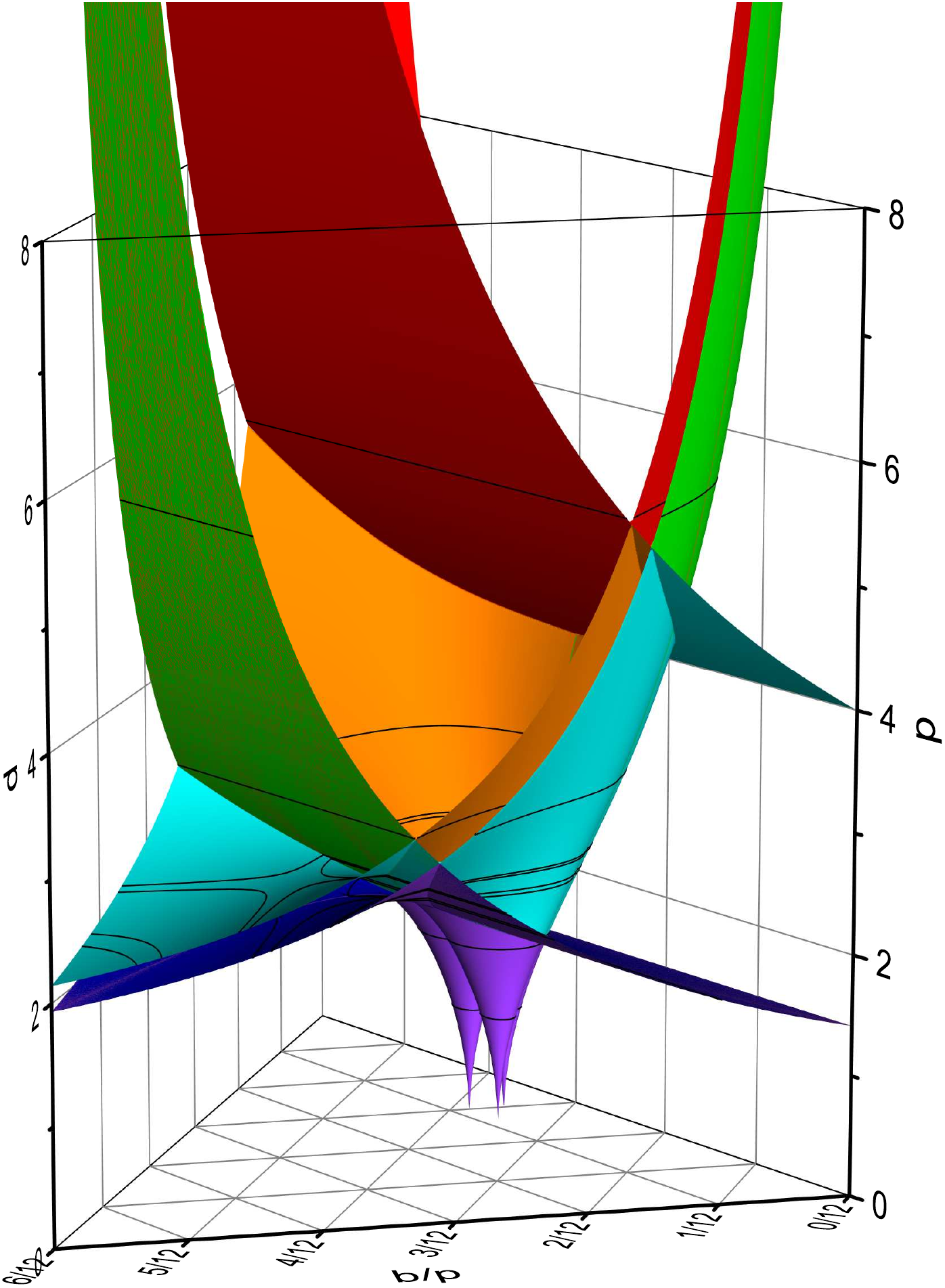}
\includegraphics[clip,scale=0.35]{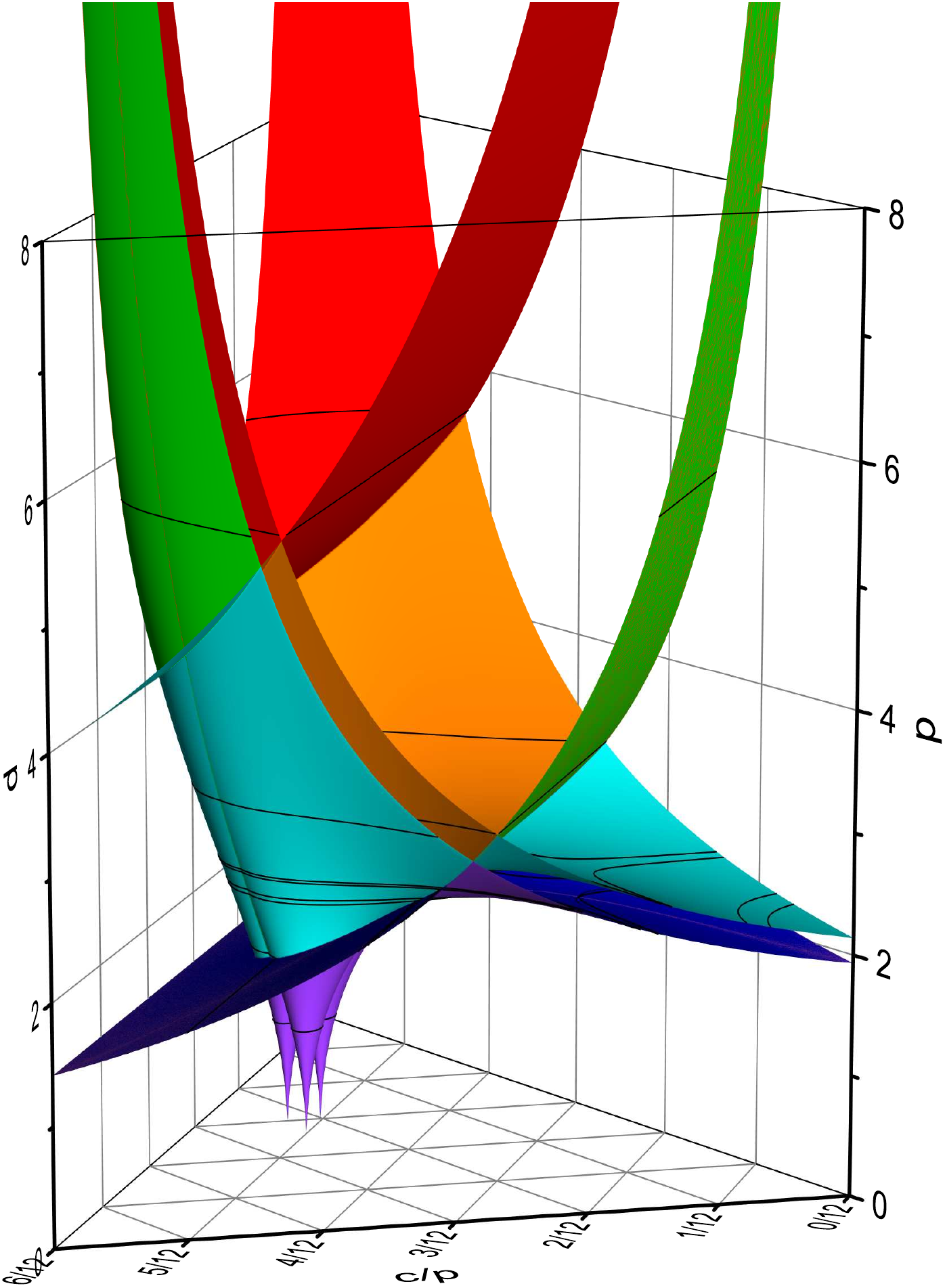}
\caption{Planet in Binary: boundaries in the lens configuration parameter space separating different critical-curve topologies. The three views of the surfaces are rotated by $2\pi/3$ around the vertical axis of the prism so that ternary axis $a_p$ (left), $b_p$ (middle), or $c_p$ (right) lies at front. Notation as in Figure~\ref{fig:EqMassSur}. Black contours identify the horizontal sections shown in Figure~\ref{fig:PiBTernaries}. An animated version of this figure is available in the online version of this article, showing the prism rotated full-circle around its vertical axis in $\pi/18$ steps.}}
\label{fig:PiBSur}
\efi

\clearpage
\bfi
{\centering
\vspace{2cm}
\includegraphics[width=16.8 cm]{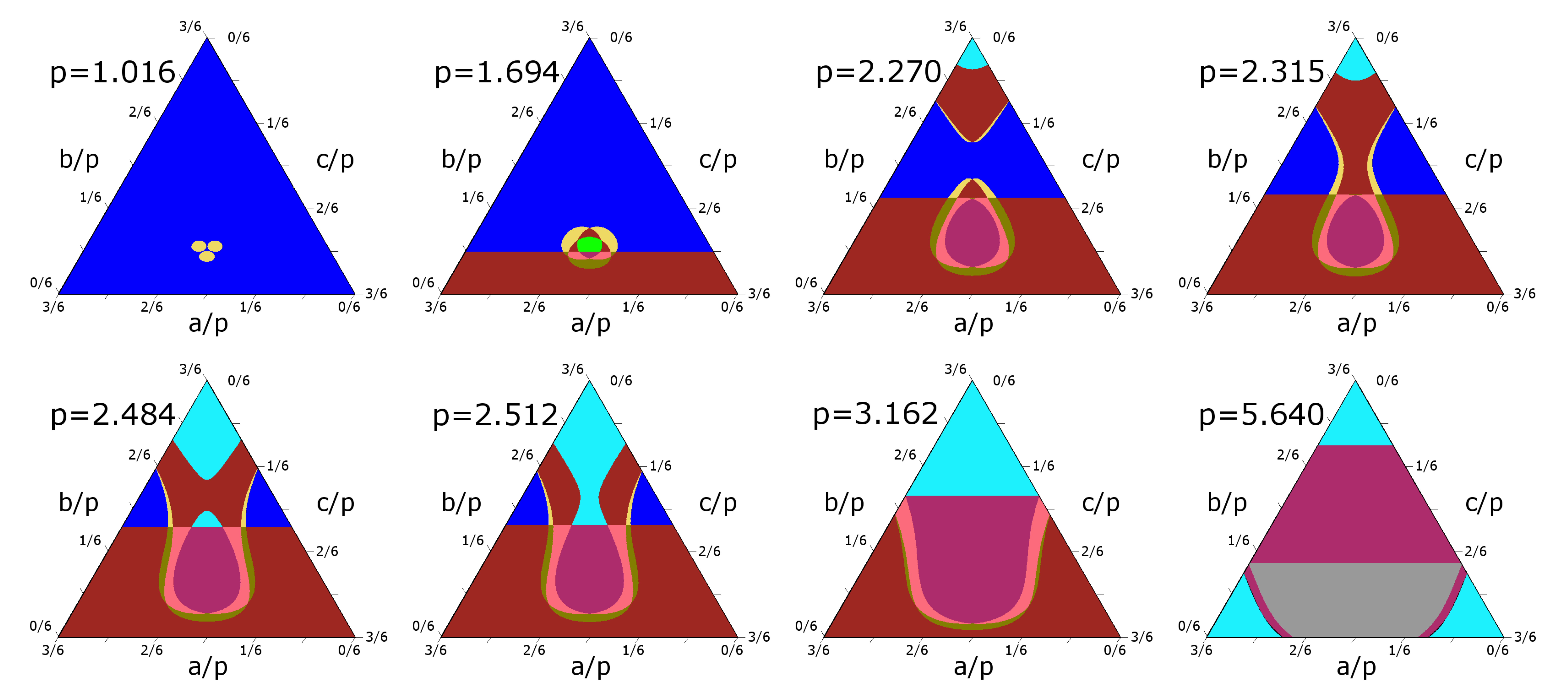}
\vspace{2mm}
\\
\begin{tabular}{ccccccccccc}
\includegraphics[height=2.10 cm]{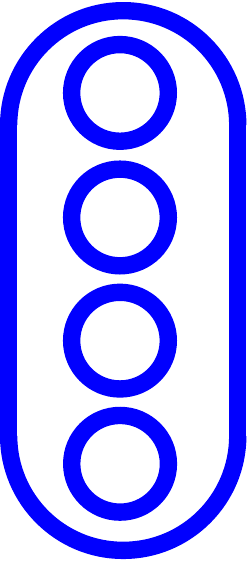}
&%\hspace{0.7mm}
\includegraphics[height=2.10 cm]{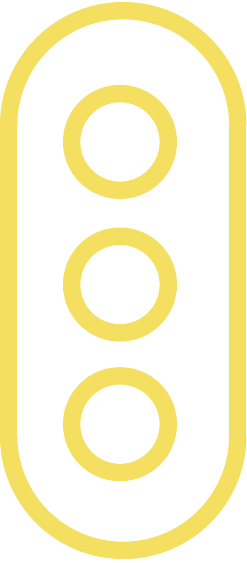}
&%\hspace{0.7mm}
\includegraphics[height=2.10 cm]{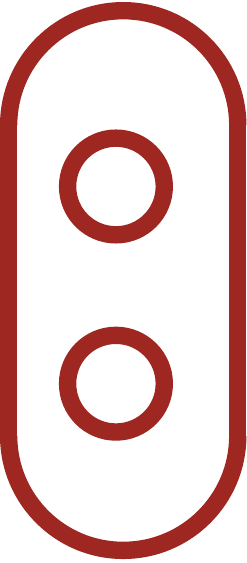}
&%\hspace{0.7mm}
\includegraphics[height=2.10 cm]{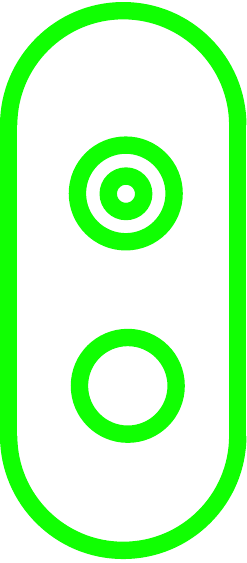}
&%\hspace{0.7mm}
\includegraphics[height=2.10 cm]{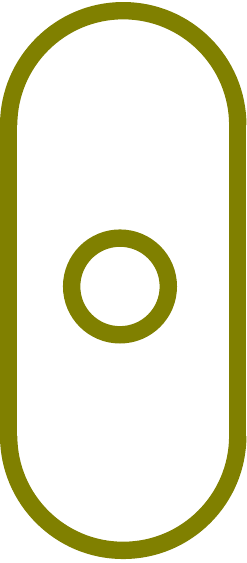}
&%\hspace{0.7mm}
\includegraphics[height=2.10 cm]{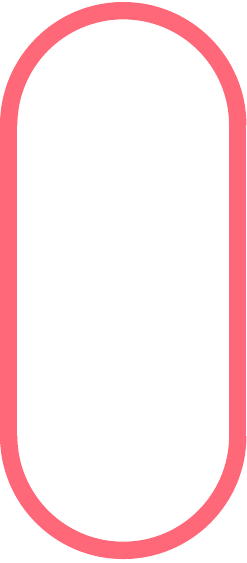}
&%\hspace{0.7mm}
\includegraphics[height=2.10 cm]{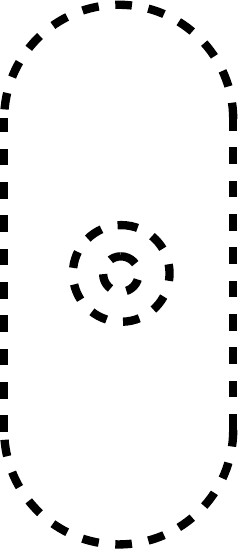}
&%\hspace{0.7mm}
\includegraphics[height=2.10 cm]{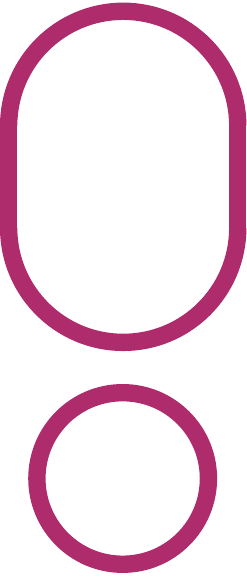}
&%\hspace{0.7mm}
\includegraphics[height=2.10 cm]{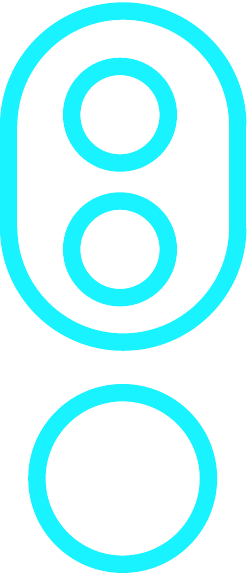}
&%\hspace{0.7mm}
\includegraphics[height=2.10 cm]{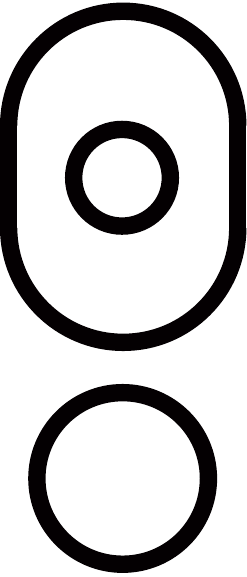}
&%\hspace{0.7mm}
\includegraphics[height=2.10 cm]{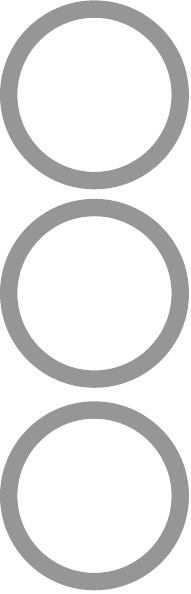}
\\
T1&T2&T3&T10&T4&T5&T11&T7&T6&T8&T9
\end{tabular}
\caption{Planet in Binary: critical-curve topologies on a sequence of horizontal ternary-plot sections of the prism from Figure~\ref{fig:PiBSur} at perimeters marked next to the individual plots. Colors correspond to the 11 different topologies T1--T11 sketched in the bottom row. The plot sequence is ordered by perimeter from the closest $p=1.016$ at top left to the widest $p=5.640$ at bottom right.}}
\label{fig:PiBTernaries}
\efi

\clearpage
\bfi
{\centering
\vspace{1cm}
\includegraphics[width=8.6 cm]{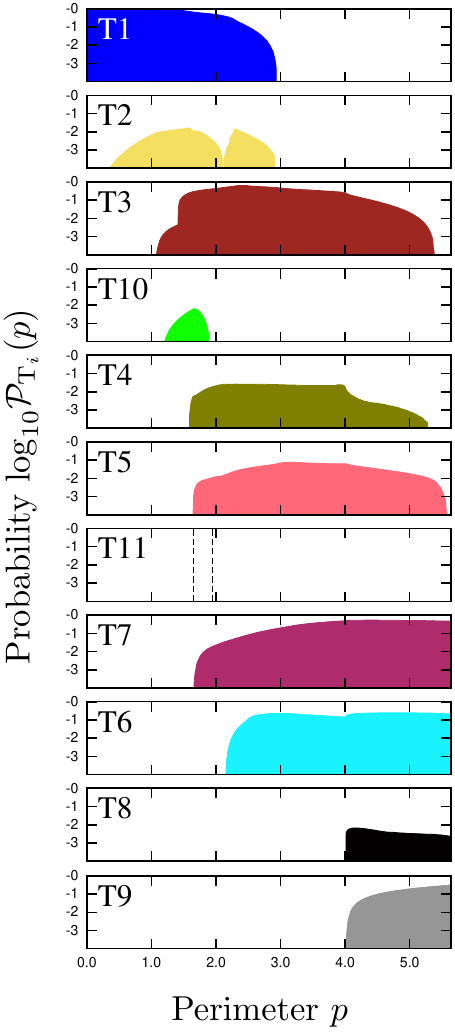}
\caption{Planet in Binary: probability $\mathcal{P}_{{\rm T}_i}(p)$ of occurrence of topology ${\rm T}_i$ at perimeter $p$. Topologies are marked and color-coded following the key in the bottom row of Figure \ref{fig:PiBTernaries}.}}
\label{fig:PiBPerPer}
\efi

\clearpage
\bfi
{\centering
\vspace{1.5 cm}
\includegraphics[clip,scale=0.35]{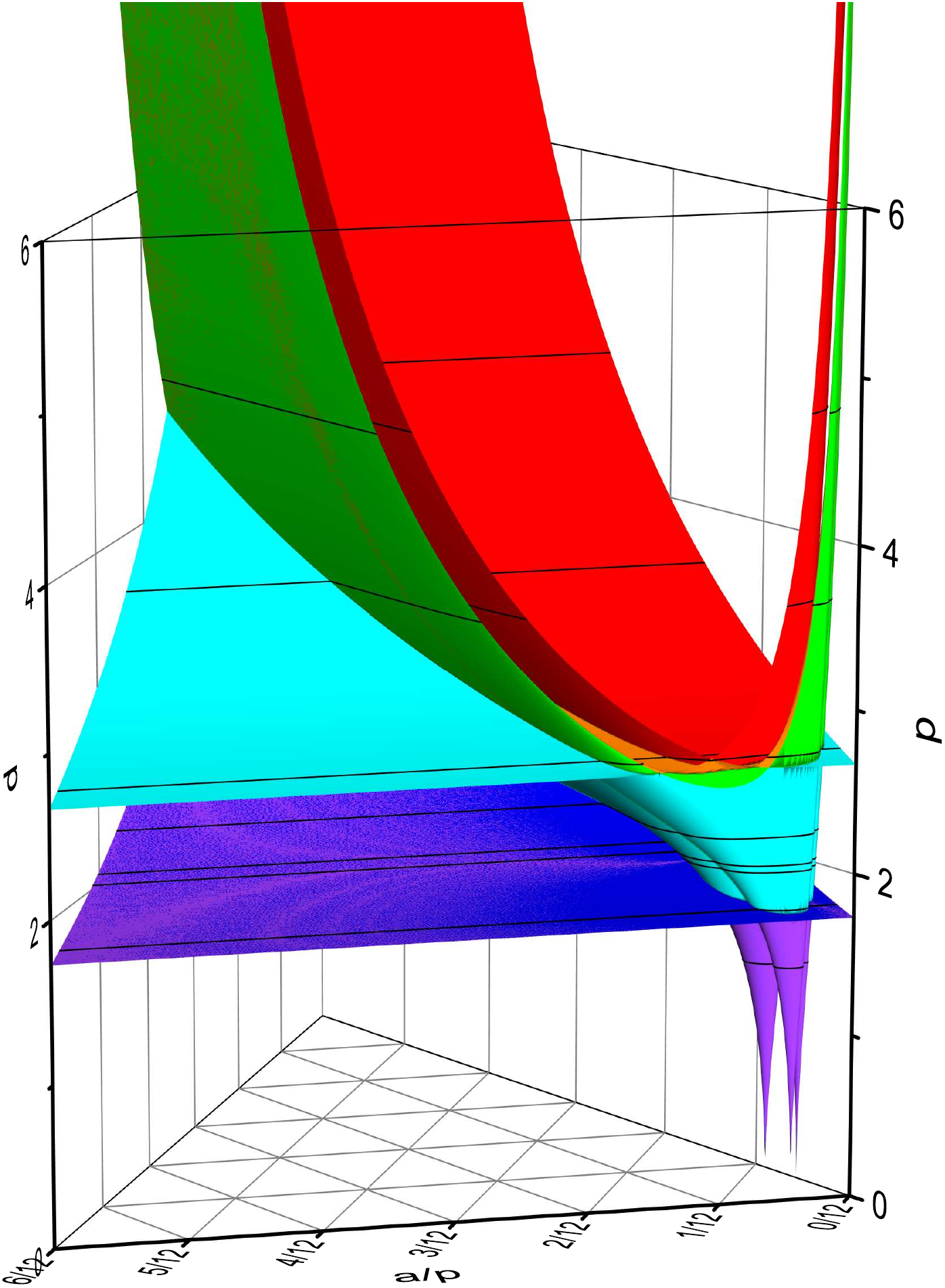}
\includegraphics[clip,scale=0.35]{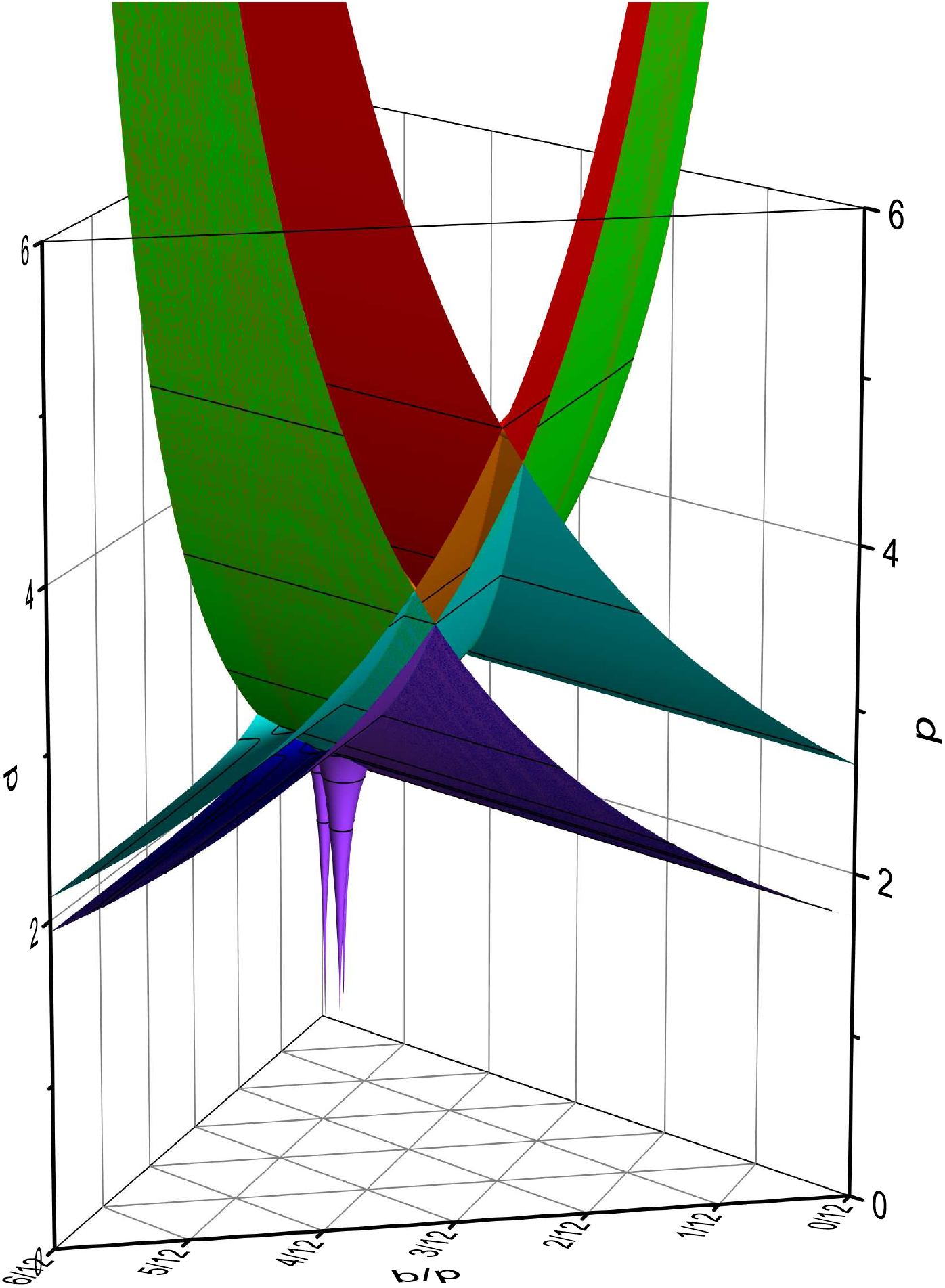}
\includegraphics[clip,scale=0.35]{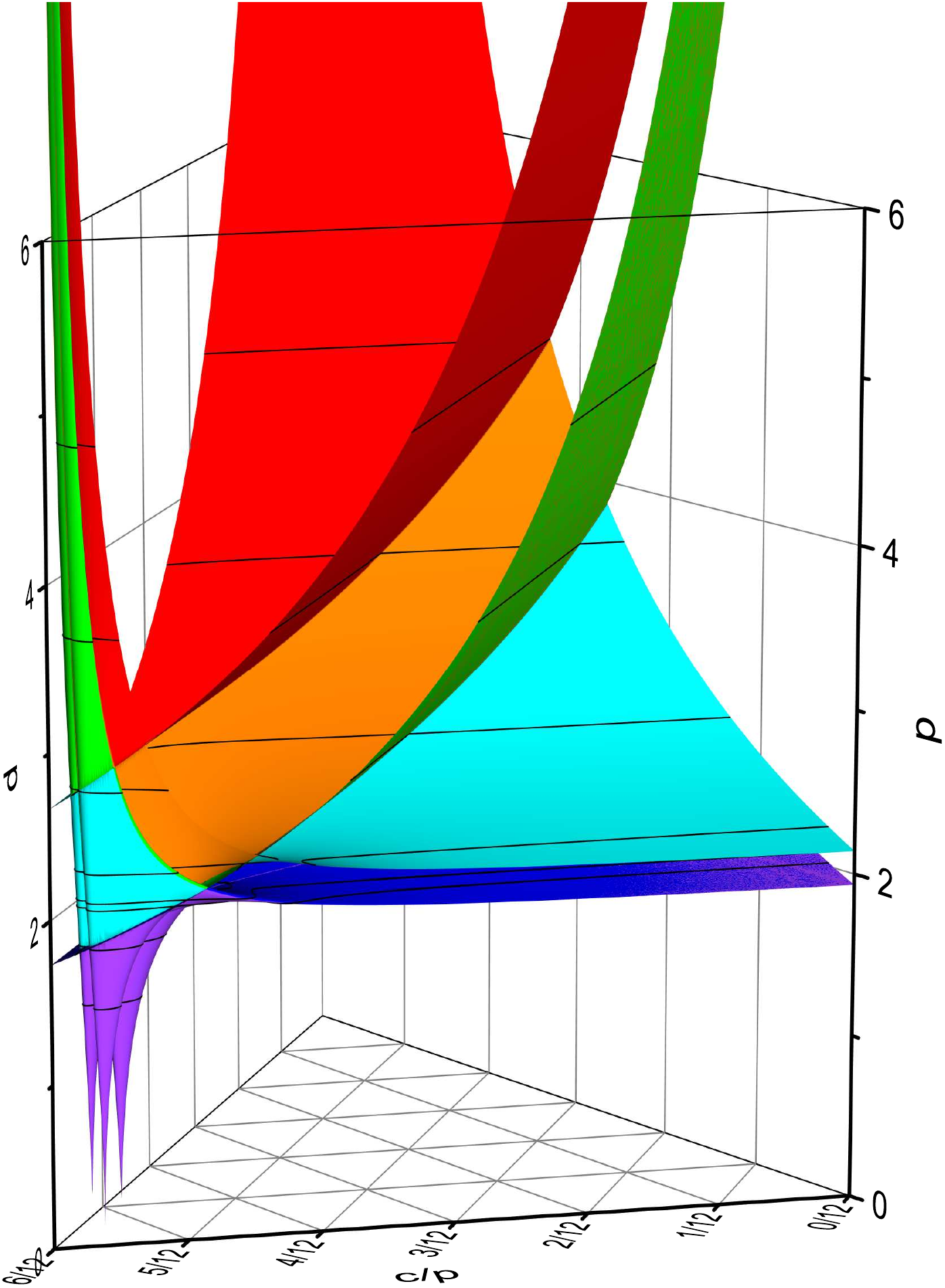}
\caption{Hierarchical Masses: boundaries in the lens configuration parameter space separating different critical-curve topologies. The three views of the surfaces are rotated by $2\pi/3$ around the vertical axis of the prism so that ternary axis $a_p$ (left), $b_p$ (middle), or $c_p$ (right) lies at front. Notation as in Figure~\ref{fig:EqMassSur}. Black contours identify the horizontal sections shown in Figure~\ref{fig:HiMTernaries}. An animated version of this figure is available in the online version of this article, showing the prism rotated full-circle around its vertical axis in $\pi/18$ steps.}}
\label{fig:HiMSur}
\efi

\clearpage
\bfi
{\centering
\vspace{2cm}
\includegraphics[width=16.8 cm]{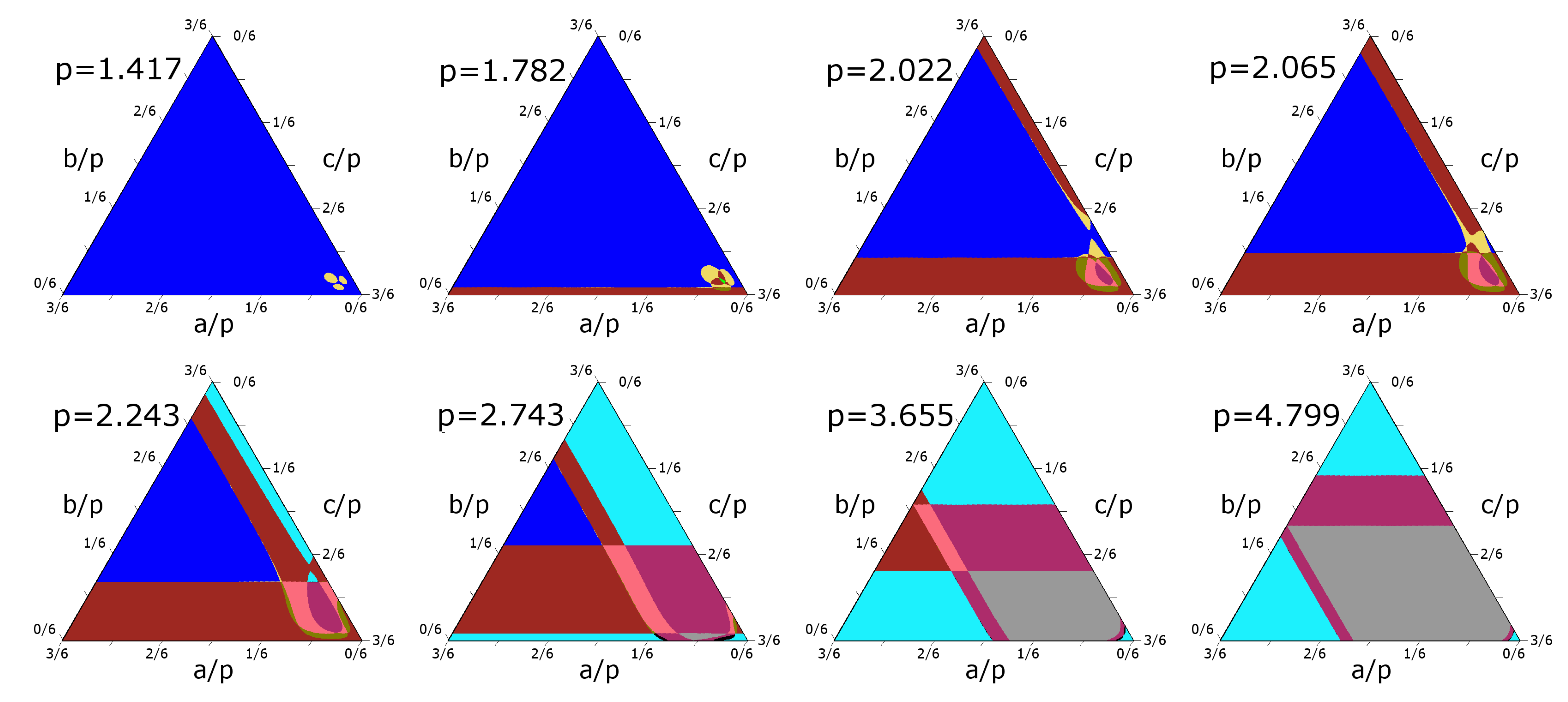} \\
\vspace{3mm}
\begin{tabular}{ccccccccccc}
\includegraphics[height=2.10 cm]{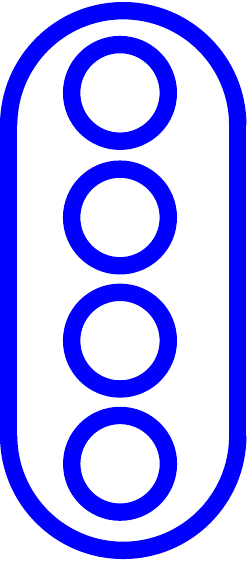}
&
\includegraphics[height=2.10 cm]{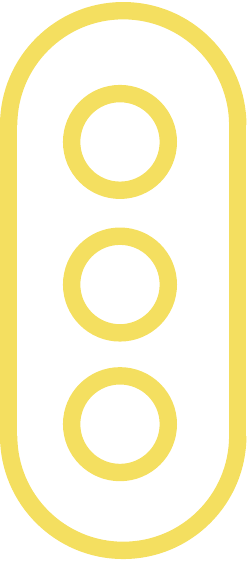}
&
\includegraphics[height=2.10 cm]{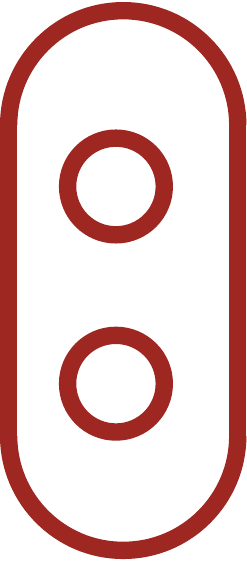}
&
\includegraphics[height=2.10 cm]{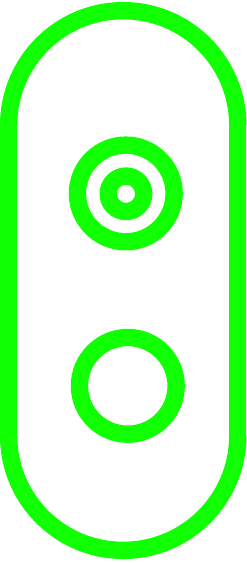}
&
\includegraphics[height=2.10 cm]{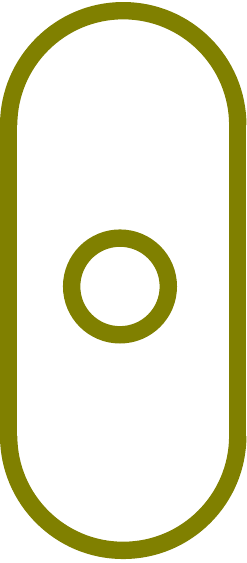}
&
\includegraphics[height=2.10 cm]{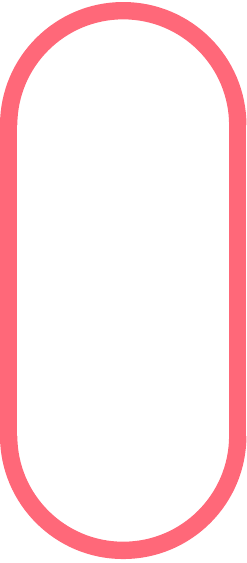}
&
\includegraphics[height=2.10 cm]{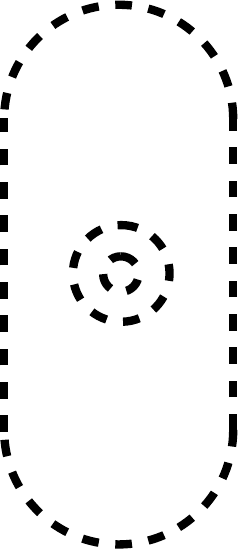}
&
\includegraphics[height=2.10 cm]{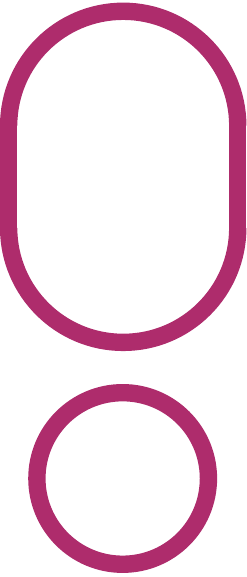}
&
\includegraphics[height=2.10 cm]{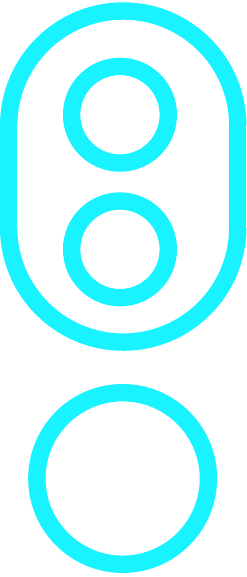}
&
\includegraphics[height=2.10 cm]{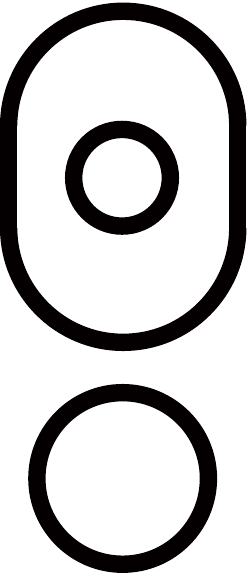}
&
\includegraphics[height=2.10 cm]{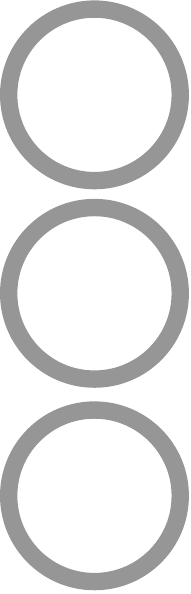}
\\
\vspace{-3mm}
T1&T2&T3&T10&T4&T5&T11&T7&T6&T8&T9
\vspace{6mm}
\end{tabular}
\caption{Hierarchical Masses: critical-curve topologies on a sequence of horizontal ternary-plot sections of the prism from Figure~\ref{fig:HiMSur} at perimeters marked next to the individual plots. Colors correspond to the 11 different topologies T1--T11 sketched in the bottom row. The plot sequence is ordered by perimeter from the closest $p=1.417$ at top left to the widest $p=4.799$ at bottom right.}}
\label{fig:HiMTernaries}
\efi

\clearpage
\bfi
{\centering
\vspace{1cm}
\includegraphics[width=8.6 cm]{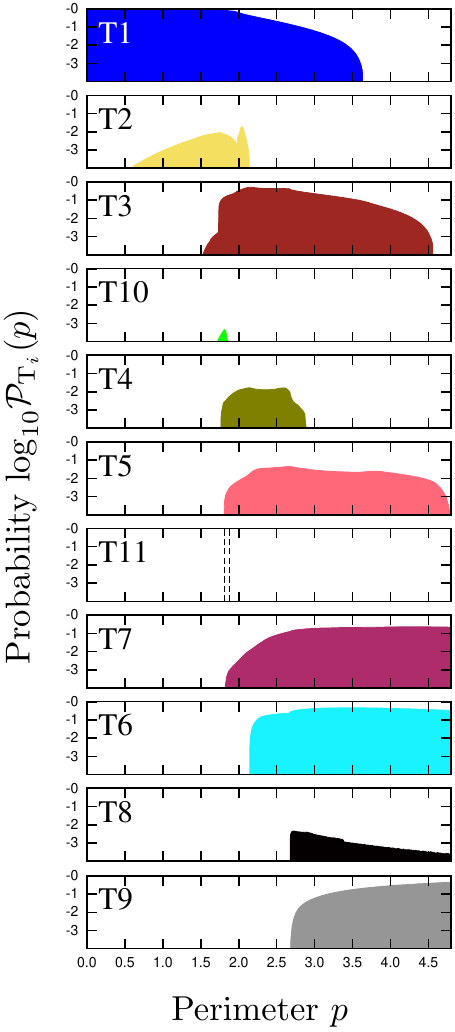}
\caption{Hierarchical Masses: probability $\mathcal{P}_{{\rm T}_i}(p)$ of occurrence of topology ${\rm T}_i$ at perimeter $p$. Topologies are marked and color-coded following the key in the bottom row of Figure \ref{fig:HiMTernaries}.}}
\label{fig:HiMPerPer}
\efi

\clearpage
\bfi
{\centering
\vspace{1cm}
\includegraphics[width=16.8 cm]{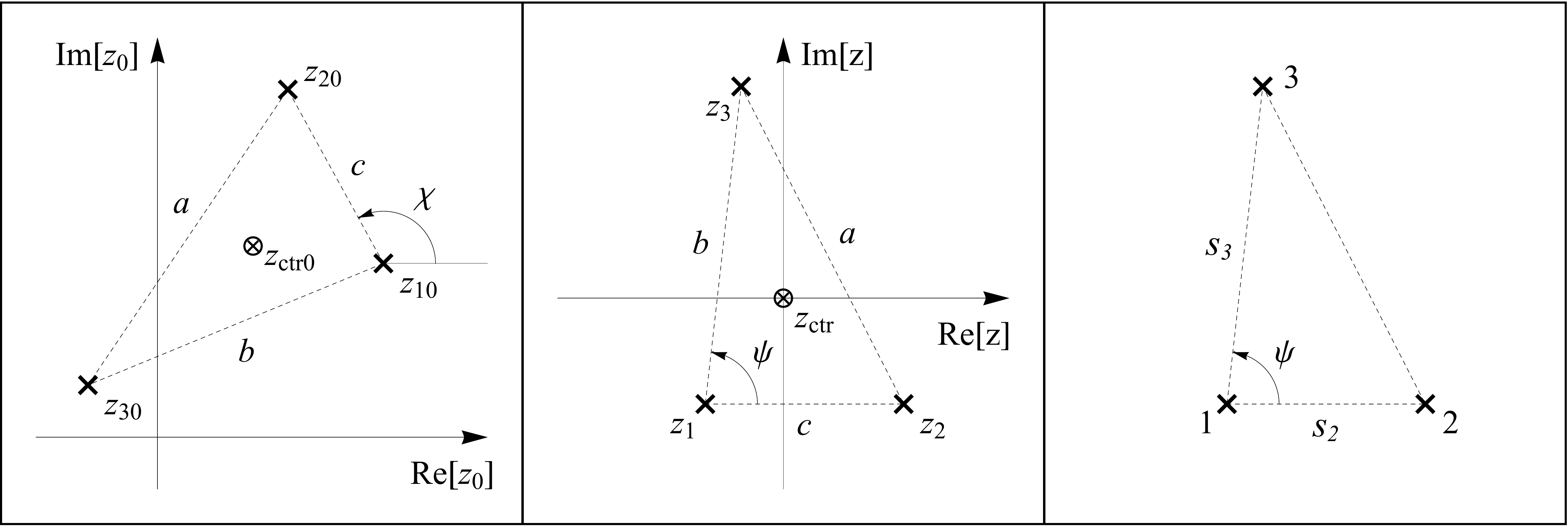}
\caption{Parameters describing the spatial configuration of the triple lens: a sample combination of components (crosses) in terms of general-frame positions (left panel), preferred-frame positions (middle panel), and microlensing parameters (right panel). The transformation from the general to the preferred frame consists of shifting the original centroid $z_{\rm ctr0}$ of the components to $z_{\rm ctr}$ at the origin and rotating the axes by $\chi$ so that component 2 lies horizontally to the right of component 1. For details see Appendix~\ref{sec:Appendix-conversion}.}}
\label{fig:parameters}
\efi

\clearpage
\bfi
{\centering
\vspace{1cm}
\includegraphics[width=5.5cm]{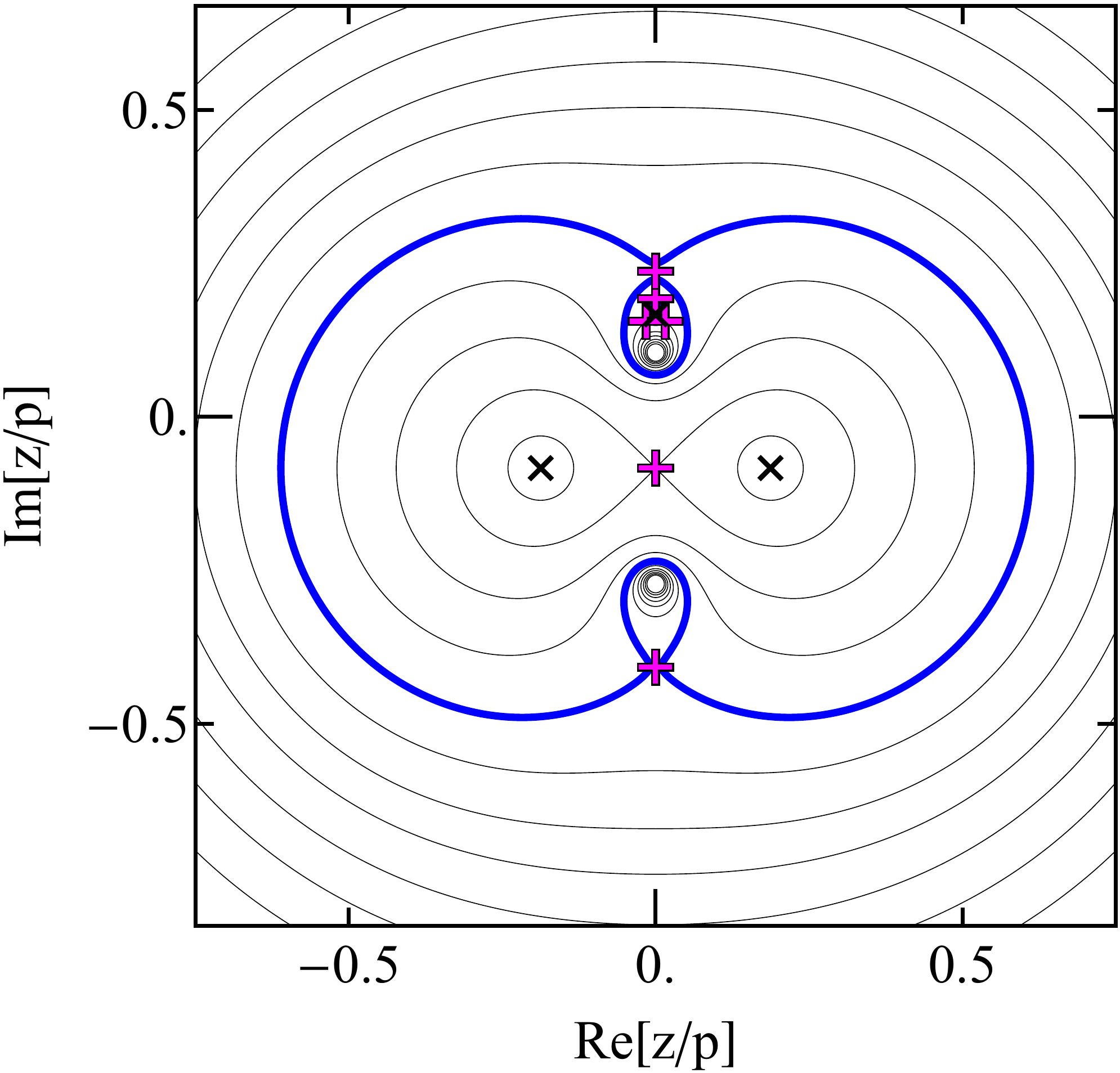}
\hspace{0.2cm}\includegraphics[width=5.5cm]{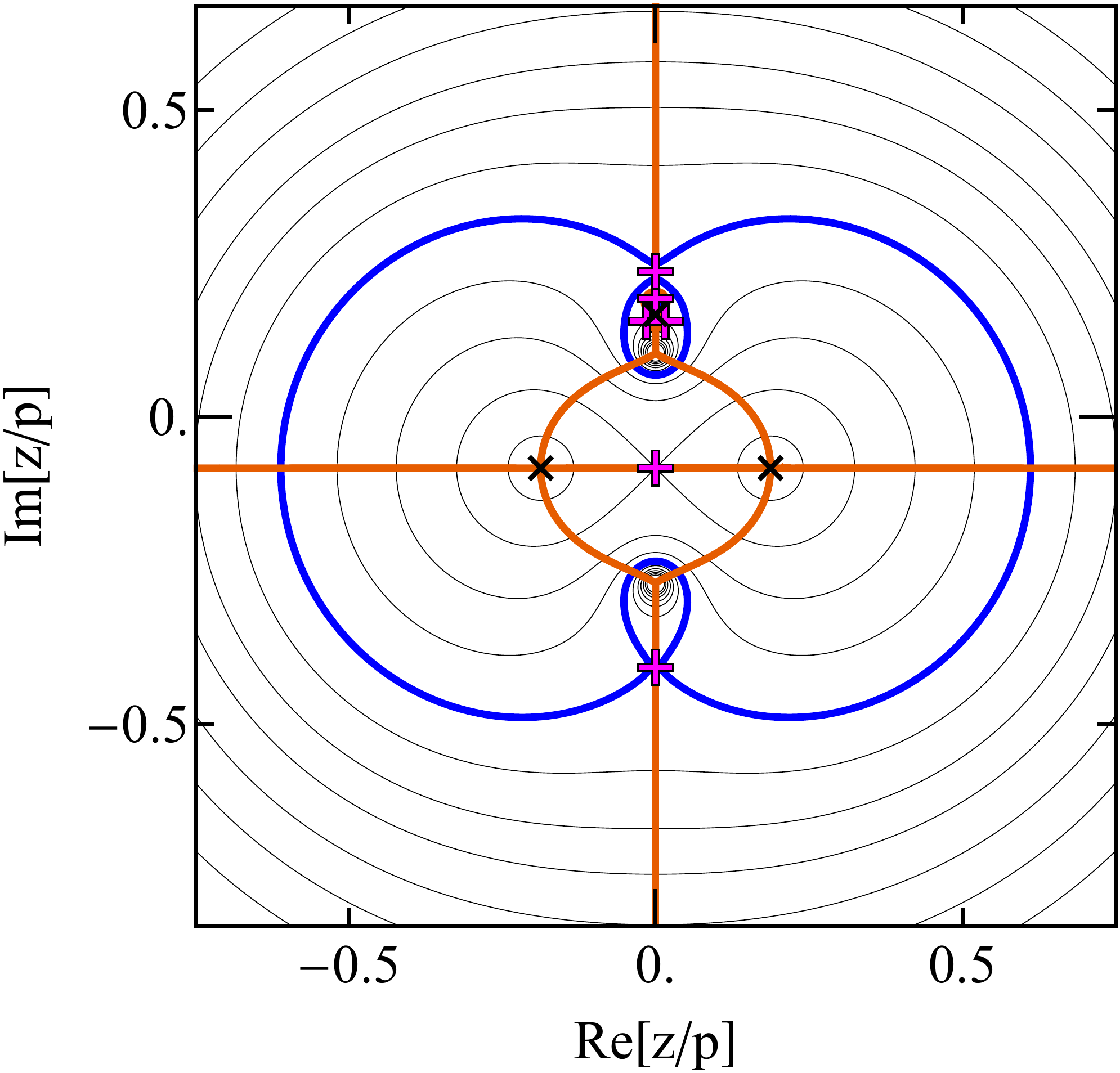}
\hspace{0.2cm}\includegraphics[width=5.5cm]{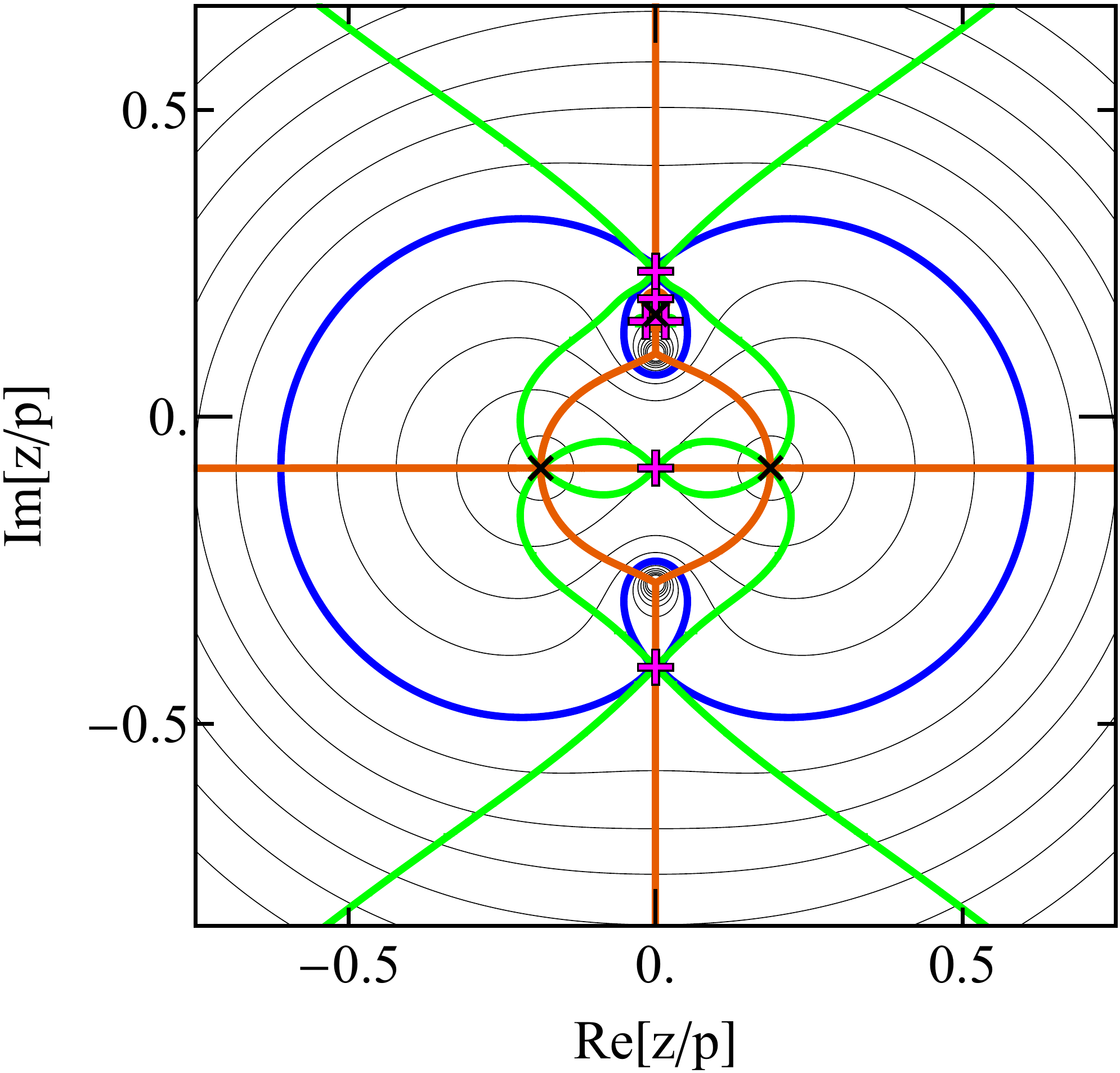}\\
\vspace{0.4cm}
\includegraphics[width=5.5cm]{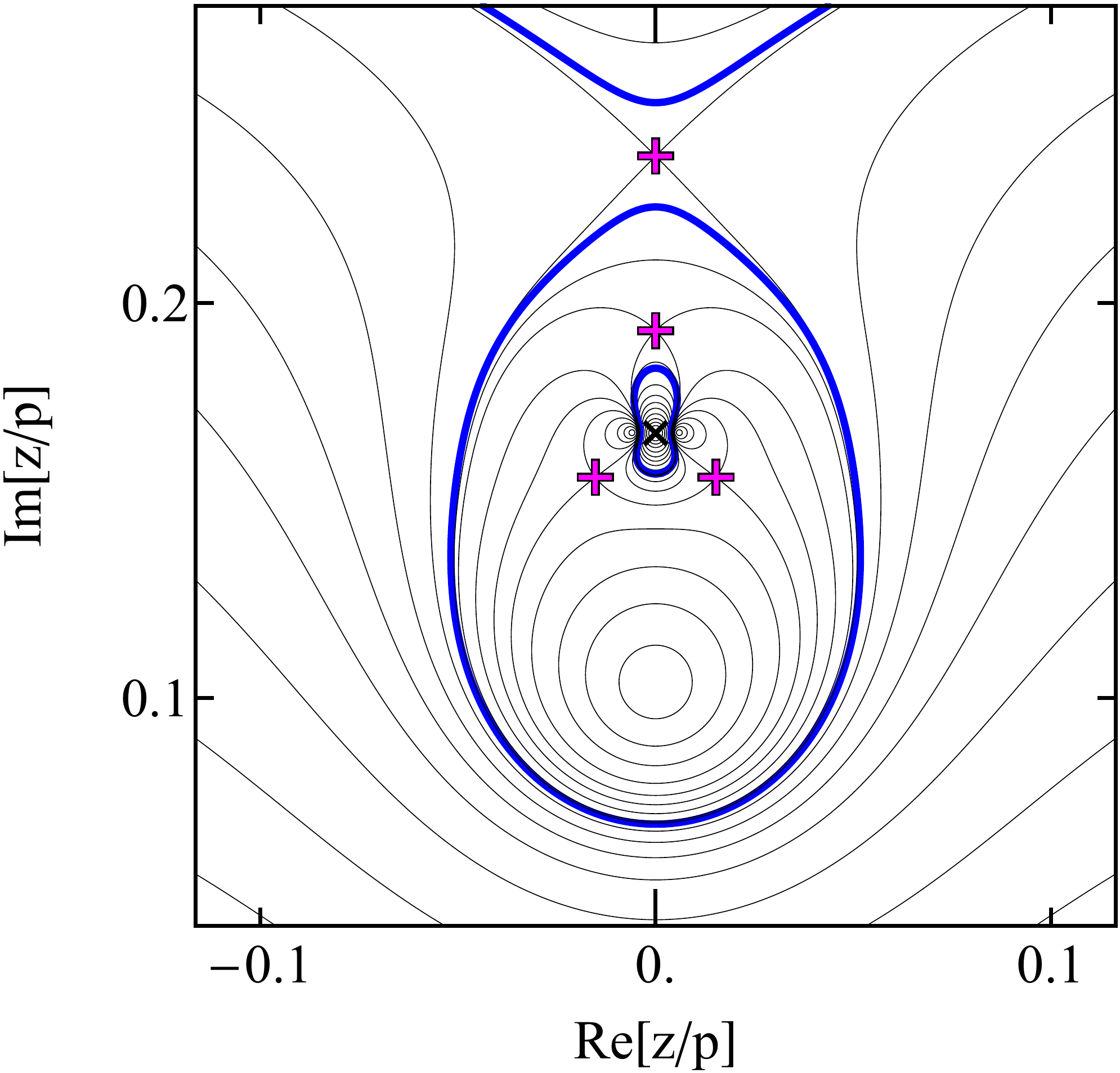}
\hspace{0.2cm}\includegraphics[width=5.5cm]{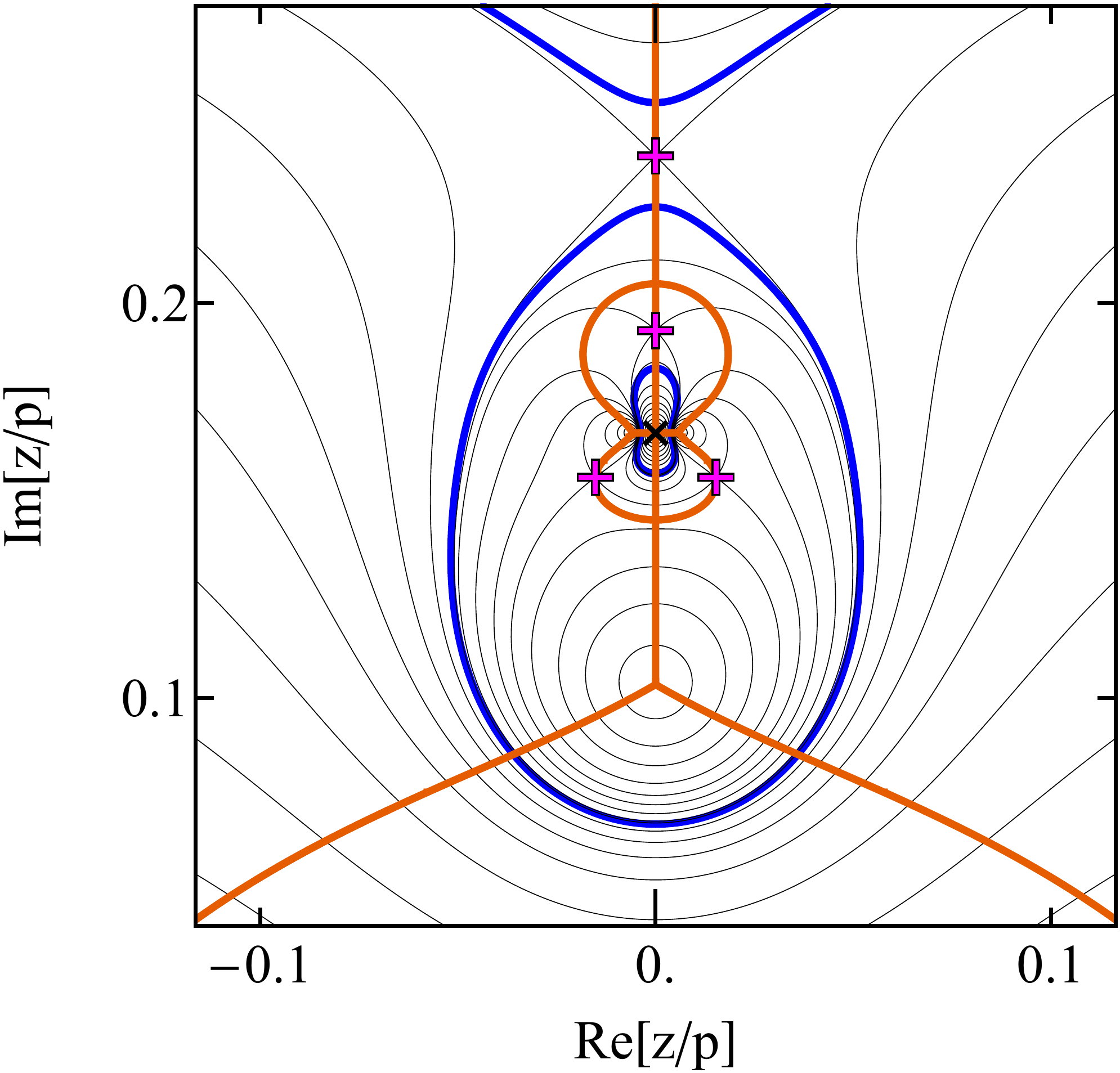}
\hspace{0.2cm}\includegraphics[width=5.5cm]{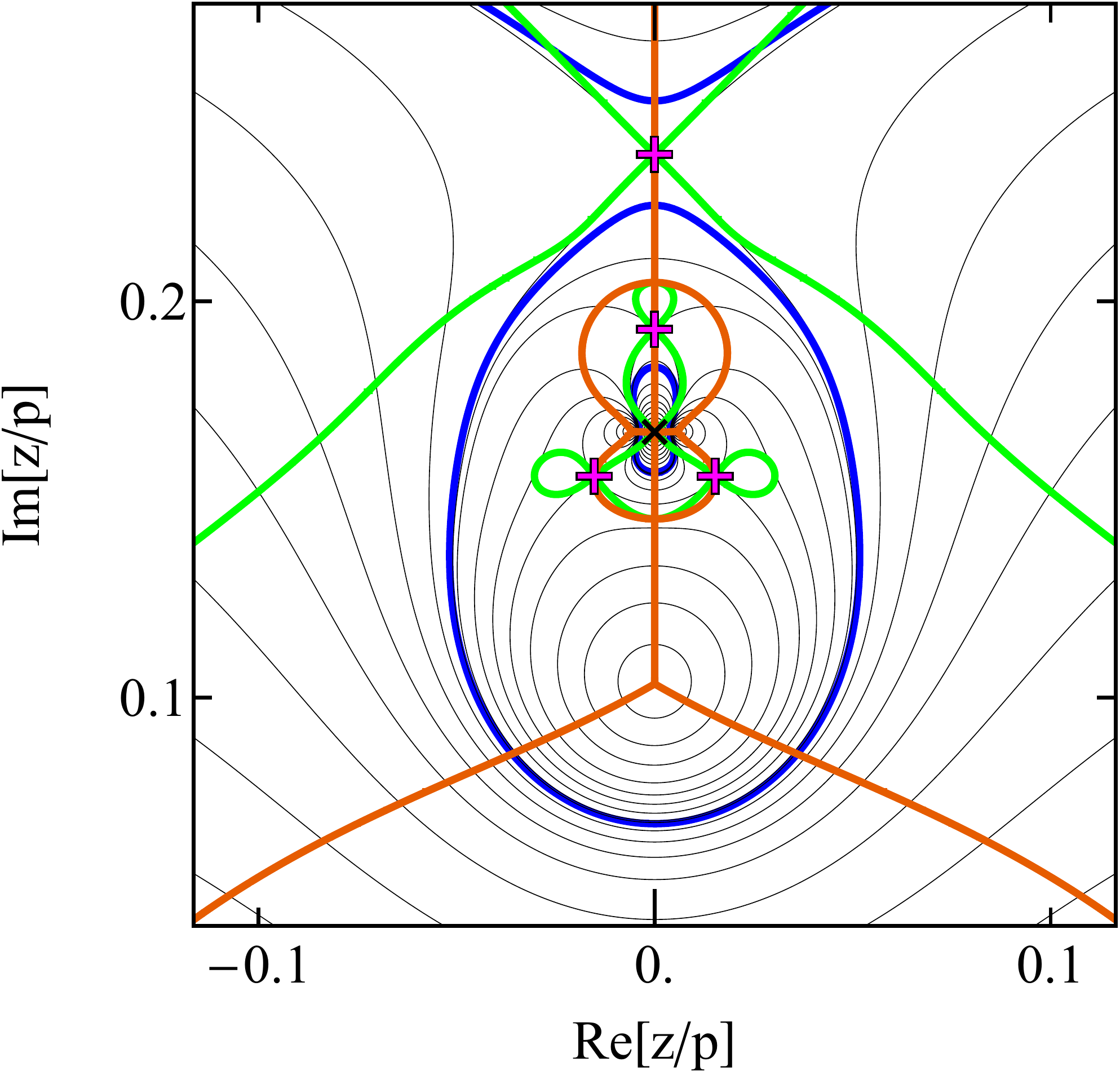}\\
\caption{Doubly nested critical-curve topology T11 for the Planet in Binary model with $(a_p, b_p, c_p)=(0.31276, 0.31276, 0.37448)$. Top row: global view; bottom row: detailed view of the vicinity of the planet. All panels show the Jacobian contours with the $p=1.89$ critical curve marked in blue; the middle and right columns include the cusp curve (orange); the right column includes the morph curve (green). Notation as in Figure~\ref{fig:EqMassSeq}.}}
\label{fig:T11curves}
\efi

\clearpage
\bfi
{\centering
\vspace{1cm}
\includegraphics[height=4.1cm]{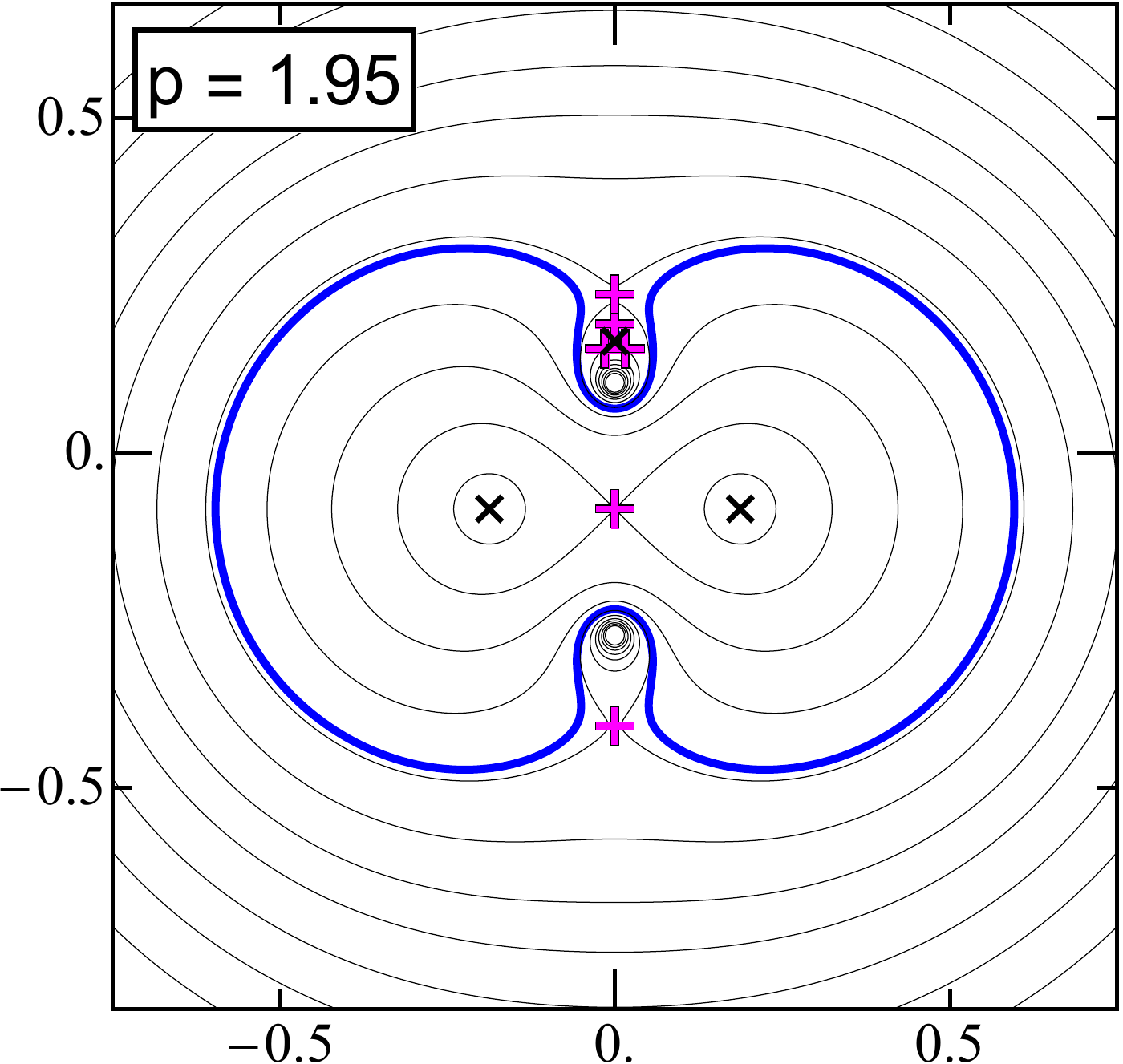}
\hspace{0.15cm}\includegraphics[height=4.1cm]{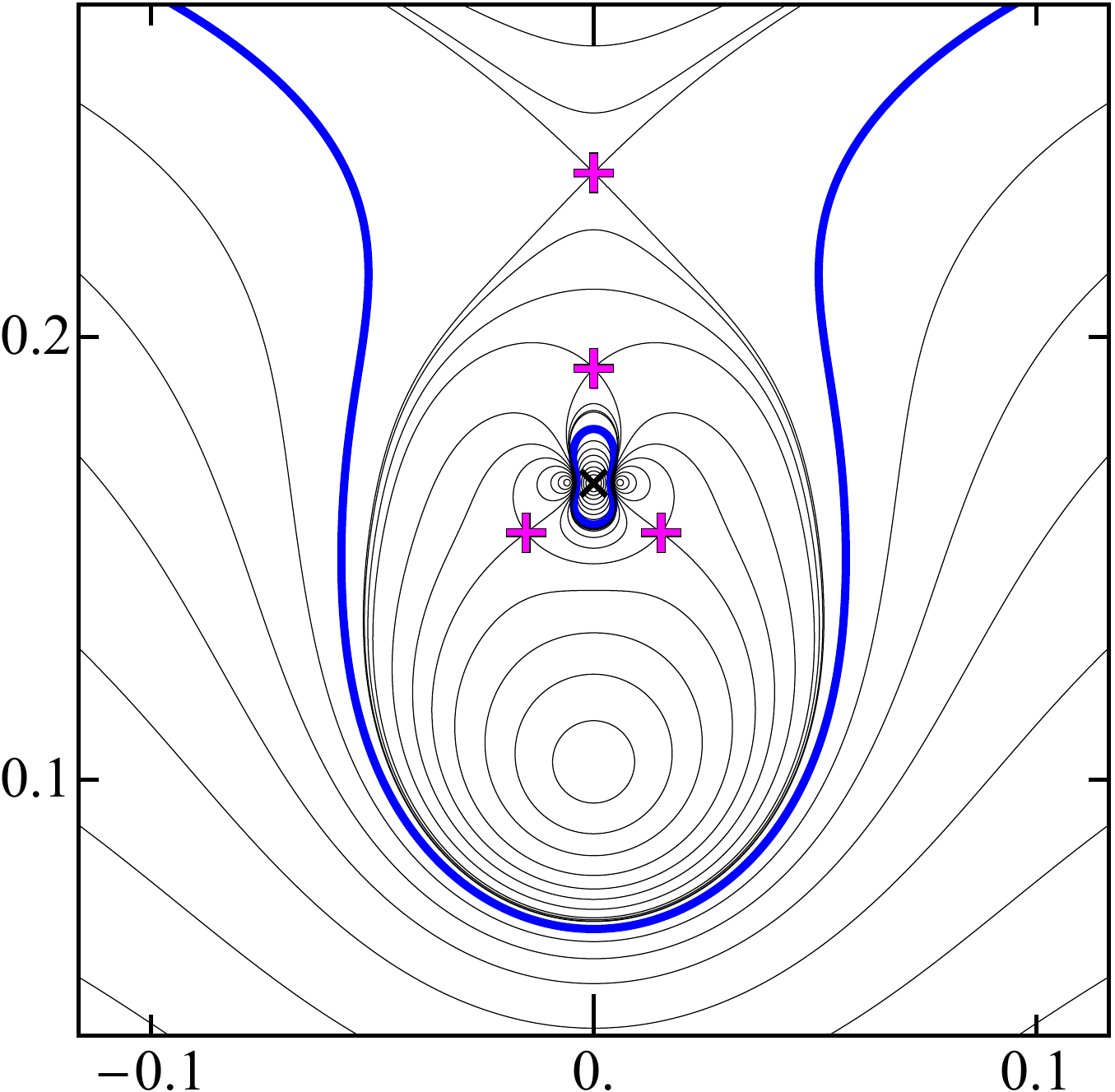}
\hspace{0.2cm}\includegraphics[height=4.1cm]{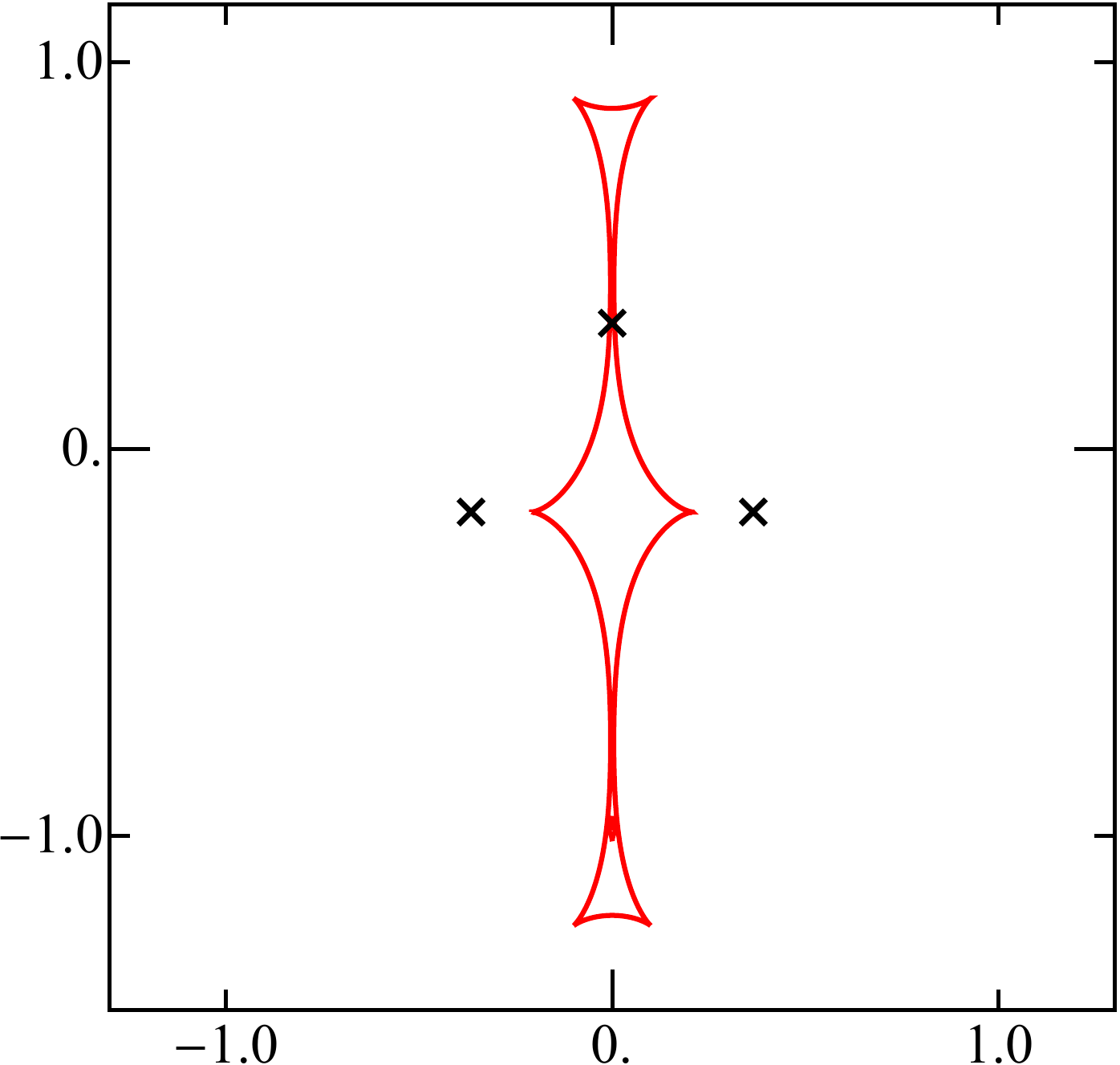}
\hspace{0.15cm}\includegraphics[height=4.1cm]{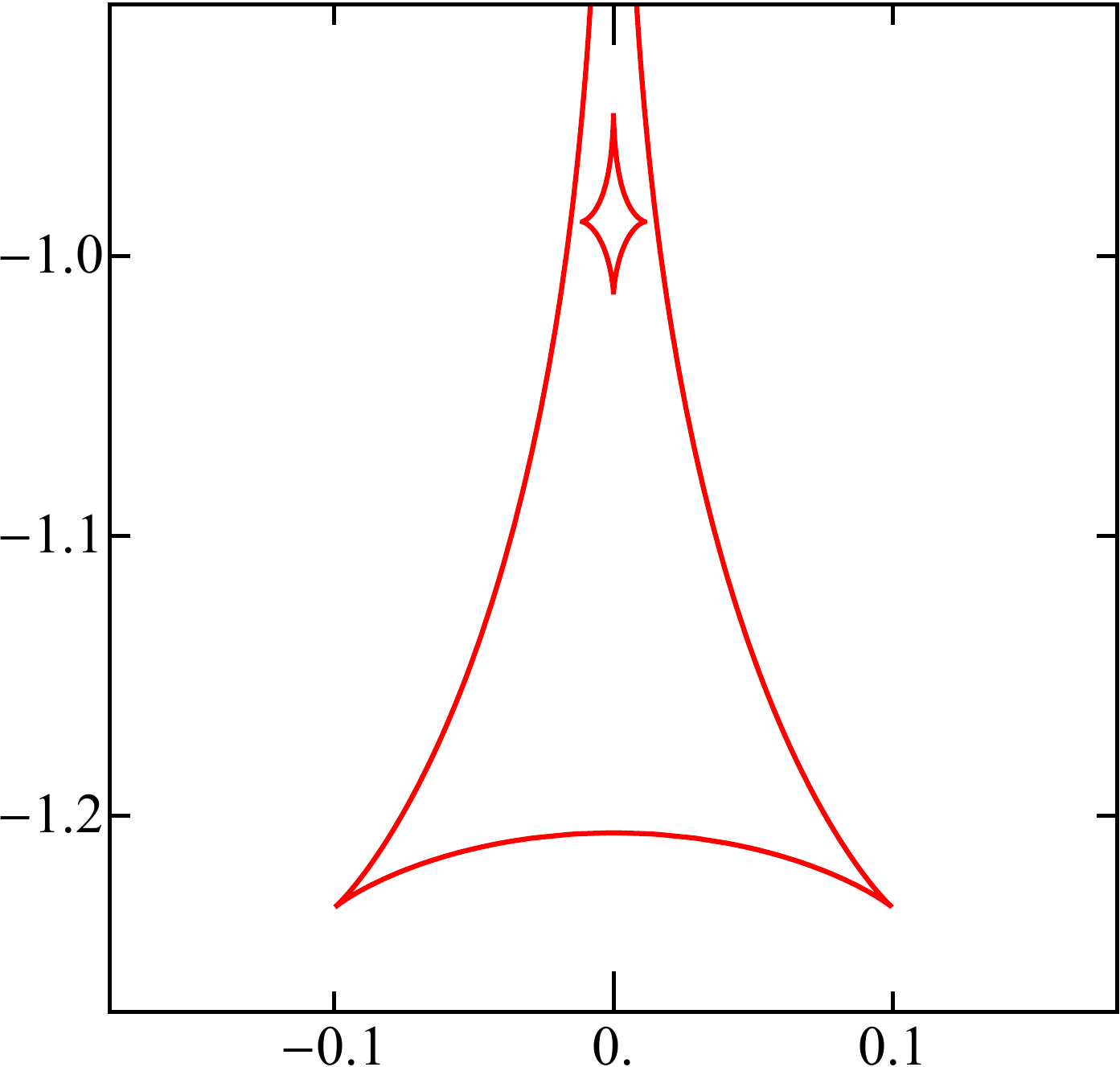}\\
\vspace{0.2cm}
\includegraphics[height=4.1cm]{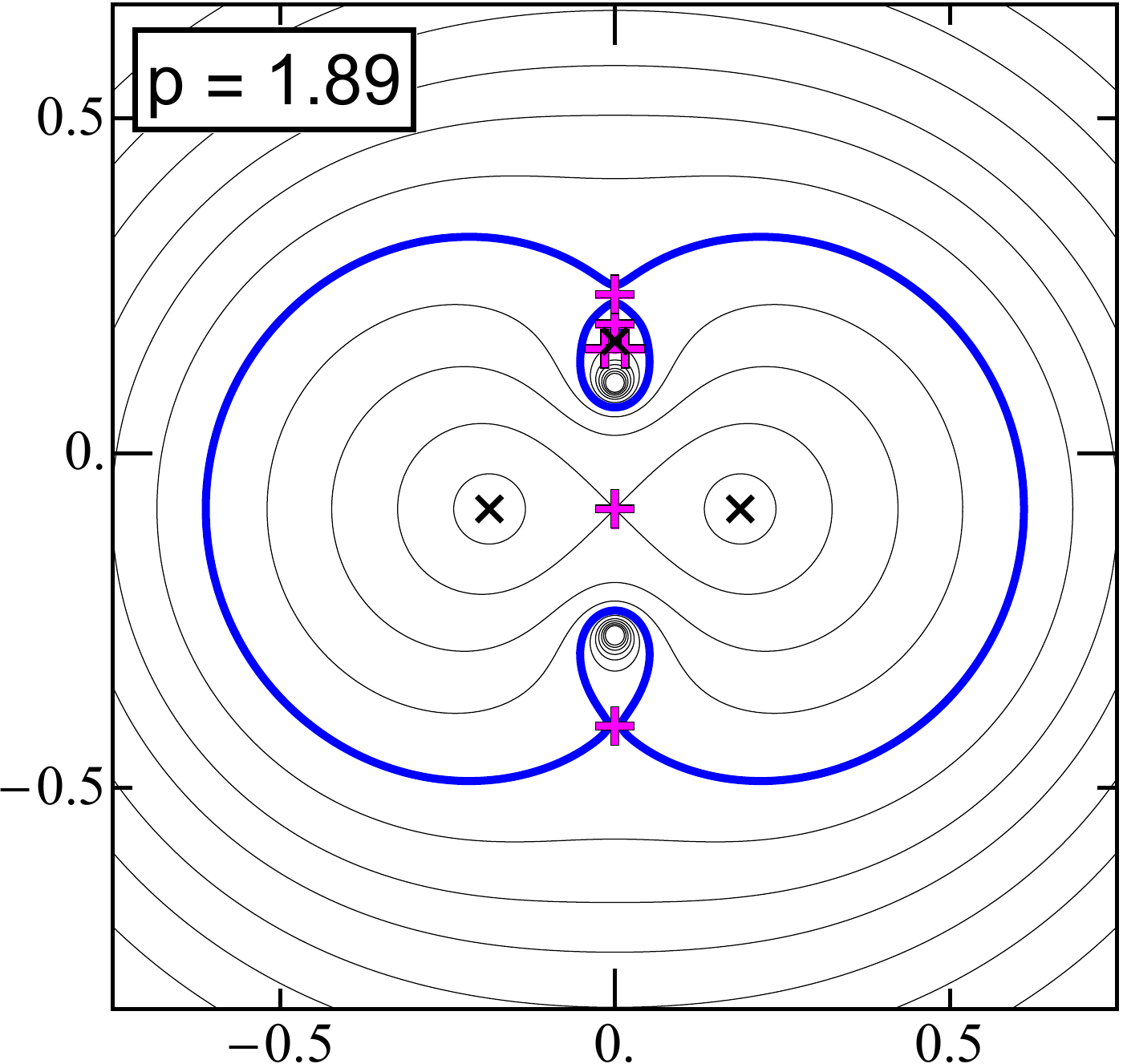}
\hspace{0.15cm}\includegraphics[height=4.1cm]{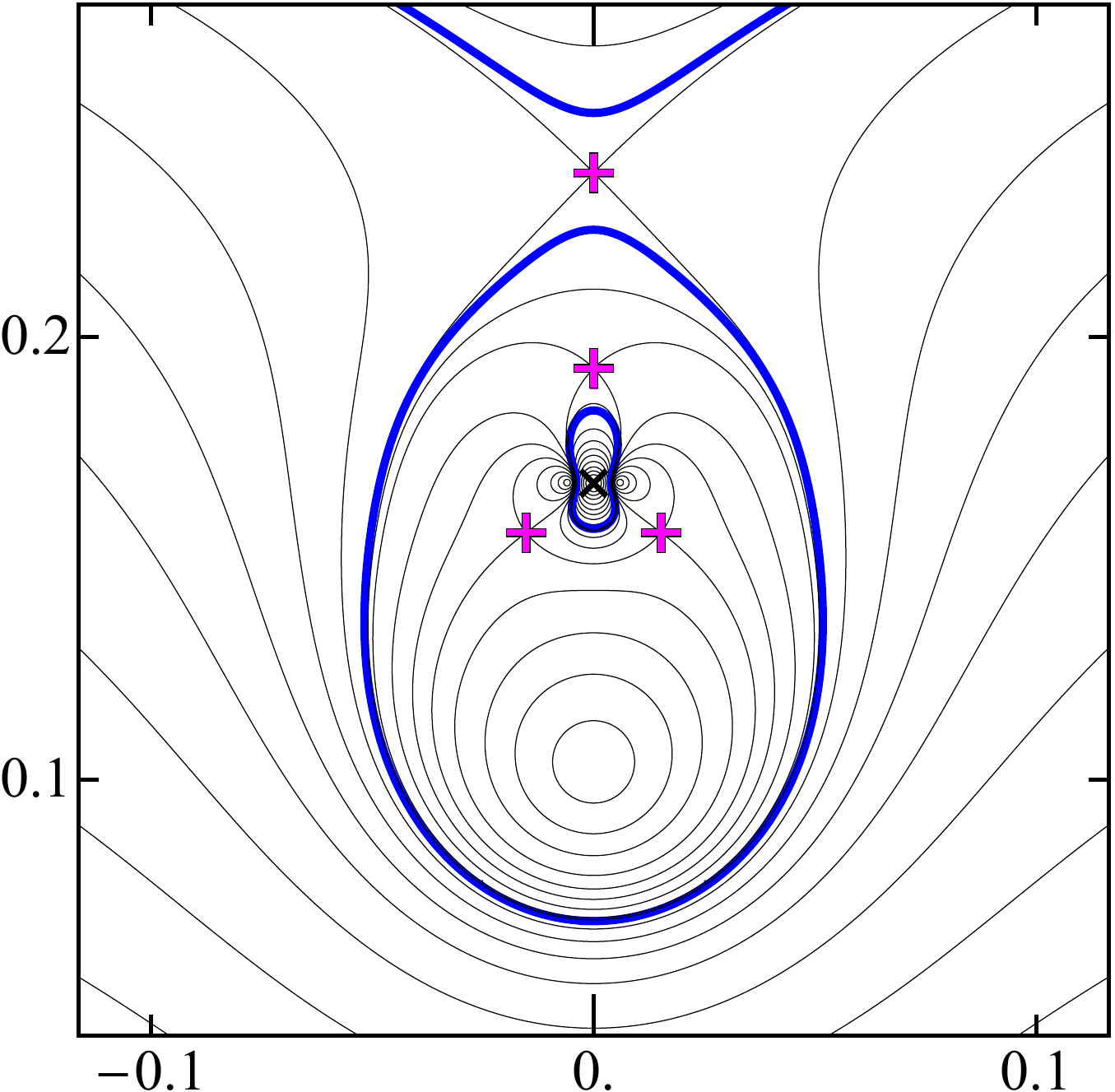}
\hspace{0.2cm}\includegraphics[height=4.1cm]{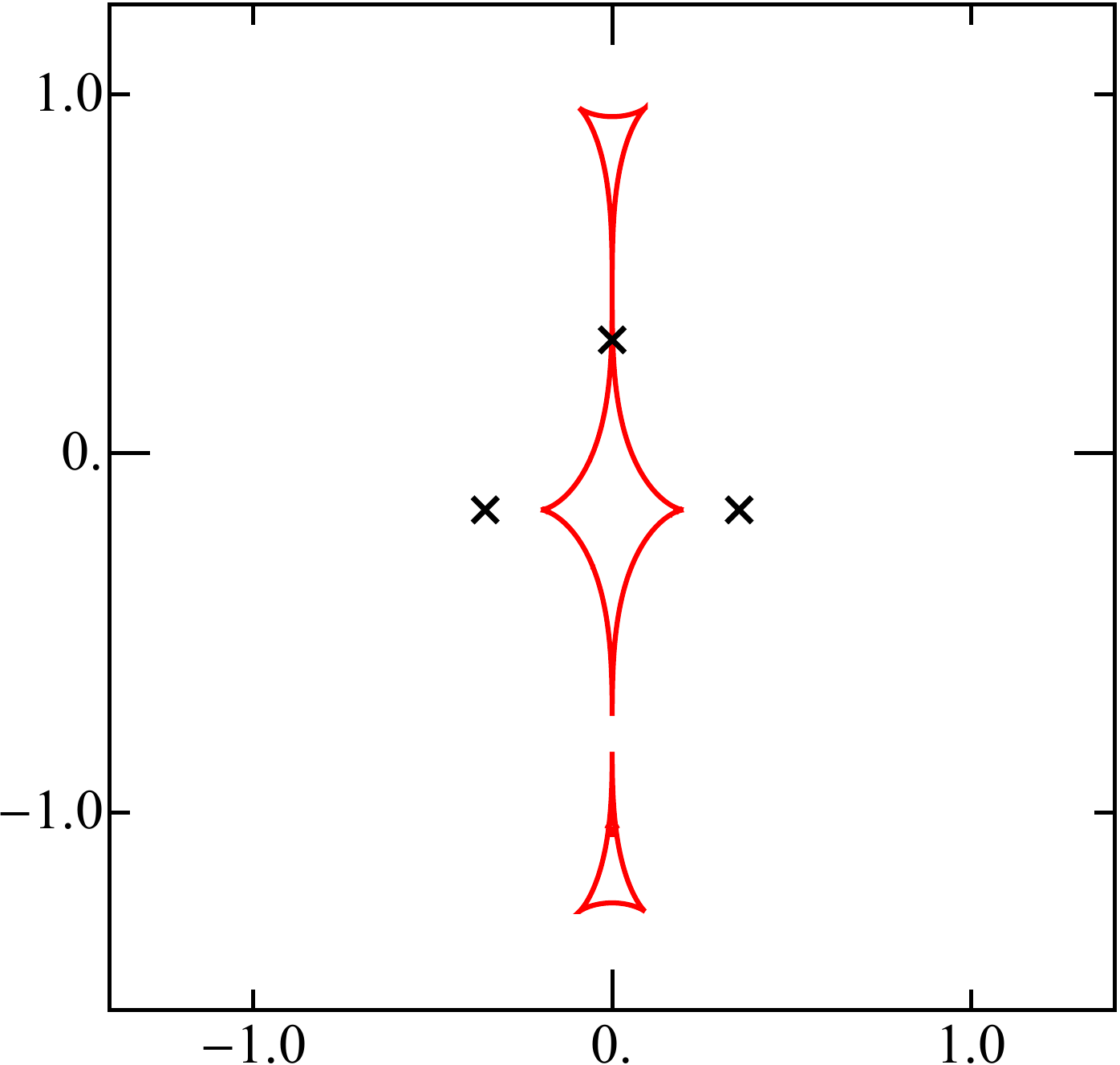}
\hspace{0.15cm}\includegraphics[height=4.1cm]{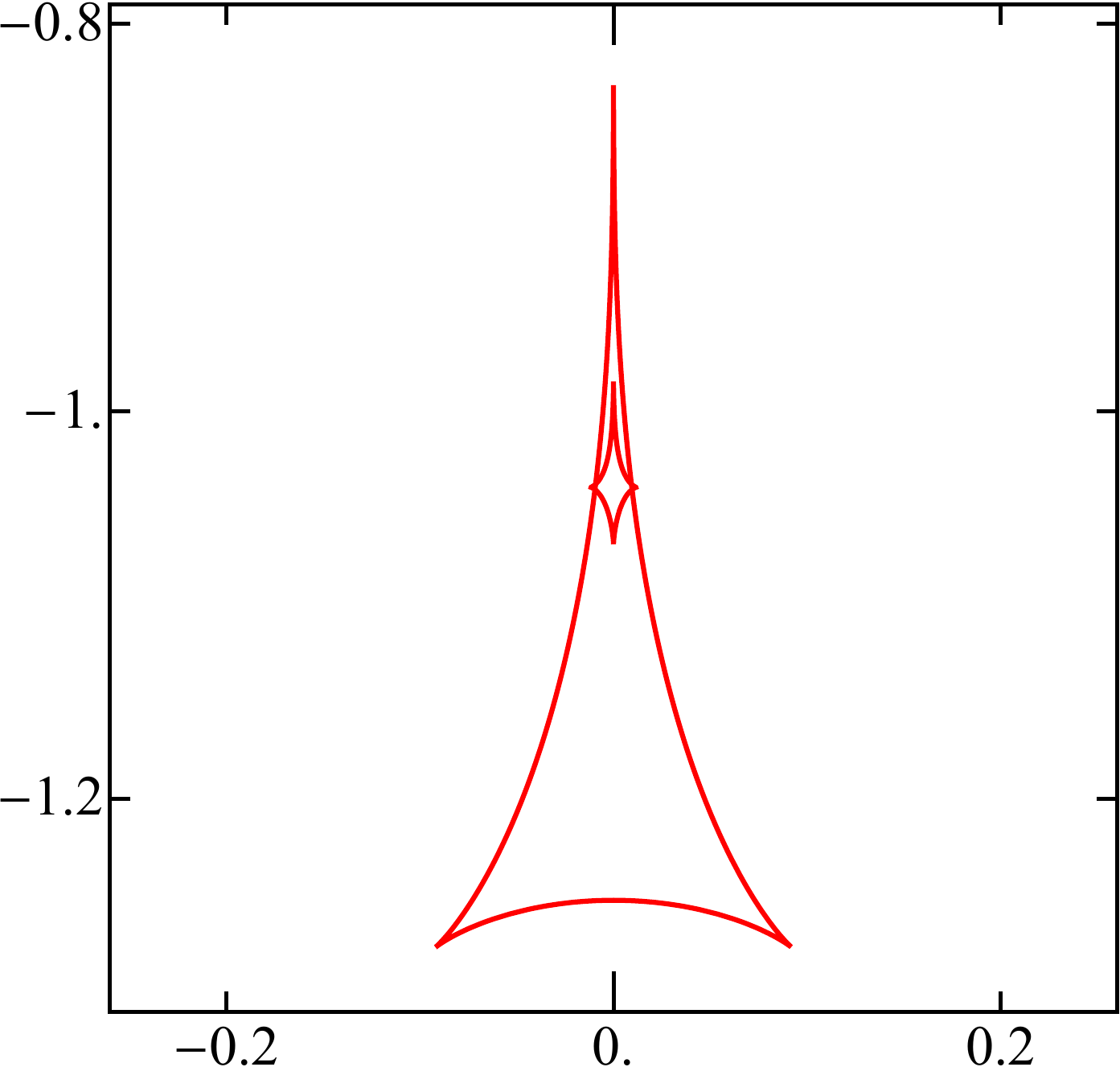}\\
\vspace{0.2cm}
\includegraphics[height=4.1cm]{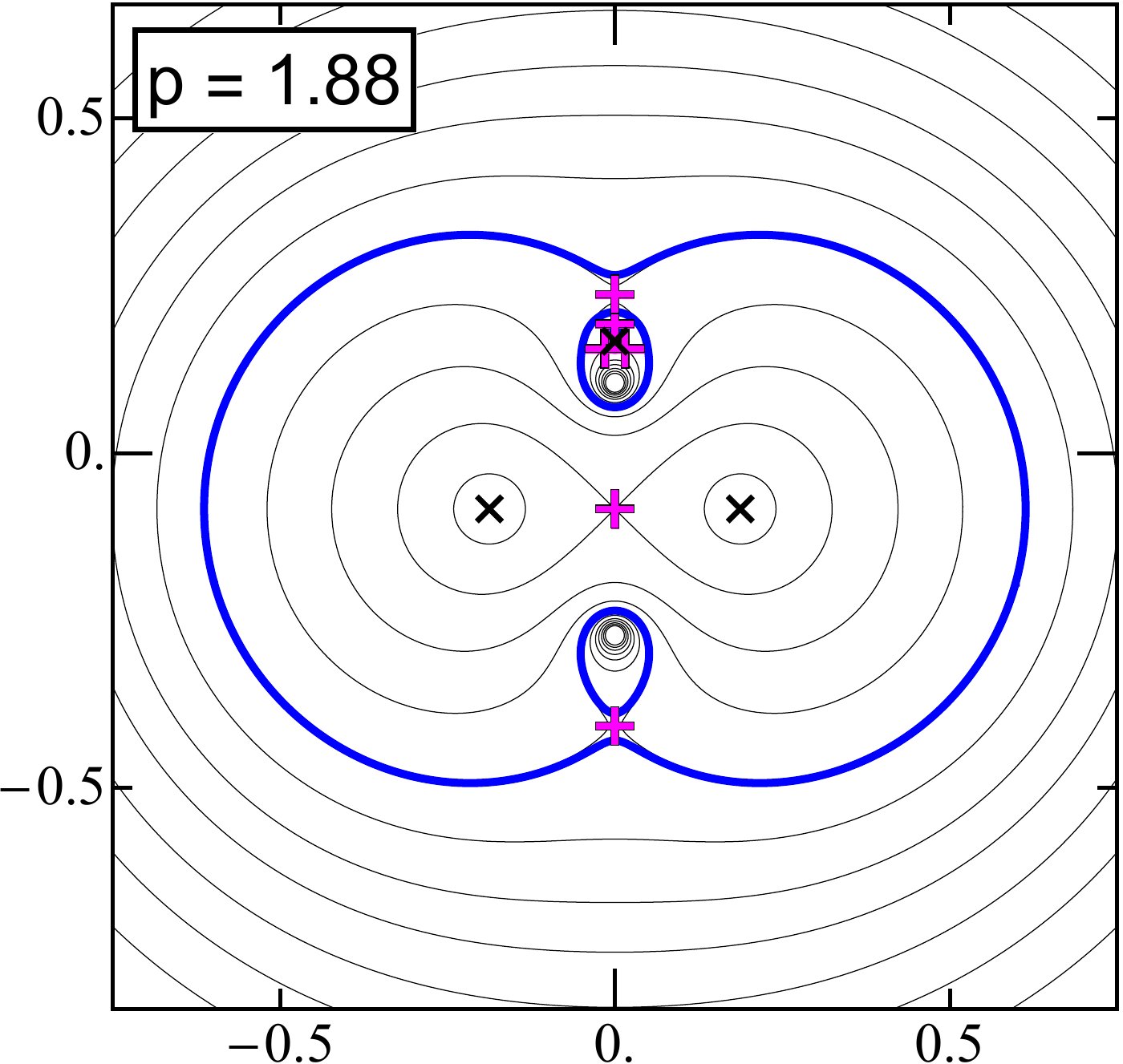}
\hspace{0.15cm}\includegraphics[height=4.1cm]{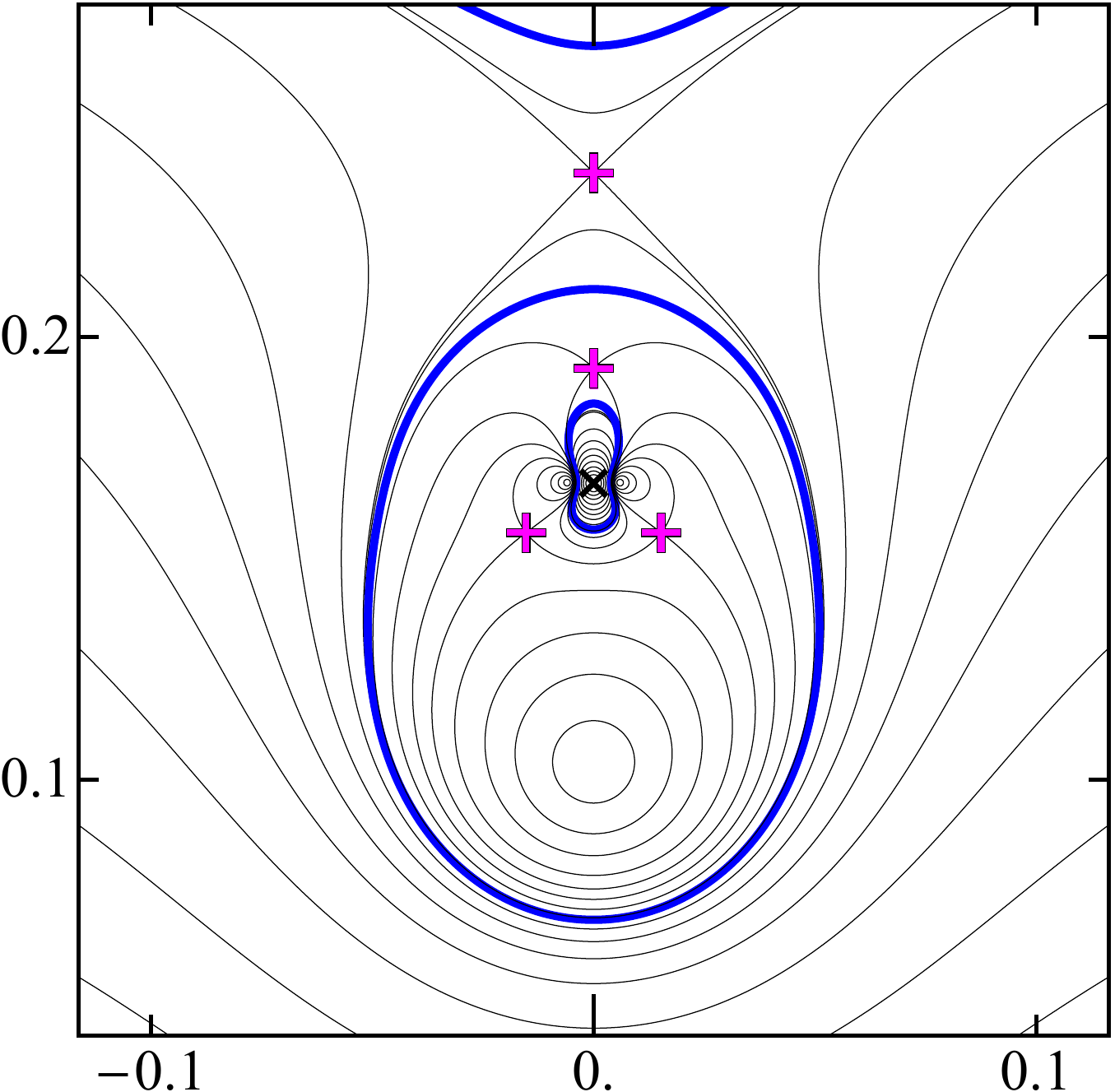}
\hspace{0.2cm}\includegraphics[height=4.1cm]{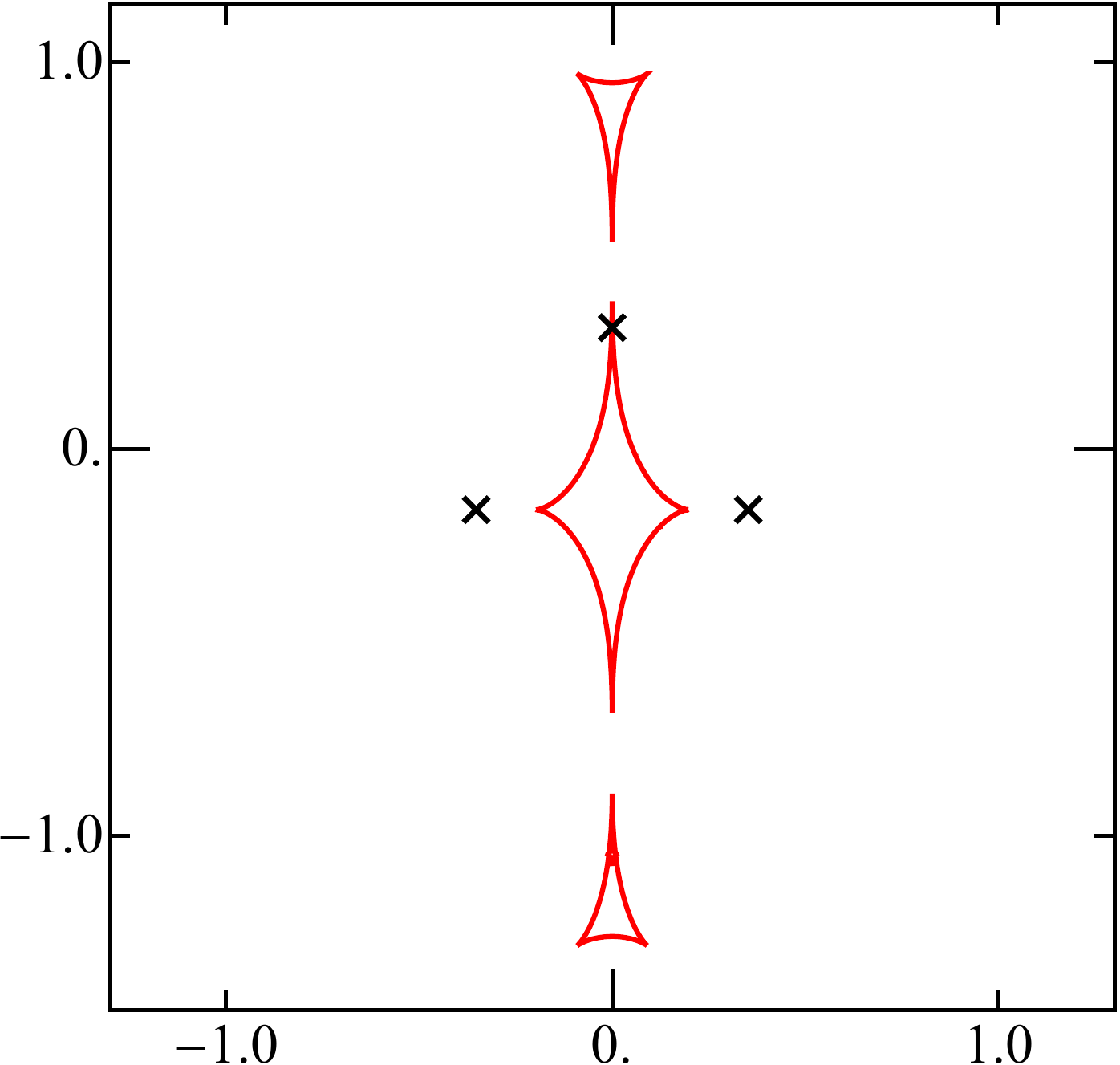}
\hspace{0.15cm}\includegraphics[height=4.1cm]{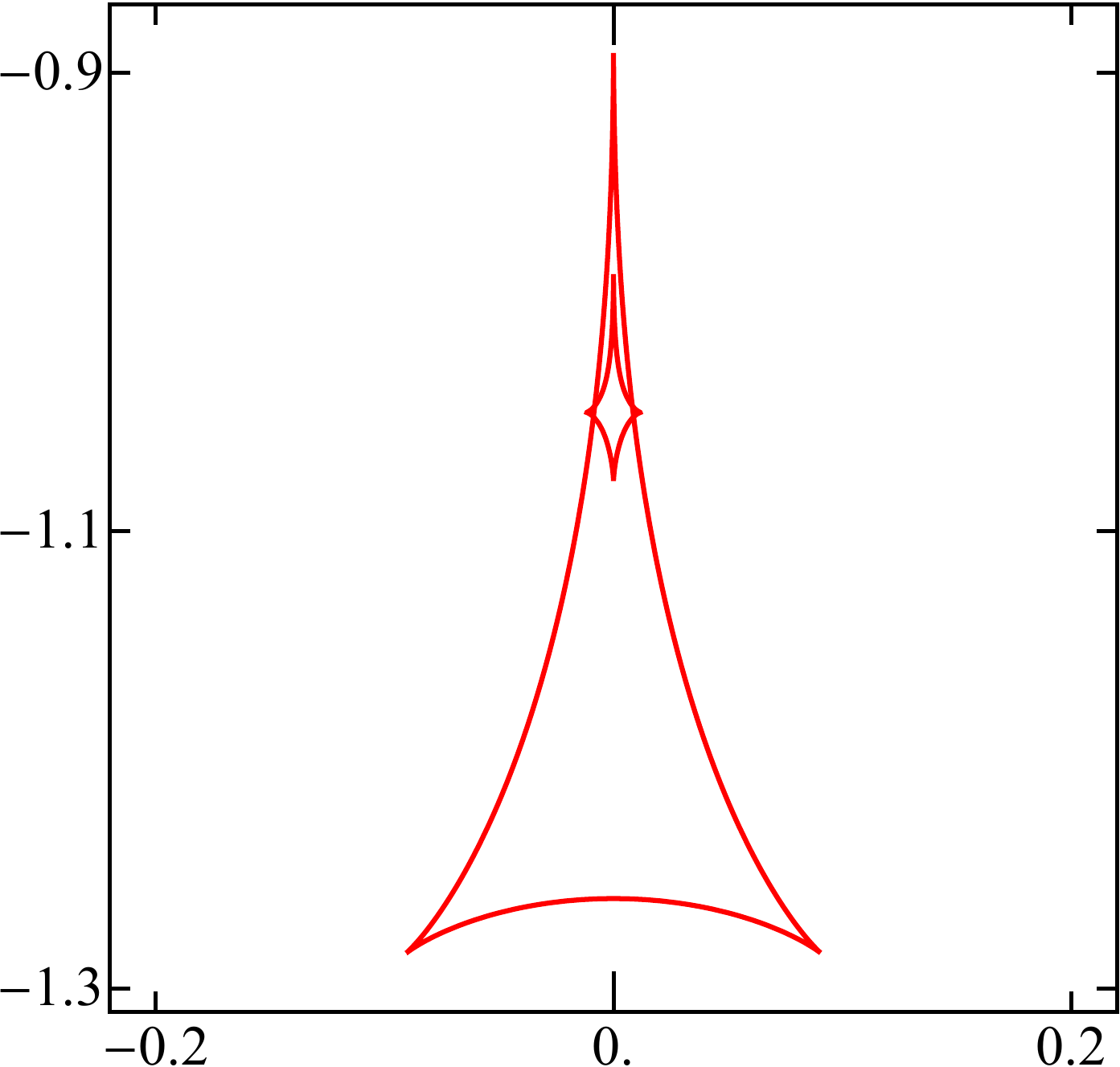}\\
\vspace{0.2cm}
\includegraphics[height=4.1cm]{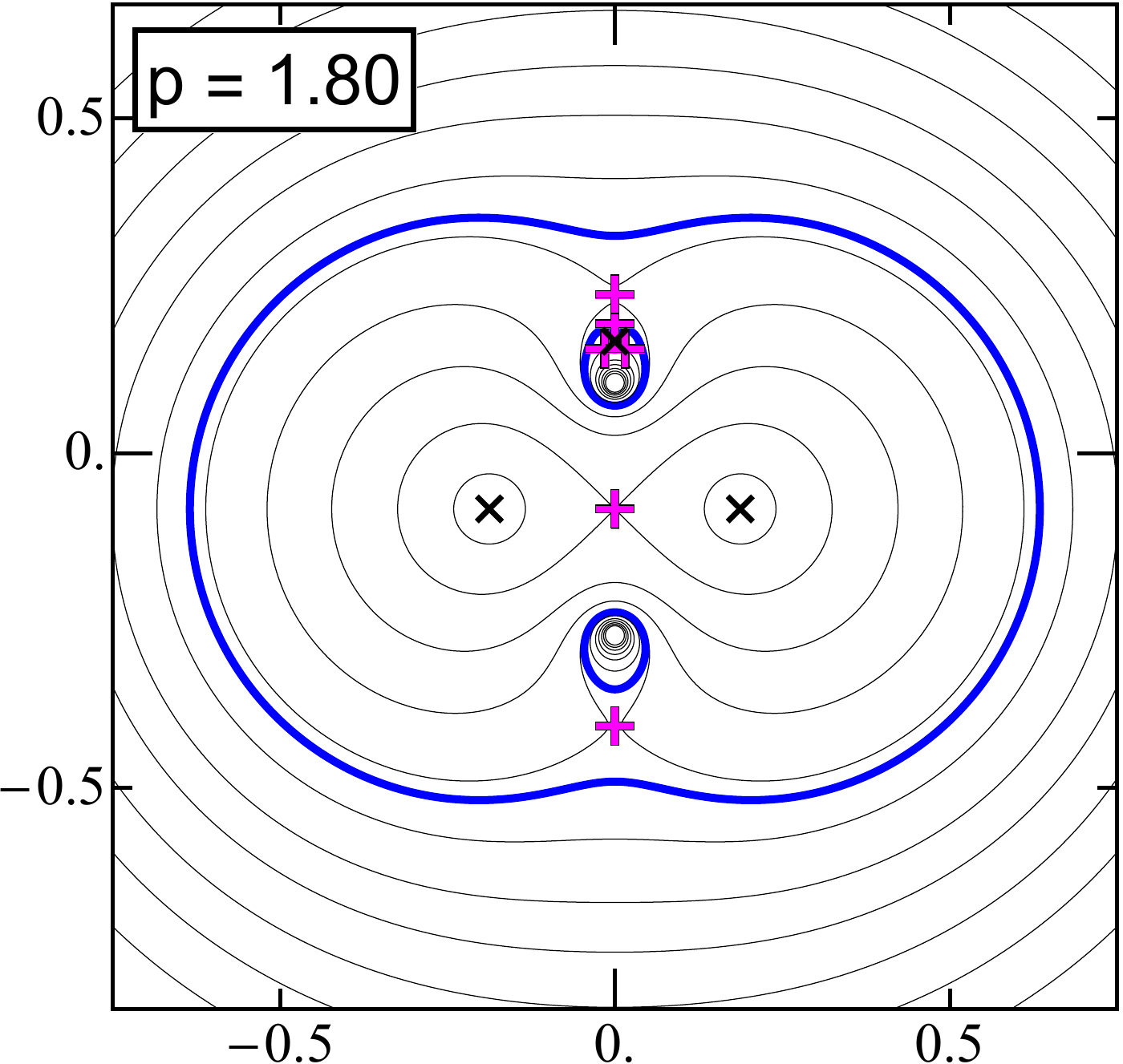}
\hspace{0.15cm}\includegraphics[height=4.1cm]{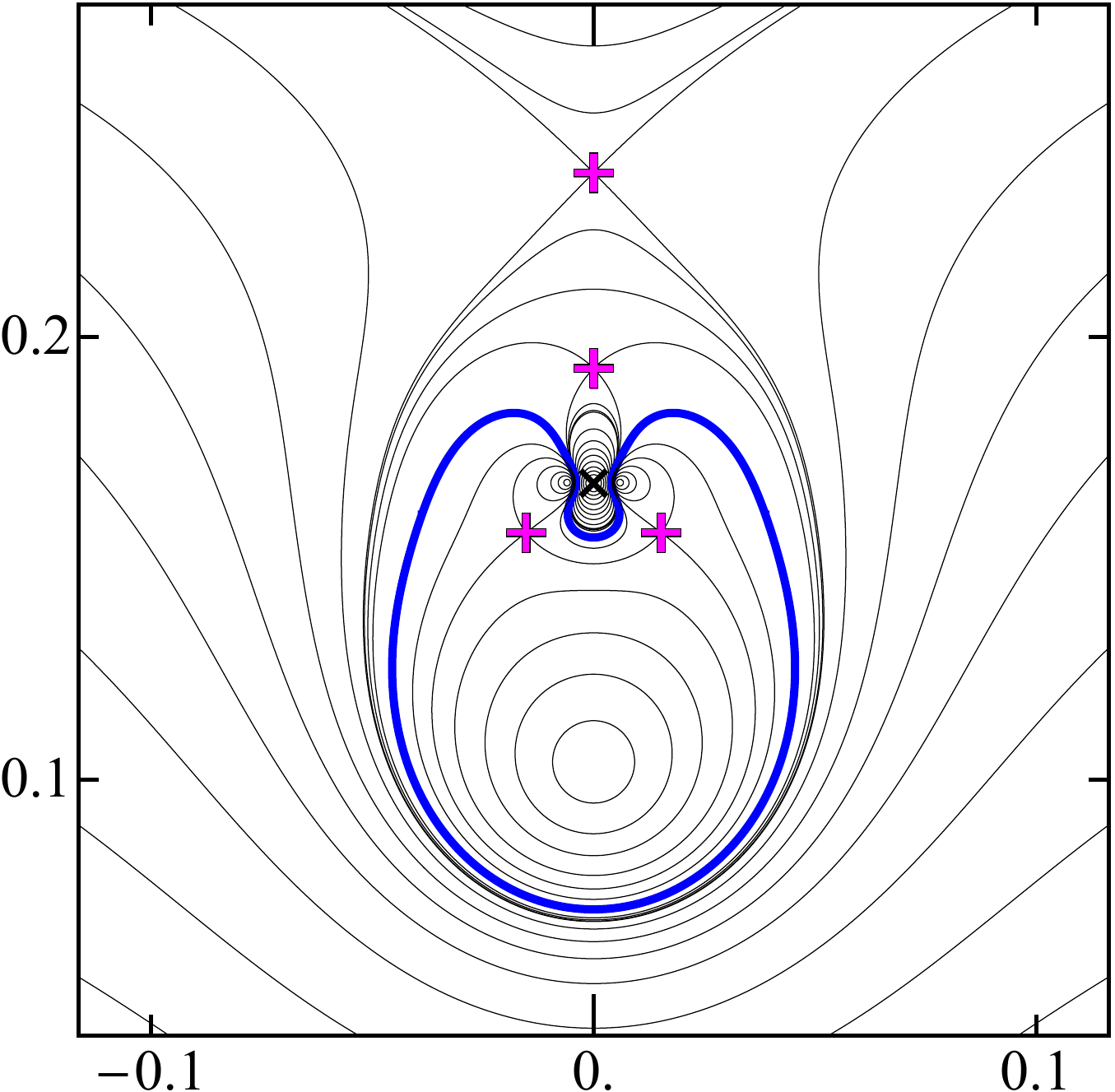}
\hspace{0.2cm}\includegraphics[height=4.1cm]{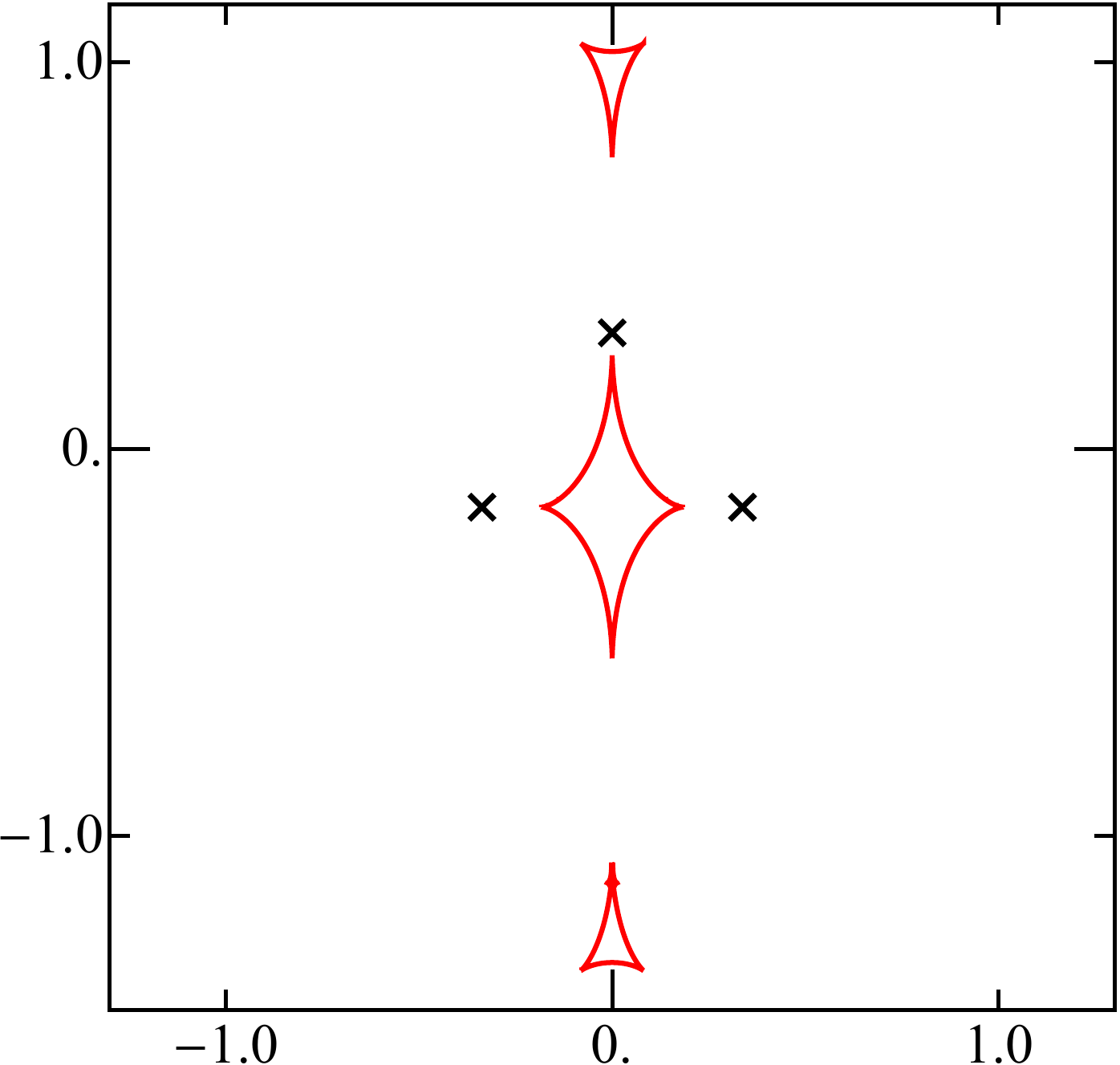}
\hspace{0.15cm}\includegraphics[height=4.1cm]{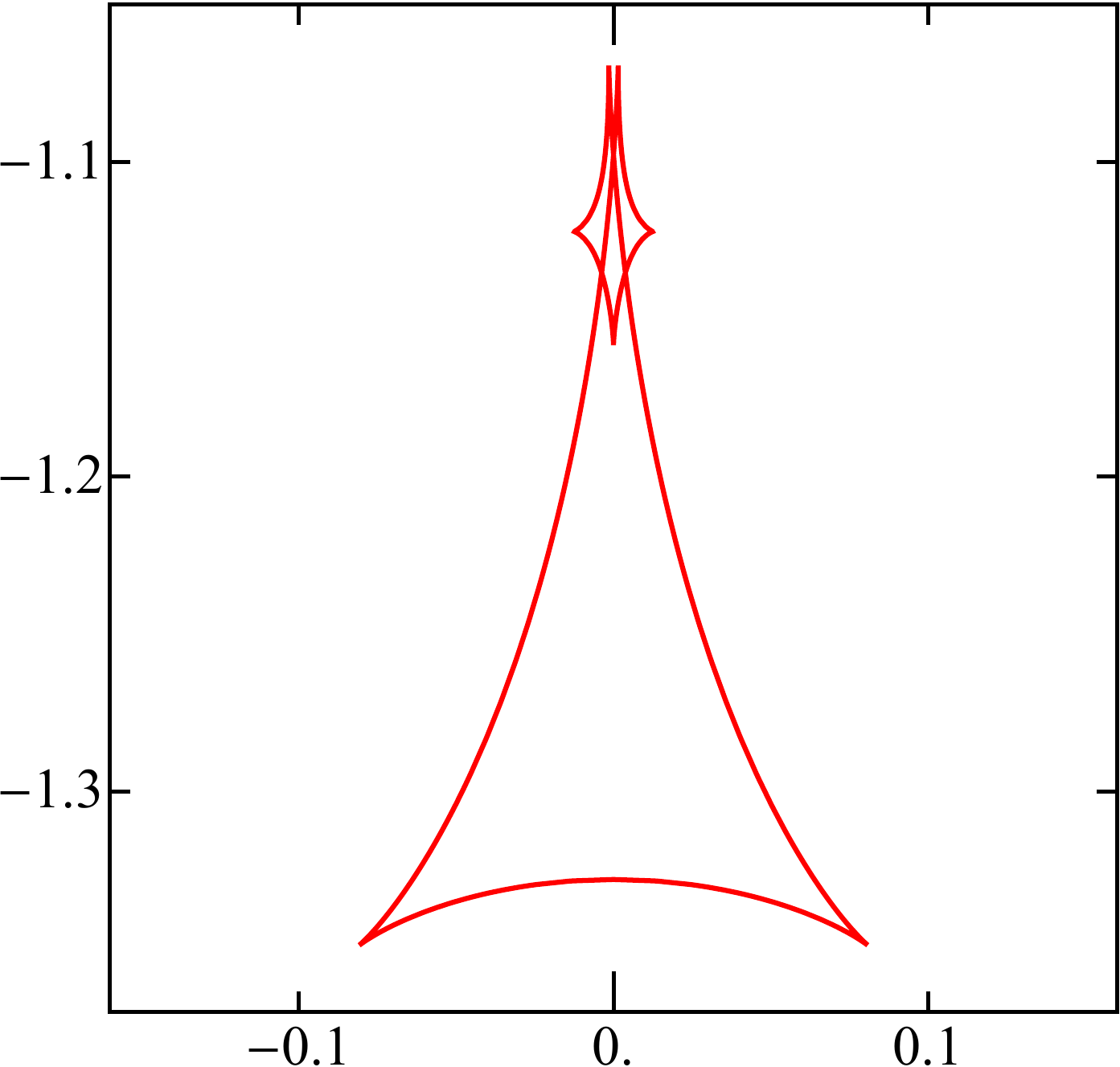}\\
\caption{Sequence of transitions leading to doubly nested critical-curve topologies T10 and T11. Same lens configuration as in Figure~\ref{fig:T11curves}, with perimeter varying from $p=1.80$ in the bottom row (topology T3), via $p=1.88$ (T10) and $p=1.89$ (T11), to $p=1.95$ in the top row (T7). Two left columns: critical curve global and detail; two right columns: caustic global and detail. Notation as in Figure~\ref{fig:EqMassSeq}.}}
\label{fig:T10T11sequence}
\efi

\clearpage
\begin{deluxetable}{ccccc}
\tabletypesize{\small}
\vspace{1cm}
\tablecaption{Probability $\mathcal{P}_{{\rm T}_i,p_{\rm max}}$ of Topology Occurrence in Studied Triples\label{tab:Volumes}}
\tablewidth{0pt}
\tablehead{\\[-0.2cm] \multicolumn{2}{c}{Topology} & Equal Masses & Planet in Binary & Hierarchical Masses \\
  & & $p_{\rm max}=6.711$ & $p_{\rm max}=5.640$ & $p_{\rm max}=4.799$ }
\startdata
\\[0.0cm]
 T1 &{\vspace{3mm}\includegraphics[height=1.11cm, bb= 15 -1 71 165, angle=90]{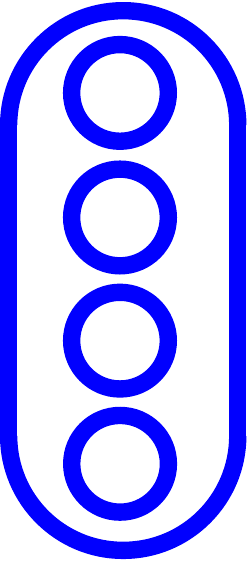}}&\hspace{5mm} 4.11e-3 & 1.87e-2 & 5.52e-2 \\
 T2 &{\vspace{3mm}\includegraphics[height=1.11cm, bb= 15 -1 71 165, angle=90]{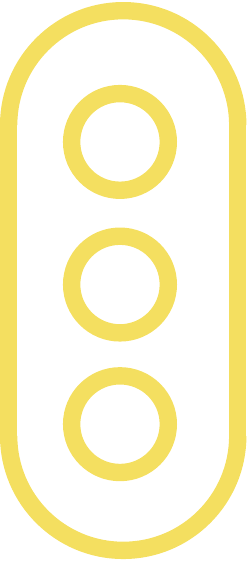}}&\hspace{5mm} 1.49e-3&3.88e-4& 2.39e-4 \\
 T3 &{\vspace{3mm}\includegraphics[height=1.11cm, bb= 15 -1 71 165, angle=90]{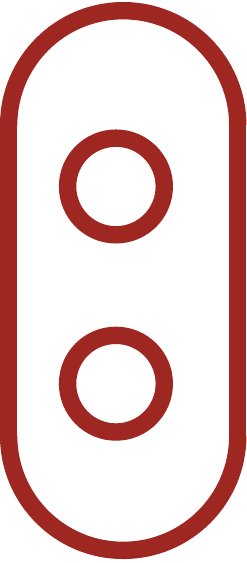}}&\hspace{5mm} 6.50e-2&1.32e-1&8.76e-2 \\
 T4 &{\vspace{3mm}\includegraphics[height=1.11cm, bb= 15 -1 71 165, angle=90]{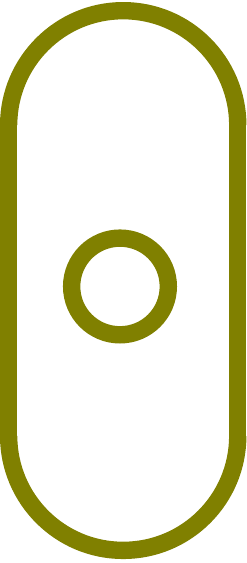}}&\hspace{5mm} 2.12e-2 & 6.64e-3 & 1.09e-3 \\
 T5 &{\vspace{3mm}\includegraphics[height=1.11cm, bb= 15 -1 71 165, angle=90]{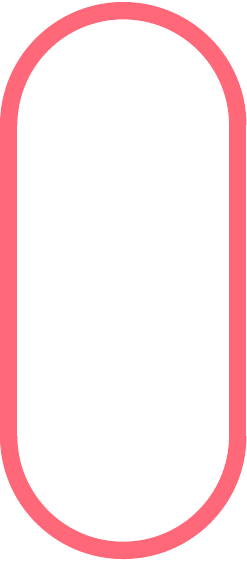}}&\hspace{5mm} 2.10e-1 & 2.95e-2 & 1.67e-2 \\
 T6 &{\vspace{3mm}\includegraphics[height=1.11cm, bb= 15 -1 71 165, angle=90]{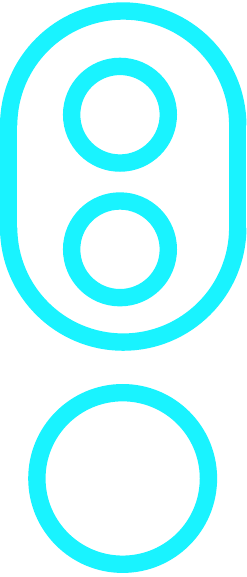}}&\hspace{5mm} 1.25e-1 & 2.18e-1 & 3.79e-1 \\
 T7 &{\vspace{3mm}\includegraphics[height=1.11cm, bb= 15 -1 71 165, angle=90]{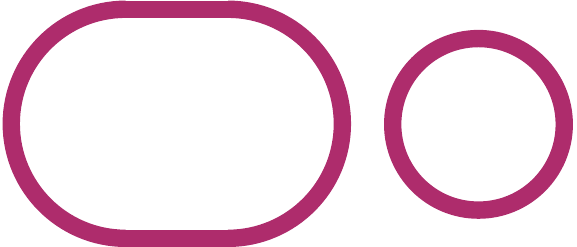}}&\hspace{5mm} 4.90e-1 & 4.54e-1 & 2.01e-1  \\
 T8 &{\vspace{3mm}\includegraphics[height=1.11cm, bb= 15 -1 71 165, angle=90]{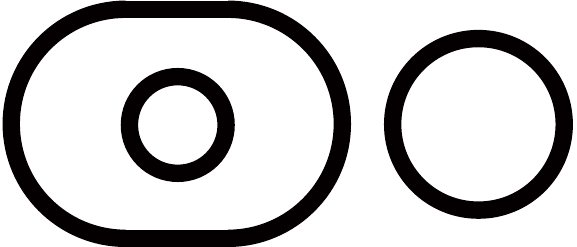}}&\hspace{5mm} 3.57e-3 & 2.57e-3 & 7.69e-4 \\
 T9 &{\vspace{3mm}\includegraphics[height=1.11cm, bb= 15 -1 71 165, angle=90]{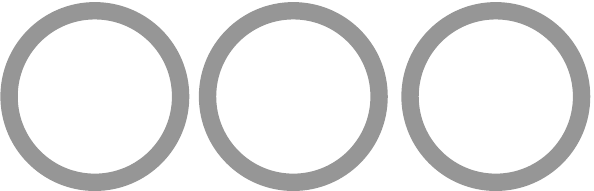}}&\hspace{5mm} 7.99e-2 & 1.38e-1 & 2.59e-1 \\
 T10 &{\vspace{3mm}\includegraphics[height=1.11cm, bb= 15 -1 71 165, angle=90]{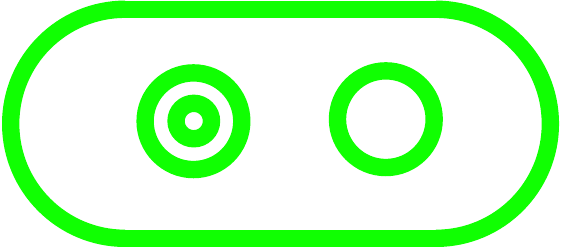}}&\hspace{5mm} N/A & 3.08e-5 & 1.66e-6  \\
 T11 &{\vspace{3mm}\includegraphics[height=1.11cm, bb= 15 -1 71 165, angle=90]{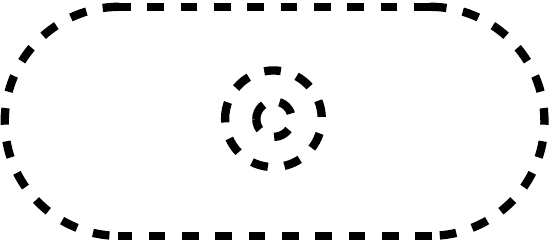}}&\hspace{5mm} N/A & $<$3e-6 & $<$1e-7 \\
\hline
\enddata
\vspace{-0.5cm}
\end{deluxetable}

\end{document}